\def\fileversion{v2.25} \def\filedate{2005/03/25}
\documentclass[fullpage,12pt]{uiucthesis}
\usepackage{graphicx}
\usepackage{amsmath}
\usepackage[numbers]{natbib}
\begin{document}

\title{A Theory for Market Impact: How Order Flow Affects Stock Price}
\author{Austin Nathaniel Gerig}
\department{Physics}
\schools{B.S., University of Notre Dame, 2000\\
         M.S., University of Illinois at Urbana-Champaign, 2002\\
         M.S., University of Illinois at Urbana-Champaign, 2006}
\phdthesis
\advisor{Alfred H\"{u}bler}
\degreeyear{2007}
\maketitle

\frontmatter

\begin{abstract}
It is known that the impact of transactions on stock price (market impact) is a concave function of the size of the order, but there exists little quantitative theory that suggests why this is so.  I develop a quantitative theory for the market impact of hidden orders (orders that reflect the true intention of buying and selling) that matches the empirically measured result and that reproduces some of the non-trivial and universal properties of stock returns (returns are percent changes in stock price).  The theory is based on a simple premise, that the stock market can be modeled in a mechanical way - as a device that translates order flow into an uncorrelated price stream.  Given that order flow is highly autocorrelated, this premise requires that market impact (1) depends on past order flow and (2) is asymmetric for buying and selling.  I derive the specific form for the dependence in (1) by assuming that current liquidity responds to information about all currently active hidden orders (liquidity is a measure of the price response to a transaction of a given size).  This produces an equation that suggests market impact should scale logarithmically with total order size.  Using data from the London Stock Exchange I empirically measure market impact and show that the result matches the theory.  Also using empirical data, I qualitatively specify the asymmetry of (2).  Putting all results together, I form a model for market impact that reproduces three universal properties of stock returns - that returns are uncorrelated, that returns are distributed with a power law tail, and that the magnitude of returns is highly autocorrelated (also known as clustered volatility).

\end{abstract}


\chapter*{Acknowledgments}

I express my deepest gratitude to the following people who have made this project possible.  To my research group at the University of Illinois, I thank current and previous group members for their support - this includes Colin Breen, Joe Brewer, Benny Brown, Peter Fleck, Glenn Foster, Vadas Gintautaus, Anne Hanna, Paul Melby, Davit Sivil, Chris Strelioff, Tim Wotherspoon, and Jian Xu.  To my fellow econophysics researchers at the Santa Fe Institute, thank you for your thoughtful comments on my work and for the stimulating discussions: Luwen Huang, Adlar Kim, Szabolcs Mike, Anuj Parwar, Michael Twardos, Javier Vicente, and Neda Zamani.

To my advisors, Alfred H\"{u}bler and Doyne Farmer, thank you for your guidance and for allowing me to join you in your quest - I only hope I have been faithful in completing what you started.  To my collaborator Fabrizio Lillo, thank you for contributing to this project and for all of your thoughtful suggestions along the way.

Finally, to my family.  To Mom and Dad, thank you for your constant support and for your example of wisdom.  To Tony, I have always looked to your lead in these matters - thank you for blazing the trail.  To Adam, thank you for showing me how to fully enjoy life.  Last and most importantly, to my wife Molly and son Elliot - thank you for enduring everything that comes with an undertaking such as this.  I love you dearly for it.

This work was funded in part by the National Science Foundation Grant Nos. NSF PHY 01-40179, NSF DMS 03-25939 ITR, NSF DGE 03-38215, and NSF SES 06-24351; and by a Graduate Fellowship from the Santa Fe Institute.


\tableofcontents
\setcounter{tocdepth}{2}
\listoftables
\listoffigures



\chapter{List of Symbols}

\begin{symbollist}[0.7in]
\item[$i$] Transaction time or index.
\item[$p_i$] Midpoint price.
\item[$\xi$] Tail exponent of cumulative distribution function.
\item[$\epsilon_i$] Transaction sign.
\item[$f(\cdot)$] Revealed order price impact function.
\item[$v_i$] Transaction volume.
\item[$\eta_i$] Uncorrelated noise.
\item[$\lambda_i$] Liquidity parameter.
\item[$\tilde{\lambda}_i$] Liquidity parameter.
\item[$G_0(\cdot)$] Bare impact propagator.
\item[$\Omega$] A set of publicly available or discernable historical financial data.
\item[$a_k$] Autoregressive coefficient at lag $k$.
\item[$H$] Hurst exponent.
\item[$\gamma$] Decay exponent for the autocorrelation function of $\epsilon_i$.
\item[$\phi$] Decay exponent of the bare impact propagator.
\item[$p_+(k)$] The probability of a transaction at time $(i+k)$ to have the same sign as $\epsilon_i$.
\item[$p_-(k)$] The probability of a transaction at time $(i+k)$ to have the opposite sign as $\epsilon_i$.
\item[$r_+(k)$] The expected return (measured in the direction of the transaction) of a transaction at time $(i+k)$ given it has the same sign as $\epsilon_i$.
\item[$r_-(k)$] The expected return (measured in the direction of the transaction) of a transaction at time $(i+k)$ given it has the same sign as $\epsilon_i$.
\item[$l_i$] The initial impact cause by a transaction at time $i$.  This is measured from directly before the transaction to directly after the transaction.
\item[$q_i$] The return cause by quote revisions at time $i$.  This is measured from directly after the transaction at time $i$ to directly before the transaction at time $(i+1)$.
\item[$l_+(k)$] The expected initial impact (measured in the direction of the transaction) of a transaction at time $(i+k)$ given it has the same sign as $\epsilon_i$.
\item[$l_-(k)$] The expected initial impact (measured in the direction of the transaction) of a transaction at time $(i+k)$ given it has the same sign as $\epsilon_i$.
\item[$I(T)$] The average total price response at time $(i+T)$ due to the transaction placed at time $i$.
\item[$I_N(T)$] Naively assuming that the magnitude of returns does not respond to the predictibility of transactions, the average total price response at time $(i+T)$ due to the transaction placed at time $i$.
\item[$I_L(T)$] Assuming that quote revisions can be approximated as noise, the average total price response at time $(i+T)$ due to the transaction placed at time $i$.
\item[$\alpha$] Exponent of hidden order size cumulative probability distribution.
\item[$\Omega'$] Information about all active hidden orders.
\item[$j$] Hidden order index at time $i$.
\item[$A_{i,j}$] Determines if hidden order $j$ is active at time $i$ ($A_{i,j}=1$ if active and $A_{i,j}=0$ if not).
\item[$\epsilon_j$] (Also $\epsilon$) Sign of hidden order.
\item[$N_j$] (Also $N$) Total size of hidden order (in number of pieces).
\item[$n_j$] (Also $n$) Numbered piece of hidden order.
\item[$V_j$] (Also $V$) Volume of hidden order.
\item[$v_j$] (Also $v$) Volume of revealed transactions of hidden order.
\item[$1/\theta_j$] (Also $1/\theta$) Probability per timestep of hidden order transacting.
\item[$t_j$] (Also $t$) Time when hidden order starts.
\item[$t(n_j)$] (Also $t(n)$) Time when piece $n_j$ of hidden order is transacted.
\item[$\hat{\epsilon}_i$] Predicted sign of transaction $i$ given information $\Omega$.
\item[$r^+_i$] Expected return of buyer initiated transaction.
\item[$r^-_i$] Expected return of seller initiated transaction.
\item[$F(\cdot)$] Hidden order price impact function.
\item[$\Psi$] The set of information $\{\epsilon, v, \theta, N\}$.
\item[$\Phi_k$] Binary information term, it determines if the hidden order that transacted at time $\left(i-k\right)$ has completed.
\item[$\hat{\epsilon}_i^{E1}$] The predicted transaction sign at time $i$ under (E1).
\item[$\hat{\epsilon}_i^{E2}$] The predicted transaction sign at time $i$ under (E1).
\item[$p_+^{E1}(k)$] The probability of a transaction at time $(i+k)$ to have the same sign as $\hat{\epsilon}_i^{E1}$.
\item[$p_-^{E1}(k)$] The probability of a transaction at time $(i+k)$ to have the opposite sign as $\hat{\epsilon}_i^{E1}$.
\item[$p_+^{E2}(k)$] The probability of a transaction at time $(i+k)$ to have the same sign as $\hat{\epsilon}_i^{E2}$.
\item[$p_-^{E2}(k)$] The probability of a transaction at time $(i+k)$ to have the opposite sign as $\hat{\epsilon}_i^{E2}$.
\item[$r_+^{E1}(k)$] The expected return (measured in the direction of the transaction) of a transaction at time $(i+k)$ given it has the same sign as $\hat{\epsilon}_i^{E1}$.
\item[$r_-^{E1}(k)$] The expected return (measured in the direction of the transaction) of a transaction at time $(i+k)$ given it has the opposite sign as $\hat{\epsilon}_i^{E1}$.
\item[$r_+^{E2}(k)$] The expected return (measured in the direction of the transaction) of a transaction at time $(i+k)$ given it has the same sign as $\hat{\epsilon}_i^{E2}$.
\item[$r_-^{E2}(k)$] The expected return (measured in the direction of the transaction) of a transaction at time $(i+k)$ given it has the opposite sign as $\hat{\epsilon}_i^{E2}$.

\end{symbollist}

\mainmatter

\chapter{Introduction\label{ch.intro}}

\section{Stock Price as a Stochastic Process}

In his thesis, \emph{The Theory of Speculation}, Louis Bachelier treated changes in stock price as a random variable\cite{Bachelier64}.  In so doing, he was the first person to mathematically model a random walk and was the first to notice that for simple random walks, the variance of fluctuations increases linearly with timescale.  His thesis was written in 1900 and predated Einstein's work on Brownian motion by several years\cite{Einstein05}.  Sixty years passed before the economics community took note of Bachelier's work and began in full earnest to treat stock prices as a stochastic process.  At this time, it was determined that prices move in percentage increments (called returns) rather than absolute increments, and these increments were assumed drawn independently and identically from a Gaussian distribution so that prices follow geometric Brownian motion.  Using this model for stock prices, the two most important theories in modern finance were developed, the Capital Asset Pricing Model (CAPM)\cite{Sharpe64, Lintner65} and the Black-Scholes-Merton Model (BSM)\cite{Black73,Merton73}.  These theories explained, respectively, the relative pricing of stocks and the absolute pricing of options on stocks.

Why is it that stock prices can be modeled as a stochastic process?  Without considering the fundamental cause, stock prices must approximate a martingale for a very simple but powerful reason - if the next value in a price series were predictable using historic values of the series, then this predictability would be exploited until it disappears (the act of exploiting diminishes the predictability and is called arbitrage in finance).  This result is related to the weak form of the efficient market hypothesis (EMH), which I discuss in more detail later and which I use extensively in this thesis.  

As for the cause of the random fluctuations in price, this question is a matter of much debate.  The standard answer given by economists is heavily influenced by neoclassical economic thought - that prices are always in equilibrium and are determined by the intersection of the aggregate supply and the aggregate demand of fully rational agents.  This forces a very specific interpretation for the randomness of stock prices:  because new and unexpected information arrives randomly and in random increments and because the market immediately incorporates this information into the stock price (through the updating of fully rational agents' supply and demand functions) then the randomness of stock prices is just a reflection of the randomness of new and unexpected information\cite{Fama70,Samuelson65}.  There are many problems with this interpretation.  First, the large size and large number of stock price movements seems unexplainable in terms of new information arrival\cite{Shiller81}  Second, that agents are fully rational (even if considered in aggregate rather than individually) is highly debated\cite{Shleifer00, Thaler05}.  Third, it is known that the act of trading causes price movements and there are many circumstances where a market participant will trade, and therefore influence the price, without introducing any new information about the value of the company.  

Towards the other extreme of agent rationality, there is another possible reason why stock prices can be modeled as a stochastic process.  If agents are assumed to act randomly and unintelligently and if their actions lead to price changes in a deterministic way, then random price changes result from the random actions of agents.  Models that simulate the exact structure of modern electronic markets, but that incorporate the random actions of agents with `zero intelligence' have been shown to reproduce many of the properties of stock prices\cite{Farmer05,Bouchaud02}.  These models suggest that the market is not an aggregator of information per se, but a mere translater that takes random order flow as input and outputs a price series that reflects this randomness.

As stated in the paper by Bouchaud\cite{Bouchaud05} when discussing these two extremes, ``of course, reality should lie somewhere in the middle''.  We shouldn't expect the stock price of a company to be entirely detached from the fundamental value of that company.  Alternatively, because the fundamental value of a company can be so difficult to quantify, it should be altogether irrelevant on the order of days, even weeks.

In this thesis, I will take seriously the suggestion that stock markets do not act as aggregators of information, but act to translate order flow into a price stream.\footnote{This can be reconciled with the standard interpretation of economists by considering order flow and information one and the same.  Many physicists have problems with this interpretation because they argue it is unfalsifiable.}  By doing so, the market can be viewed entirely as a mechanical system and any structure in the price stream that it outputs is entirely due to the mechanics of the market and the details of the order flow at the input.  We will see that the structure of the price series is not trivial - relative price increments, or returns, are not Gaussian; and the absolute sizes of returns are autocorrelated.  We will also see that order flow is anything but random, it is highly autocorrelated and therefore very predictable.  The way in which `the market' transforms this autocorrelated order flow into the observed price sequence, although complex and involving the strategic interaction of agents, can be formulated as a simple equation.  The result is that the interesting properties of the stock price series emerge when autocorrelated order flow is translated by the market.  To the extent that the properties of order flow are universal and derivable from first principles, the properties of stock prices are then also.\footnote{See the paper by Gabaix et al. for an example of such a theory\cite{Gabaix03}.  This was the first attempt by a group of physicists in collaboration with an economist to derive the universal properties of the stock market from first principles.  Their theory contradicts the theory I develop here and is at odds with the empirical evidence I present.}  This research is at the forefront of research in the field of econophysics and can be considered part of complex systems research where complex phenomena are explained with simple models.  It is square in the realm of physics, where the universal and fundamental principles of phenomena are sought.

\section{The Properties of Stock Returns}

Because stock prices tend to move in percentage terms, and not absolute terms, the relevant variable for the study of price movement is the return, or relative price increment.  If the $i^{th}$ return is labeled $r_i$, then it is calculated:
\begin{equation}
r_i=\delta p_i/p_i \approx log(p_{i+1})-log(p_{i}),\label{eq.returns}
\end{equation}
where $p_i$ is the midpoint price - midway between the highest price bid (best bid) and the lowest price offered (best ask) in the orderbook.\footnote{See the Appendix for an explanation of how modern stock markets operate and for an overview of the terminology.}  The approximation in Eq.~\ref{eq.returns} is $99.9\%$ accurate at short timescales (and approximately $99\%$ accurate on timescales of a day) - I will implicitly use it throughout this thesis.  I will measure returns at the timescale of transactions, and therefore $i$ will be considered a measure of \emph{transaction time}, updated by one increment whenever a transaction occurs.  

\begin{figure}[htb]
\centering
\includegraphics[width=4in]{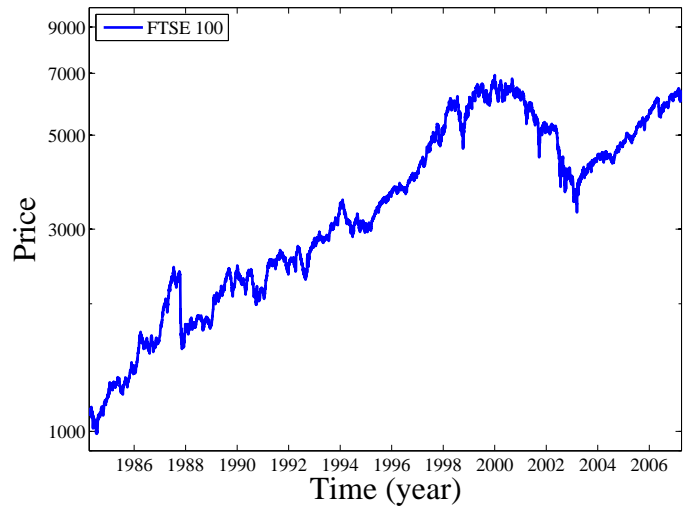}
\caption{Daily price of the FTSE 100 stock index from April, 1984 through March, 2007.  The log of the price appears to follow a random walk but with a dominant upward trend at long timescales.}
\label{fig.ftse_pricevstime}
\end{figure}
In Fig.~\ref{fig.ftse_pricevstime} I plot the daily price of the FTSE 100 stock index (a weighted aggregation of the price of 100 stocks traded on the London Stock Exchange) from April, 1984 through March, 2007 (this data was downloaded from from the website \verb|http://finance.yahoo.com/|).  As seen in the figure, the log price appears to follow a random walk but with a dominant upward trend at long timescales (this is why prices are often considered to follow geometric brownian motion with positive drift - below I show why this is incorrect).  The drift term is just the expected return of stocks - approximately $12\%$ per annum.  Because I will focus on micro-returns, returns measured on the order of minutes and seconds, I will not compensate for any long-term expected return.  Its contribution at such short timescales is negligible - on the order of $10^{-6}$ per minute.

With the advent of large financial datasets, physicists and economists have been able to document the specific properties of stock returns.  The uncorrelated nature of returns has been verified, the assumption of Gaussian distributed returns (and therefore brownian motion) has been shown incorrect\cite{Lux96, Longin96, Plerou99}, and the magnitude of returns has been found to be highly autocorrelated\cite{Lo91,Ding93}.  These properties are universal and hold across stocks and across stock markets - this by itself suggests a universal origin.  Below I show these properties for one of the stocks I will be studying.  In total, I will study 6 stocks traded on the London Stock Exchange from the period May 2, 2000 to December 31, 2002.  The dataset I use contains all on-book order flow for these securities, which represents roughly $60\%$ of all traded volume during this period.  For details about the data and for information about the 6 stocks used in this thesis, see the Appendix.

\subsection{Not Autocorrelated / Efficient\label{sec.efficient}}
\begin{figure}[!ht]
\centering
\includegraphics[width=4in]{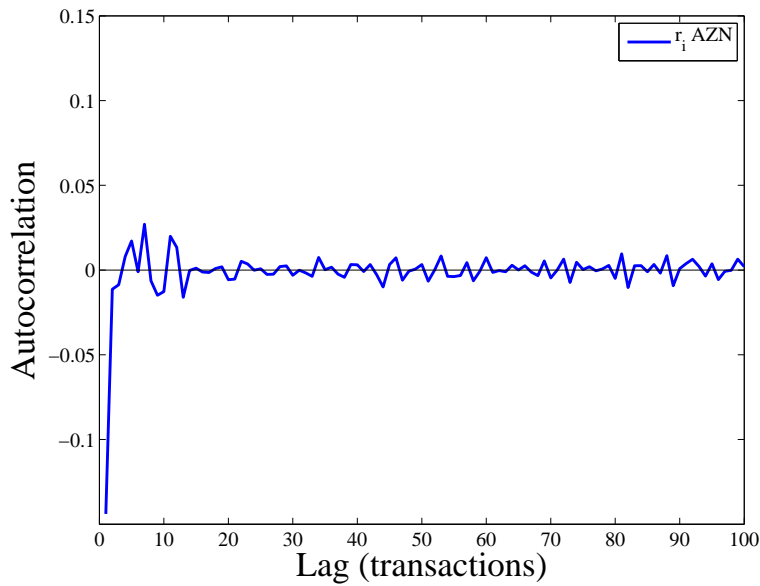}
\caption{The autocorrelation function of one-transaction returns for the stock AZN.  Returns do not show signs of autocorrelated structure, even at the timescale of transactions.}
\label{fig.AZN_ret_autocorr}
\end{figure}

In Fig.~\ref{fig.AZN_ret_autocorr} I plot the autocorrelation function of one-transaction returns, $r_i$, for the stock Astrazeneca (AZN).  
As seen in the plot, returns do not show signs of being autocorrelated, even at the timescale of a few transactions!

\subsection{Distributed with Power Law Tails\label{sec.power_tails}}
\begin{figure}[!ht]
\centering
\includegraphics[width=4in]{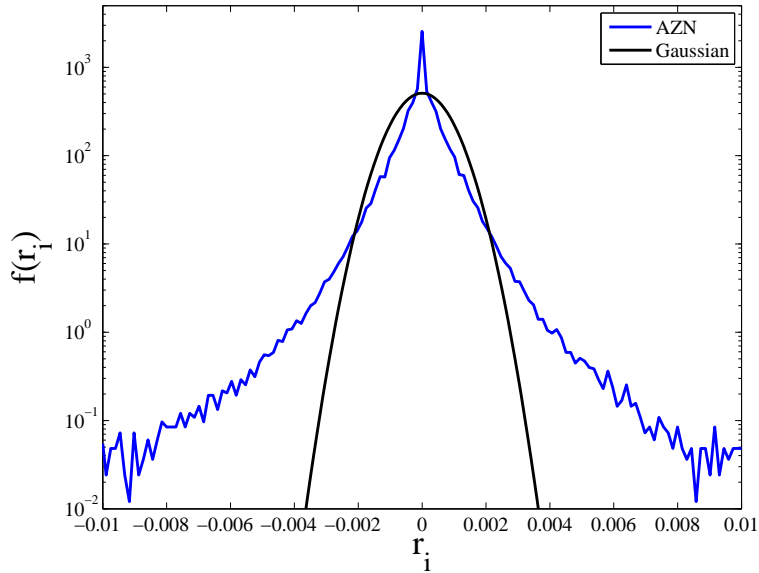}
\caption{The probability density function of one-transaction returns for the stock AZN.  A Gaussian distribution with the same variance is plotted for comparison.  The assumption that stock returns follow geometric Brownian motion is incorrect.}
\label{fig.AZN_ret_pdf}
\end{figure}

\begin{figure}[!ht]
\centering
\includegraphics[width=4in]{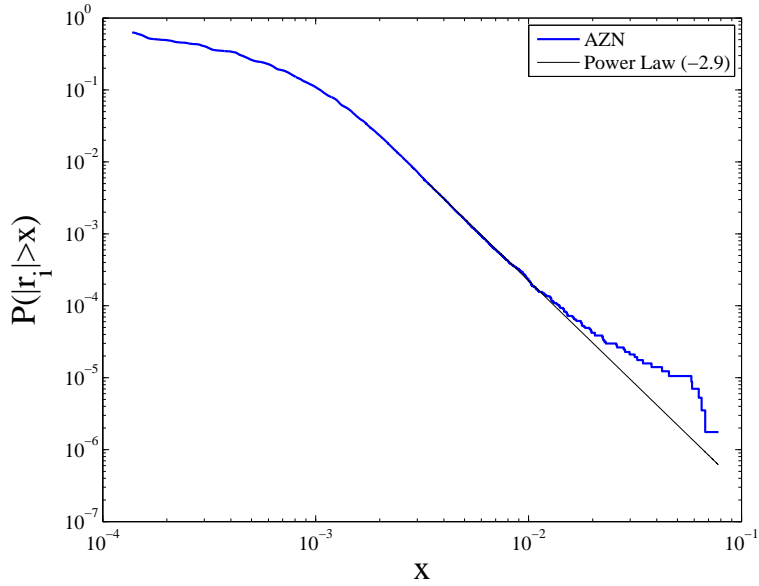}
\caption{The cumulative distribution function of one-transaction returns for the stock AZN.  The tail of this distribution decays as a power law with exponent $-2.9$.}
\label{fig.AZN_ret_ecdf}
\end{figure}

\begin{figure}[t]
\centering
\includegraphics[width=4in]{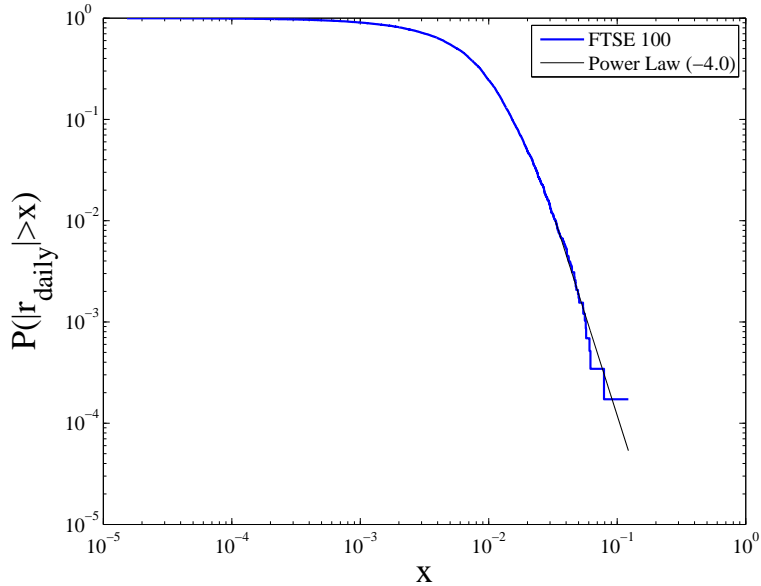}
\caption{The cumulative distribution function of daily returns for the FTSE 100 index.  The tail of this distribution decays as a power law with exponent $-4.0$.  The stock market crash of 1987 is the last point on this plot and serves as an example of the likelihood of large events for distributions with power law tails.}
\label{fig.ftse_ret_ecdf}
\end{figure}

In Figs.~\ref{fig.AZN_ret_pdf} and \ref{fig.AZN_ret_ecdf} I plot the probability density function of one-transaction returns for AZN on semilog scale and the cumulative distribution function on log-log scale.
For comparison I also plot a Gaussian distribution with the same variance in Fig.~\ref{fig.AZN_ret_pdf}.  As seen in the plots, returns are not distributed as a Gaussian.  Because the cumulative distribution is straight on log-log scale, this suggests the distribution has a power law tail, i.e., that
\begin{equation}
\lim_{x\rightarrow \infty} f(x)g(x) = K x^{-\xi},
\end{equation}
where $g(x)$ is some slowly varying function, $K$ is a positive constant, and $\xi$ is called the tail exponent.  For AZN, the tail exponent is measured $2.9$\footnote{I use the Hill estimator\cite{Hill75} to determine tail exponents throughout this thesis.  This is formulated as $\hat{\xi} = 1+n/\sum_{i=1}^n\log(x_i/x_{min})$, where $n$ is the number of observations $x_i\geq x_{min}$.  I usually set $n$ such that $0.5\%$ to $1\%$ of the data is included in the estimate.}.  To show that this behavior is not limited to micro-returns, I plot in Fig.~\ref{fig.ftse_ret_ecdf} the cumulative distribution function of the daily returns for the FTSE 100 index.  The tail exponent is measured $4.0$.
That return distributions have power law tails has severe consequences for market participants.  The stock market crash of 1987, which was a one-day return of $-12\%$ for the FTSE 100, was a $12\sigma$ event under the assumption of Gaussian distributed daily returns - an impossibility (probability $10^{-29}$).  Given that returns are power law distributed and using the observed exponent of $4.0$, such an event occurs with probability $5\times10^{-5}$, or on average about once every $70$ years.  Looking at the cumulative distribution, this event is the last point and does not appear to be an outlier - it is expected given the distribution.

\subsection{Clustered Volatility\label{sec.clust_vol}}
\begin{figure}[!ht]
\centering
\includegraphics[width=4in]{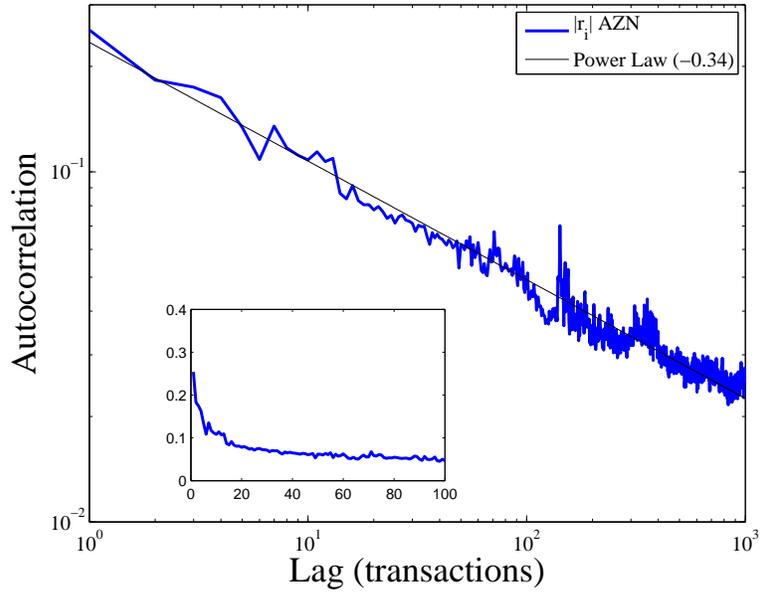}
\caption{Autocorrelation function of the magnitude of one-transaction returns for the stock AZN.  The autocorrelation function decays as a power law with exponent $-0.34$.}
\label{fig.AZN_absret_autocorr}
\end{figure}

\begin{figure}[!ht]
\centering
\includegraphics[width=4in]{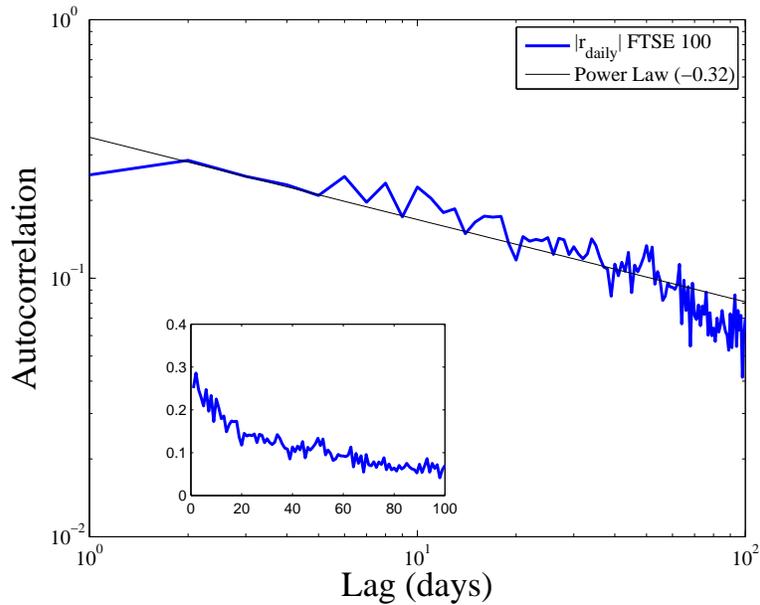}
\caption{Autocorrelation function of the magnitude of daily returns for the FTSE 100 index.  The autocorrelation function decays as a power law with exponent $-0.32$.  This shows volatility clustering is not just a property of micro-returns.}
\label{fig.ftse_absret_autocorr}
\end{figure}

In Fig.~\ref{fig.AZN_absret_autocorr}, I plot the autocorrelation function of the magnitude of one-transaction returns for the stock AZN.  This is plotted both in regular scale and in log-log scale. Notice that the autocorrelation function decays as a power law, the exponent is measured as $-0.34$.  Again, to show that this is not just a property of micro-returns, I plot in Fig.~\ref{fig.ftse_absret_autocorr} the autocorrelation function of the magnitude of daily returns for the FTSE 100 index.  The exponent is measured $-0.32$.

\begin{figure}[!ht]
\centering
\includegraphics[width=4in]{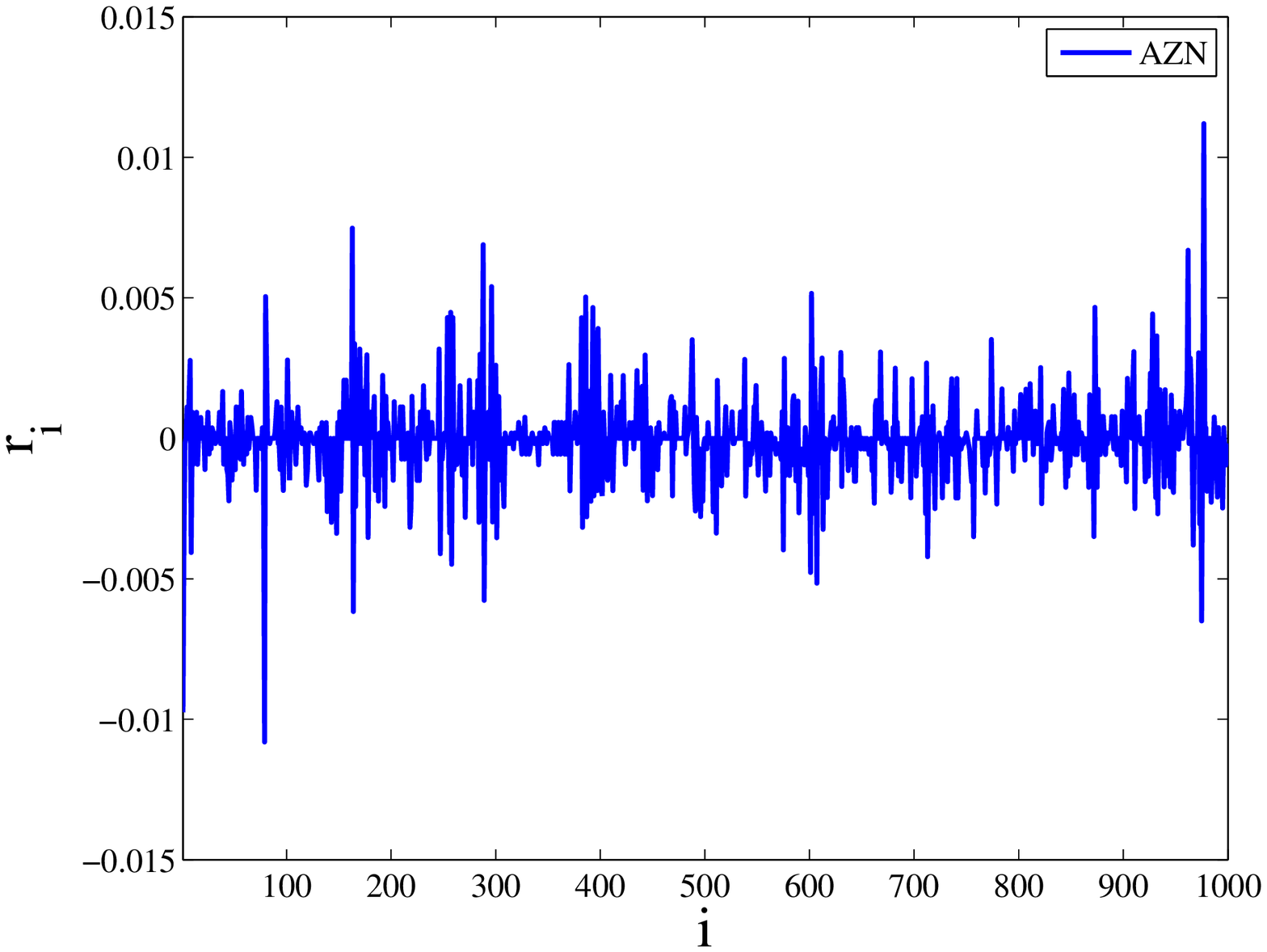}
\caption{One-transaction returns as a function of transaction time from 1 to 1000 for the stock AZN.  The high degree of autocorrelation in the magnitude of returns leads to the clustering of volatility in time.}
\label{fig.AZN_clusteredvol}
\end{figure}

This property of returns has been called `clustered volatility', where volatility is defined as the standard deviation of returns, i.e, the expected value of the magnitude of returns.  It is `clustered' because of the autocorrelation function - small values of volatility tend to be followed by small values, and large values of volatility tend to be followed by large values.  This clustering can be seen in Fig.~\ref{fig.AZN_clusteredvol}, where I've plotted the returns of the first 1000 transactions for AZN from my dataset.

\section{A Simple Model for Stock Returns}
This thesis has three goals.  First, to determine how order flow is transformed into a price series.  Second, to develop a quantitative theory for the market impact of individual participants' order flow.  Third, to show that the statistical properties of stock returns (shown in the previous section) result from the market's action on order flow.  To begin this process, I present below a simple model for stock returns,
\begin{equation}
r_i = \epsilon_i f(v_i) + \eta_i.\label{eq.simple_model}
\end{equation}
This equation connects order flow information, specifically transaction information, to the price return.  Here, $i$ indexes transactions and is updated by one increment whenever a transaction occurs - as stated before, this is also known as \emph{transaction time}.  A transaction occurs whenever two market participants decide on an agreeable price to exchange shares of stock for cash.  Every transaction that occurs on an electronic order book exchange (which describes the data I am analyzing) is initiated by one of the parties.  One party has placed a limit order - an order that specifies the minimum/maximum agreeable price for selling/buying a specified quantity of stock.  Another party finds this price agreeable and decides to initiate the transaction \dots this requires that there are no other limit orders at a more agreeable price or that specify the same price and were placed earlier; otherwise the initiator transacts with the other limit order.  In Eq.~\ref{eq.simple_model}, $\epsilon_i$ is called the sign of the transaction and is $+1$ if initiated by a buyer and is $-1$ if initiated by a seller.  $v_i$ is the volume (or quantity) transacted and is measured in GBP ($\pounds$).  The function $f(\cdot)$ is called the price impact function - an empirically determined function that specifies the expected price impact, or response, due to a transaction of size $v_i$.  $\eta_i$ is an uncorrelated noise term that models the specific details of the orderbook at time $i$ that are assumed unimportant for the modeling of returns.

The motivation for this model is the following.  Given the structure and rules of the orderbook, we expect that buy (sell) transactions will tend to cause positive (negative) returns - this is represented by $\epsilon_i$.  We also expect that transactions with larger size should have a larger impact on the price.  In the model this is represented by $f(\cdot)$, which is some monotonically increasing function, empirically determined and as yet undefined.  Unlike other `zero intelligence' models, which include complete order flow information containing limit order placements and cancellations as well as transactions\cite{Farmer05,Mike05,Hopman02}, all order flow is condensed into transaction information.  We should not expect, then, that this model will generate all of the interesting features of price formation.  It will be surprising how well this model, slightly modified, reproduces stock price dynamics.  

That the model might reproduce the properties of the return series is surprising for another reason.  If price is determined in the minds of market participants regardless of what transactions are occuring, then transactions should have no long term impact on price.  Under this scenario, participants tend to cluster limit orders around a collectively agreed price - transactions might impact this price, but as limit orders are placed around the original reference point, the impact disappears.  The model suggests the opposite, that transactions impact the price permanently and that this impact can be measured exactly as shown in Eq.~\ref{eq.simple_model}.  That market participants accept the updated price due to a transaction as their new reference price (as suggested by the model) is a strong hypothesis - this will be tested towards the end of Chapter~\ref{ch.long_mem}.

Eq.~\ref{eq.simple_model} is in the tradition of market impact models found in the market microstructure\footnote{Market microstructure is a field within finance that attempts to connect the mechanics of trading to price formation.} literature\cite{Kyle85,Hasbrouck91,dufour99}.  These models, although phenomenologically identical to Eq.~\ref{eq.simple_model}, are framed in such a way that permanent impact results from the informational content of trades and not ipso facto.  Many papers in this literature find empirically that $f(\cdot)$ is a concave and monotonically increasing function with volume\cite{Hasbrouck91,Algert90,Kempf99}, but theoretically they suggest $f(\cdot)$ should be a linear function with intercept zero\cite{Kyle85,Huberman04g} - this results because a linear function is the only function that does not allow arbitrage opportunities.  The disconnect results from the malformed assumption that price impact is independent of past order flow - I will show that it is dependent on past order flow, this will show up immediately when testing Eq.~\ref{eq.simple_model} with real data and is the subject of the next chapter.  That liquidity - the price response to a transaction of a given size - is dependent on past order flow was anticipated, although unformalized, by Hasbrouck when he suggested that it is the \emph{unanticipated} component of order flow that impacts the price\cite{Hasbrouck88b,Hasbrouck91}.  The modifications to Eq.~\ref{eq.simple_model} that will be presented in Chapter~\ref{ch.long_mem} can be interpreted in this way.  Finally, I should mention two confusing parts of this literature.  First, sometimes $f(\cdot)$ is measured for individual transactions, and other times it is measured for a collection of transactions all part of the same large order.  These functions are not the same and I notate the latter as $F(\cdot)$ when discussing it in this thesis.  Second, throughout the market microstructure literature, it is assumed that order flow impact consists of a permanent and transient component - the former is said to be the information content of the trade and the latter is due to market `friction' (the cost of transacting in the market).  It is ambiguous if these components should be measured from the unconditional impact of a transaction, or if they should be measured when impact is conditioned on the specific order flow of a market participant.  As I show later in this thesis, the unconditional expectation of an impact remains permanent, and it is only when conditioned on very specific things that the impact appears transient.\footnote{It turns out that the conditioning variable that makes impacts (partially) transient is quite meaningful.  The condition is that the market participant who initiated that trade is no longer active in the market.  This means that in the eyes of a market participant, her impact decays when she stops trading - to everyone else this information is fuzzy and therefore all impacts appear as permanent (in one state of the world the participant continues to trade and the impact grows, in another state she stops and the impact decays \dots this averages to an overall permanent impact).}


\subsection{Results of the Model}

\begin{figure}[htb]
\centering
\includegraphics[width=4in]{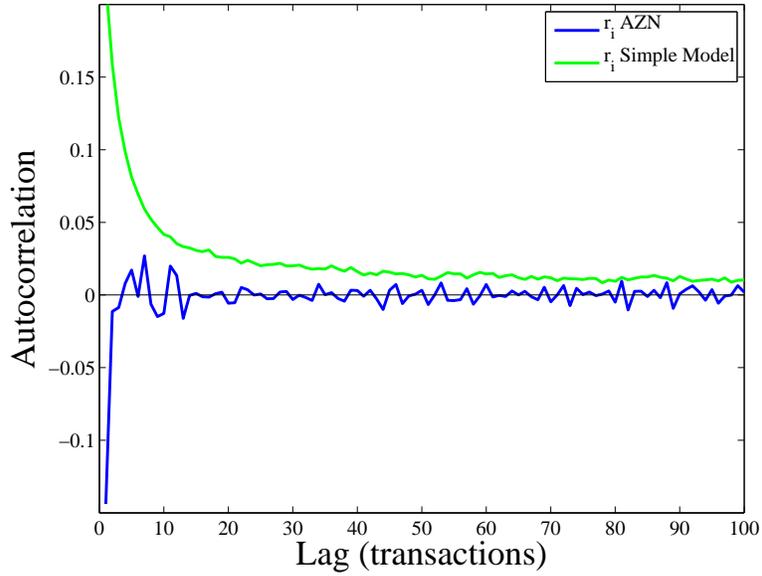}
\caption{The autocorrelation function of one-transaction returns generated by the simple model of Eq.~\ref{eq.simple_model}.  The result is compared to the autocorrelation function of empirical returns.  The model produces returns that are correlated, whereas the empirical returns are not.}
\label{fig.AZN_simple_ret_autocorr}
\end{figure}

In Fig.~\ref{fig.AZN_simple_ret_autocorr}, I plot the autocorrelation function for returns produced by the model for the stock AZN.
As seen in the figure, the model fails because it produces returns that are highly autocorrelated.  In the next chapter, I show that this results from the autocorrelation of transaction sign, $\epsilon_i$.  That $\epsilon_i$ is autocorrelated has been mentioned in the finance literature for quite some time\cite{Hasbrouck91,hasbrouck91b}, but the extent of this autocorrelation was grossly underestimated until discovered by two groups of physicists\cite{Lillo03c,Bouchaud04}.  I will discuss the results of these papers in detail in the next chapter.

\chapter{Long Memory of Supply and Demand \label{ch.long_mem}}

\section{Introduction}
\begin{figure}[htb]
\centering
\includegraphics[width=4in]{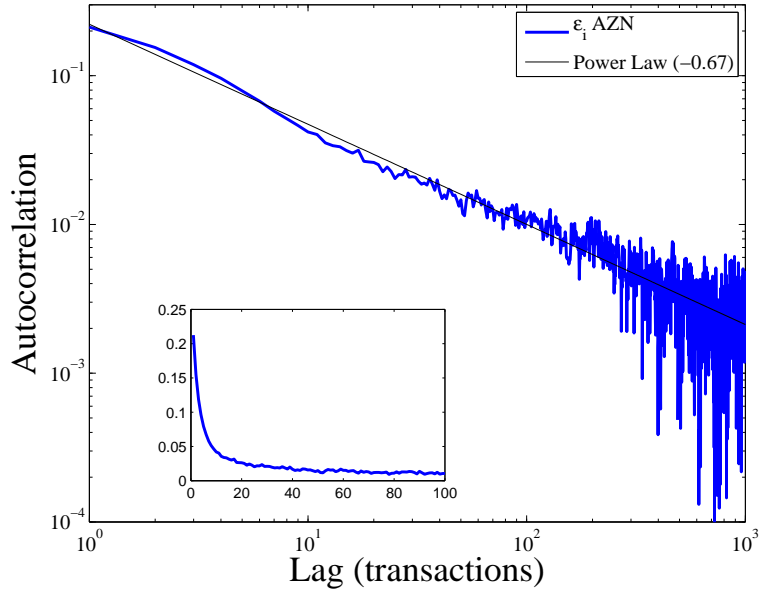}
\caption{Autocorrelation function of the transaction sign series $\epsilon_i$ for the stock AZN.  The autocorrelation function decays as a power law with exponent $-0.67$. }
\label{fig.AZN_sign_autocorr}
\end{figure}

The simple model for stock returns presented in Chapter~\ref{ch.intro} was,
\begin{equation}
r_i = \epsilon_i f(v_i) + \eta_i.
\label{eq.simple_mod}
\end{equation}
When testing this model with empirical data, the return was found to be autocorrelated.  This results because the sign of transactions, $\epsilon_i$, is highly autocorrelated.  In Fig.~\ref{fig.AZN_sign_autocorr}, I plot the autocorrelation function of $\epsilon_i$ for the stock AZN.

The autocorrelation function decays as a power law with an exponent of $-.67$, making the curve unintegrable.  Series that exhibit this property are said to have long memory - this is because values from the distant past can significantly affect present values of the series.  The long memory of $\epsilon_i$ appears to be universal, it has been verified for stocks traded on the London Stock Exchange, the Paris Bourse, and the New York Stock Exchange\cite{Lillo03c,Bouchaud04}.  It is called ``The Long Memory of Supply and Demand''\cite{Lillo05b,Farmer06} because it implies that fluctuations in individual supply functions (leading to seller initiated transactions) and demand functions (leading to buyer initiated transactions) also exhibit long memory.  There is an interesting consequence of this \dots buying and selling can be highly predictable and remains so for long periods of time.  Taking AZN as an example, using a simple autoregressive model for $\epsilon_i$ there are periods of time when $\epsilon_i$ can be determined with $80\%$ accuracy.  For some stocks, predictability remains above $50\%$ for several weeks!

The predictability of $\epsilon_i$ leads to a very intriguing puzzle.  If buying tends to push the price up and selling tends to push the price down, and we know that buying and selling are highly autocorrelated (and therefore predictable) - how is it that prices remain uncorrelated and unpredictable?  There were two papers published in 2004 that independently discovered the long memory of $\epsilon_i$, Lillo and Farmer \cite{Lillo03c} (hereafter known as LF) and Bouchaud, Gefen, Potters, and Wyart \cite{Bouchaud04} (hereafter known as BGPW).  These two groups posit two different answers to this puzzle.  Because the uncorrelated and unpredictable nature of asset prices is also called \emph{market efficiency}, I will call this puzzle the \emph{efficiency puzzle}.

\section{Review}
\subsection{Lillo and Farmer (LF)}
LF suggest that the efficiency puzzle is explained by \emph{permanent} price impacts that \emph{fluctuate} in size.  These size fluctuations depend on liquidity, where liquidity is defined as the price response to a transaction of a given size.  Their model is presented as a modification of Eq.~\ref{eq.simple_mod},
\begin{equation}
r_i = \frac{\epsilon_i f(v_i)}{\lambda_i} + \eta_i.
\end{equation}
where $\epsilon_i f(v_i)$ is divided by a liquidity parameter $\lambda_i$.  LF state that $\lambda_i$ is different for buyer initiated vs. seller initiated transactions, i.e., is dependent on $\epsilon_i$.  The idea is that when buying or selling is predictable, the $\lambda_i$ term increases or decreases the price response of buying and selling such that returns remain zero on average.  For example, if it is highly likely a buyer initiated transaction will occur, i.e., $\epsilon_i=1$, then $\lambda_i$ is increased for $\epsilon_i=1$ and decreased for $\epsilon_i=-1$, such that the expected return is zero. 

The exact form of $\lambda_i$ is not discussed in LF, but they show plots suggesting the liquidity term is acting in the correct direction when conditioned on the predictability of $\epsilon_i$ (damping buys when buys are predictable, etc.).

\subsection{Bouchaud, Gefen, Potters, and Wyart (BGPW)}
BGPW suggest that the efficiency puzzle is explained by \emph{decaying} price impacts with \emph{fixed} size.  They state that impacts are originally fixed in size at $\epsilon_i f(v_i)$ (on average) but vary in time with the propagator $G_0(\tau)$, also called a bare impact function.  Here, $\tau$ is the time since transaction $i$ occurred and is measured in transaction time (transaction time is updated by one whenever a transaction occurs).  The total price impact measured at the time of transaction $i$ is,
\begin{equation}
r_i= \epsilon_i f(v_i) - \sum_{k>0} \frac{G_0(k+1)-G_0(k)}{G_0(1)} \epsilon_{i-k} f(v_{i-k}) + \eta_i,
\end{equation}
with every transaction into the infinite past contributing (Note: I have slightly modified their notation).  They find that $G_0(\tau)$ decays as a power law, and is tuned just-so such that the autocorrelation of $\epsilon_i$ is completely canceled and returns remain unpredictable.  As stated in BGPW: ``Therefore, the seemingly trivial diffusive behavior of price changes in fact results from a fine-tuned competition between opposite effects, one leading to super-diffusion (the autocorrelation of trades) and the other leading to sub-diffusion (the decay of the bare impact function).''

BGPW state that the decay of $G_0(\tau)$ is caused by mean-reverting (in price) orderflow of \emph{liquidity providers} or \emph{market makers}.  These are market participants who are almost always willing to buy or to sell but do not want to take an overall position.  They place limit orders and make money by the tendency of limit orders to buy low and sell high.  Although market makers can influence the impact of a transaction both through liquidity and by changing the prevailing quotes (best bid and best offer prices), BGPW emphasize the second option and say it is the mean reversion of quotes that causes the decay of $G_0(\tau)$.  In a later paper by Bouchaud, Kockelkoren, and Potters \cite{Bouchaud04b} this interpretation is changed such that market makers use both liquidity and quote revision to influence the price, and a combination of the two causes $G_0(\tau)$ to decay - the exact ratio of the two is unspecified.


We can represent the BGPW model in a more general form,
\begin{equation}
r_i = \epsilon_i f(v_i) - \tilde{\lambda}_i + \eta_i,
\end{equation}
where BGPW posit that,
\begin{equation}
\tilde{\lambda}_i = \sum_{k>0} \frac{G_0(k+1)-G_0(k)}{G_0(1)} \epsilon_{i-k} f(v_{i-k}).
\end{equation}

\section{Equivalence of the Models}
We have two competing models for returns, 
\begin{eqnarray}
r_i & = & \frac{\epsilon_i f(v_i)}{\lambda_i}  + \eta_i,\label{eq.lillo}\\
r_i & = & \epsilon_i f(v_i) - \tilde{\lambda}_i  + \eta_i \label{eq.bouchaud}.
\end{eqnarray}
The first was proposed by LF and the second by BGPW.  These two models are equivalent with a simple change of variable,
\begin{equation}
\frac{1}{\lambda_i} \equiv 1 - \frac{\tilde{\lambda}_i}{\epsilon_i f(v_i)},
\label{eq.change_var_lamb}
\end{equation}
and any differences between BGPW and LF are entirely in the assumed structure of the $\tilde{\lambda}_i$ or $\lambda_i$ term and in the physical interpretation of the model.

LF do not posit a form for $\lambda_i$ but state it is a liquidity term (the price response to a transaction of a given size).  BGPW posit a specific form for $\tilde{\lambda}_i$,
\begin{equation}
\tilde{\lambda}_i = \sum_{k>0} \frac{G_0(k+1)-G_0(k)}{G_0(1)} \epsilon_{i-k} f(v_{i-k}).
\label{eq.bouch_tlamb}
\end{equation}
Because their interpretation is based on a propagator $G_0(\tau)$, this result is quite natural.  Given that the general form of the BGPW and LF models are identical, however, this can also be interpreted as a model for the variable $\lambda_i$.  Specifically, this is,
\begin{equation}
\frac{1}{\lambda_i} = 1 - \frac{\sum_{k>0} \frac{G_0(k+1)-G_0(k)}{G_0(1)} \epsilon_{i-k} f(v_{i-k})}{\epsilon_i f(v_i)}.
\end{equation}

\subsection{Deriving $\tilde{\lambda}_i$}

I would like to derive the form of $\tilde{\lambda}_i$ posited in BGPW, Eq.~\ref{eq.bouch_tlamb}, starting with the general form of the model shown in Eq.~\ref{eq.bouchaud} (this is identical to Eq.~\ref{eq.lillo} with a change of variable).  In so doing, we will understand the implicit assumptions of BGPW.
\begin{equation}
r_i = \epsilon_i f(v_i) - \tilde{\lambda}_i + \eta_i.
\label{eq.bouch_liq}
\end{equation}
Suppose that $\Omega$ is a set of publicly available historical financial data - possibly the entire past history of $\epsilon_i$ and $v_i$.  The efficient market hypothesis (EMH) in its weak form can be interpreted as,
\begin{equation}
E[r_i|\Omega] = E[r_i] \approx 0.
\end{equation}
This hypothesis is somewhat intuitive - we should not expect historical financial data to influence current returns.  If historical data was predictive for returns, then we would expect arbitrageurs to exploit it away.  I show evidence further in this chapter that it does hold - at least when using historical $\epsilon_i$ data that is several transactions old.  LF and BGPW show similar evidence.  For the time being I will assume it is true, so that,
\begin{equation}
E[r_i|\Omega] = 0.
\end{equation}
Applying this to Eq.~\ref{eq.bouch_liq} gives,
\begin{equation}
E\left[\tilde{\lambda}_i\middle|\Omega\right] = E\left[\epsilon_i f(v_i)\middle|\Omega\right].
\end{equation}
If we take that,
\begin{equation}
\tilde{\lambda}_i\left(\Omega\right) = E\left[\tilde{\lambda}_i\middle|\Omega\right] = E\left[\epsilon_i f(v_i)\middle|\Omega\right],
\end{equation}
this amounts to assuming that $\tilde{\lambda}_i$ is independent of $\epsilon_i$ and $f(v_i)$.  This is not trivial and is a main assumption of BGPW.  Doing this gives,
\begin{equation}
r_i = \epsilon_i f(v_i) - E\left[\epsilon_i f(v_i)\middle|\Omega\right] + \eta_i.
\label{eq.bouch_autoregress}
\end{equation}
We would like to specify an equation for $E\left[\epsilon_i f(v_i)\middle|\Omega\right]$, given some set of historical information $\Omega$.  That $\epsilon_i$ is highly autocorrelated suggests an autoregressive model might determine $E\left[\epsilon_i f(v_i)\middle|\Omega\right]$:  
\begin{equation}
E\left[\epsilon_i f(v_i)\middle|\Omega\right] = \sum_{k>0} a_k \epsilon_{i-k} f(v_{i-k}).
\label{eq.ar_model}
\end{equation}
$a_k$ are the coefficients of the autoregressive model.  BGPW empirically fit the coefficients $a_k$.  We will see this below, and compare their results to the results that I derive next.  But first I show the final result of the derivation,
\begin{equation}
\tilde{\lambda}_i = \sum_{k>0} a_k \epsilon_{i-k} f(v_{i-k}).
\label{eq.final_tlamb}
\end{equation}

Because $\epsilon_i$ sets the sign and $v_i$ sets the scale of $\epsilon_i f(v_i)$, $E\left[\epsilon_i f(v_i)\middle|\Omega\right]$ is more sensitive to changes in $\epsilon_i$ than changes in $f(v_i)$ ... that $f(\cdot)$ is a concave function only reinforces this (BGPW postulates a log and LF postulates a power law with exponent less than one).  This means that $a_k$ can be determined using an AR model for $\epsilon_i$ alone.  Assuming that $\epsilon_i$ can be modeled as a FARIMA$(0,H-1/2,0)$ process with Hurst exponent $H$, then in the large $k$ limit\cite{Beran94},
\begin{equation}
a_k \sim k^{-H-1/2}.
\end{equation}
If $\gamma$ is the decay exponent for the autocorrelation of $\epsilon_i$ when fit by a power law, then $H = 1-\gamma/2$\cite{Beran94}.  Using this, we have,
\begin{equation}
a_k \sim k^{(-3+\gamma)/2}.
\label{eq.ar_scaling}
\end{equation}
Below, I compare this result with the result derived in BGPW, which was verified empirically there and in the later paper by Bouchaud, Kockelkoren, and Potters \cite{Bouchaud04b}.

\subsection{Comparing to BGPW}

Comparing the hypothesis of BGPW, Eq.~\ref{eq.bouch_tlamb}, to the result above, Eq.~\ref{eq.final_tlamb} gives,
\begin{equation}
\sum_{k>0} \frac{G_0(k+1)-G_0(k)}{G_0(1)} \epsilon_{i-k} f(v_{i-k}) = \sum_{k>0} a_k \epsilon_{i-k} f(v_{i-k}).
\end{equation}
We can now interpret the main result of BGPW as an autoregressive model for $\epsilon_i f(v_i)$ with coefficients $(G_0(k+1)-G_0(k))/G_0(1)$.  BGPW determine $G_0(\tau)$ by empirically estimating the entire function and then fitting this to obtain\footnote{The original BGPW paper fit a slightly different function that was updated to this by Bouchaud, Kockelkoren, and Potters \cite{Bouchaud04b}.},
\begin{equation}
G_0(\tau) = \frac{\Gamma_0}{(\tau_0^2-\tau^2)^{\phi/2}},
\label{eq.g_fit}
\end{equation}
where I have changed some of their notation to fit with the rest of this thesis.  They derive the following relation between $\gamma$ and $\phi$,
\begin{equation}
\phi = (1-\gamma)/2,\label{eq.bb_gam}
\end{equation}
and verify that it holds empirically.  I would like to compare this result with the result of Eq.~\ref{eq.ar_scaling}.  If instead of deriving the coefficients, $a_k$, we empirically fit them in an identical way to BGPW, then we have
\begin{equation}
a_k = \frac{G_0(k+1)-G_0(k)}{G_0(1)}.
\end{equation}
From Eq.~\ref{eq.g_fit}, we know that in the large $k$ limit,
\begin{eqnarray}
G_0(k) & \sim & k^{-\phi} \\
     & \sim & k^{(-1+\gamma)/2}.
\end{eqnarray}
Taking the derivative of this relationship gives,
\begin{equation}
\left(G_0(k+1)-G_0(k)\right) \sim k^{(-3+\gamma)/2},
\end{equation}
which means that,
\begin{equation}
a_k \sim k^{(-3+\gamma)/2}.
\end{equation}
This is the same result derived above in Eq.~\ref{eq.ar_scaling}.

\subsection{Conclusion\label{sec.bouch_assumptions}}
The general form of the two models posited by LF and BGPW are equivalent.  LF leave the model in its most general form, whereas BGPW use empirical data to fit a specific form of the model that is based on several assumptions.

The main results of BGPW can be derived using the general model of Eq.~\ref{eq.bouchaud} with the following assumptions:
\begin{itemize}
\item[(A1)] The weak form of the efficient market hypothesis (EMH) holds.  That is,
\begin{equation}
E[r_i|\Omega] = 0.
\end{equation}

\item[(A2)] $\tilde{\lambda}_i$ is independent of $\epsilon_i$ and $f(v_i)$.

\item[(A3)] $E\left[\epsilon_i f(v_i)\middle|\Omega\right]$ can be approximated by an infinite autoregressive model for $\epsilon_i$ and $f(v_i)$ with coefficients determined by treating $\epsilon_i$ as a FARIMA process.
\end{itemize}
Because I reference these extensively, I have also added them to the Appendix.  (A1) was empirically verified in LF and BGPW.  I will study assumptions (A2) and (A3) in more depth in the next chapter.

\section{Interpretation of the Model\label{sec.I1I2}}
Although the models of LF and BGPW are identical, the interpretation of these models is still in question.  As stated earlier, LF suggest that impacts are permanent but fluctuating and BGPW suggest that impacts are transient but fixed.  If we take the following form of the updated model for returns,
\begin{equation}
r_i = \epsilon_i f(v_i) - \tilde{\lambda}_i + \eta_i,
\end{equation}
we can interpret this equation in two ways:
\begin{itemize}
\item[(I1)] The impact of the transaction at time $i$ is $\left(\epsilon_i f(v_i) - \tilde{\lambda}_i\right)$ with noise and is permanent.
\item[(I2)] The impact of the transaction at time $i$ is $\epsilon_i f(v_i)$ with noise and is transient.  $\tilde{\lambda}_i$ is the instantaneous decay at time $i$ of past transactions.
\end{itemize}
Because I reference these extensively, I have also added them to the Appendix.  The first interpretation is that of LF, and the second is that of BGPW.  Given that the terms $\epsilon_i f(v_i)$ and $\tilde{\lambda}_i$ are applied contemporaneously, it is impossible to determine which of these interpretations is formally correct.

These models, however, are not complete models of orderflow.  We can look inside the mechanics of a transaction, separating it into one component including the initial transaction impact (determined by liquidity) and another component including the limit order placements and cancellations that occur before the next transaction (which determines quote revisions).  LF suggest that $\lambda_i$ (and therefore also $\tilde{\lambda}_i$) is a liquidity term rather than a quote revision term.  BGPW suggest that $\tilde{\lambda}_i$ (and therefore also $\lambda_i$) is a quote revision term - although in a later paper they say it combines both liquidity and quote revision effects\cite{Bouchaud04b}.  By looking at entire orderflow, we can determine which of these is correct - I will determine this in the next section.  The answer will have ramifications for determining which interpretation above, (I1) or (I2), is more natural - this is discussed in Chapter~\ref{ch.price_impact}.


\section{Evidence for Asymmetric Liquidity and Not Mean Reversion\label{sec.imbalance}}

Suppose a transaction with known sign, $\epsilon_i$, occurs at time, $i$.  The probability that future transactions have this same sign is defined:
\begin{equation}
p_+(k) \equiv P\left(\epsilon_{i+k} = \epsilon_i \middle|\epsilon_i\right).
\end{equation}
The probability that future transactions have the opposite sign is defined:
\begin{equation}
p_-(k) \equiv P\left(\epsilon_{i+k} \neq \epsilon_i \middle|\epsilon_i\right).
\end{equation}
The expected return of future transactions (measured in the direction of the transaction) with the same and opposite sign are defined analogously:
\begin{eqnarray}
r_+(k)  & \equiv & E\left[\epsilon_{i+k} r_{i+k} \middle| \epsilon_{i+k} = \epsilon_i \right],\\
r_-(k)  & \equiv & E\left[\epsilon_{i+k} r_{i+k} \middle| \epsilon_{i+k} \neq \epsilon_i \right].
\end{eqnarray}
The price response of the market at time $(i+k)$ to a transaction of sign $\epsilon_i$ at time $i$ is $\left(\epsilon_i E\left[r_{i+k}\middle|\epsilon_i\right]\right)$.  This measures the expected return in the direction of the original transaction at time $(i+k)$ and if nonzero, means the market is still responding to the transaction (if positive the impact is still growing and if negative the impact is decaying).  This is measured:
\begin{eqnarray}
E\left[\epsilon_i r_{i+k}\middle|\epsilon_i\right] & = & \sum_{\epsilon_{i}} E\left[\epsilon_i r_{i+k}\middle|\epsilon_i, \epsilon_{i+k} \right] P\left(\epsilon_{i+k}\middle|\epsilon_i\right),\\
 & = & p_+(k)r_+(k) - p_-(k)r_-(k).
\end{eqnarray}
For the EMH to hold, the response should be zero \dots otherwise returns would be predictable,
\begin{equation}
\epsilon_i E\left[r_{i+k}\middle|\epsilon_i\right] = 0,
\end{equation}
and therefore,
\begin{equation}
p_+(k)r_+(k) - p_-(k)r_-(k) = 0.
\end{equation}
Rearranging this into ratios, the final result is,
\begin{equation}
\frac{r_-(k)}{r_+(k)} = \frac{p_+(k)}{p_-(k)}.\label{eq.imb_ch2}
\end{equation}
I call the term on the left hand side of Eq.~\ref{eq.imb_ch2} the \emph{return imbalance}, and the term on the right hand side of Eq.~\ref{eq.imb_ch2} the \emph{transaction imbalance}.

\begin{figure}[!ht]
\centering
\includegraphics[width=4.5in]{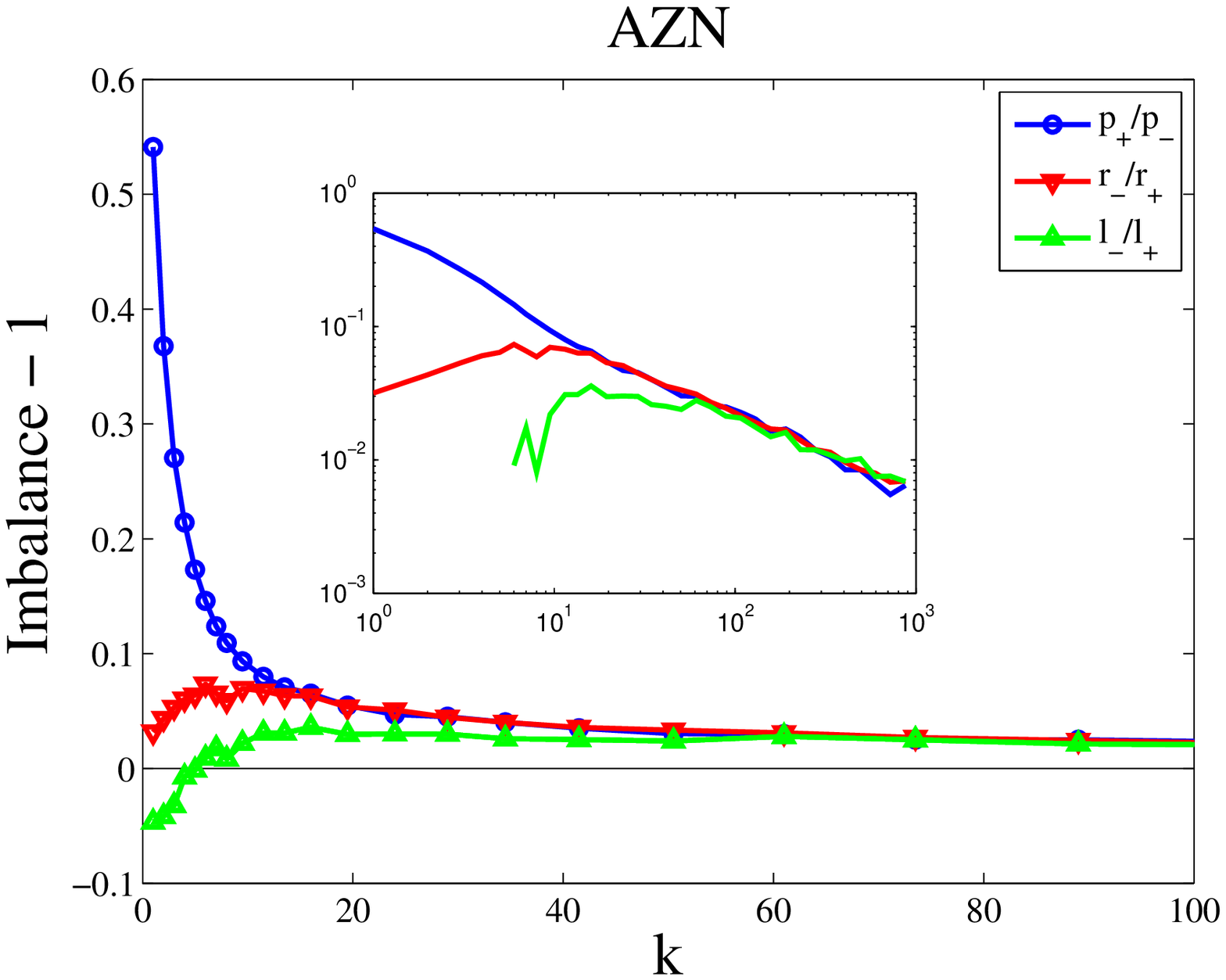}
\includegraphics[width=4.5in]{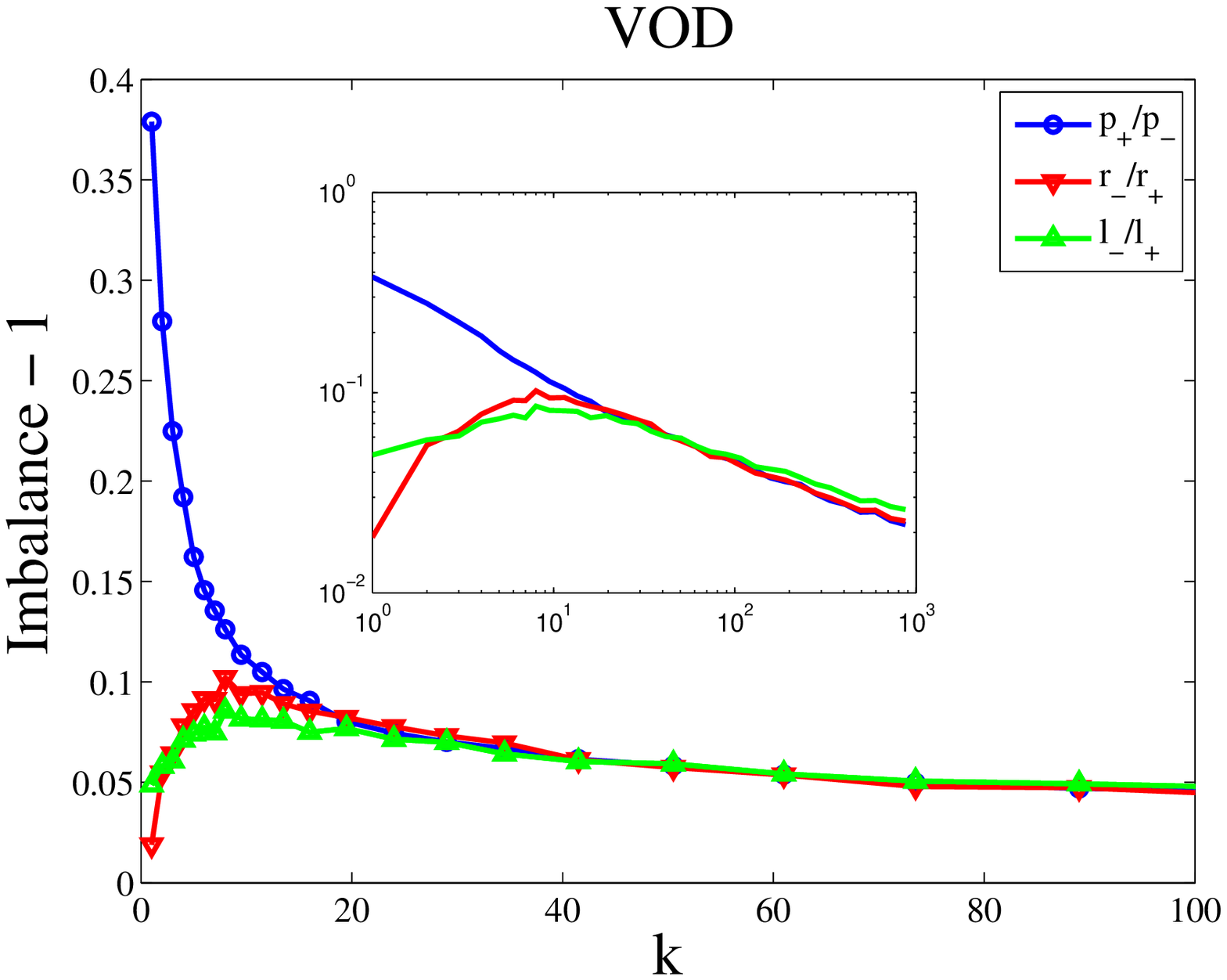}
\caption{The transaction imbalance ($p_+/p_-$), return imbalance ($r_-/r_+$), and liquidity imbalance ($l_-/l_+$) as a function of lag $k$ for the stocks AZN and VOD.  The $y$-axis is measured in units of the average spread.  The transaction imbalance measures the predictability of transaction sign - to the extent that the return imbalance matches the transaction imbalance, this cannot be used to predict returns.  To the extent the liquidity imbalance matches the transaction imbalance, it is asymmetric liquidity and not quote revisions that explains why returns remain unpredictable.}
\label{fig.BOTH_s_imb_avg}
\end{figure}

If we decompose a return into two components, one component including the initial transaction impact $l_i$ and another component including the limit order placements and cancellations that occur before the next transaction $q_i$, we have,
\begin{equation}
r_{i+k} = l_{i+k} + q_{i+k}.\label{eq.r_decomp}
\end{equation}
If we hypothesis that $q_{i+k}$ can be treated as a noise term, such that the EMH holds without its addition, then we can analogously follow the definitions and calculations above and end with,
\begin{equation}
\frac{l_-(k)}{l_+(k)} = \frac{p_+(k)}{p_-(k)}.\label{eq.imb2_ch2}
\end{equation}
I call the term on the left hand side of Eq.~\ref{eq.imb2_ch2} the \emph{liquidity imbalance}.  To the extent this equation holds, it is liquidity and not quote revisions that keeps returns unpredictable.  In Fig.~\ref{fig.BOTH_s_imb_avg}, I plot all three of these imbalances as a function of $k$ for the stocks AZN and VOD.  As seen in the figure, there is a delay before the return imbalance matches the transaction imbalance.  This means the EMH does not hold until $k \approx 15$ for AZN and VOD, but thereafter holds indefinitely (at least as far as we can measure it).  There is a further delay before the liquidity imbalance reaches them both, but it is not a bad approximation to consider the quote revision term, $q_{i+k}$, as noise.  This is because the return imbalance is almost fully reproduced by the liquidity imbalance, especially for $k>50$ for AZN and $k>20$ for VOD, the contribution of quote revisions to the return imbalance is therefore almost negligible.

To see how this looks in terms of market impact, the average total price response (or impact) at time $(i+T)$ due to the transaction placed at time $i$ with sign $\epsilon_i$ is defined:
\begin{eqnarray}
I(T) & \equiv & \sum_{k=0}^T E\left[\epsilon_i r_{i+k}\middle|\epsilon_i\right],\\
 & = & \sum_{k=0}^T \left(p_+(k)r_+(k) - p_-(k)r_-(k)\right).
\end{eqnarray}
If we naively assume that the magnitude of returns does not respond to the transaction imbalance, then we can determine the average impact under this assumption, $I_N$, as follows:
\begin{eqnarray}
I_N(T) & \equiv & \sum_{k=0}^T E\left[\epsilon_i r_{i+k}\middle|\epsilon_i\right],\\
 & = & I_o\sum_{k=0}^T \left(p_+(k) - p_-(k)\right),
\end{eqnarray}
where $I_o$ is a constant, the unconditional absolute midprice impact measured from immediately before to immediately after a transaction.  If we assume in the return decomposition of Eq.~\ref{eq.r_decomp} that quote revisions can be approximated as a noise term and disregarded, then under this assumption the average impact, $I_L$, is,
\begin{eqnarray}
I_L(T) & \equiv & \sum_{k=0}^T E\left[\epsilon_i l_{i+k}\middle|\epsilon_i\right],\\
 & = & \sum_{k=0}^T \left(p_+(k)l_+(k) - p_-(k)l_-(k)\right).
\end{eqnarray}

\begin{figure}[!ht]
\centering
\includegraphics[width=4.5in]{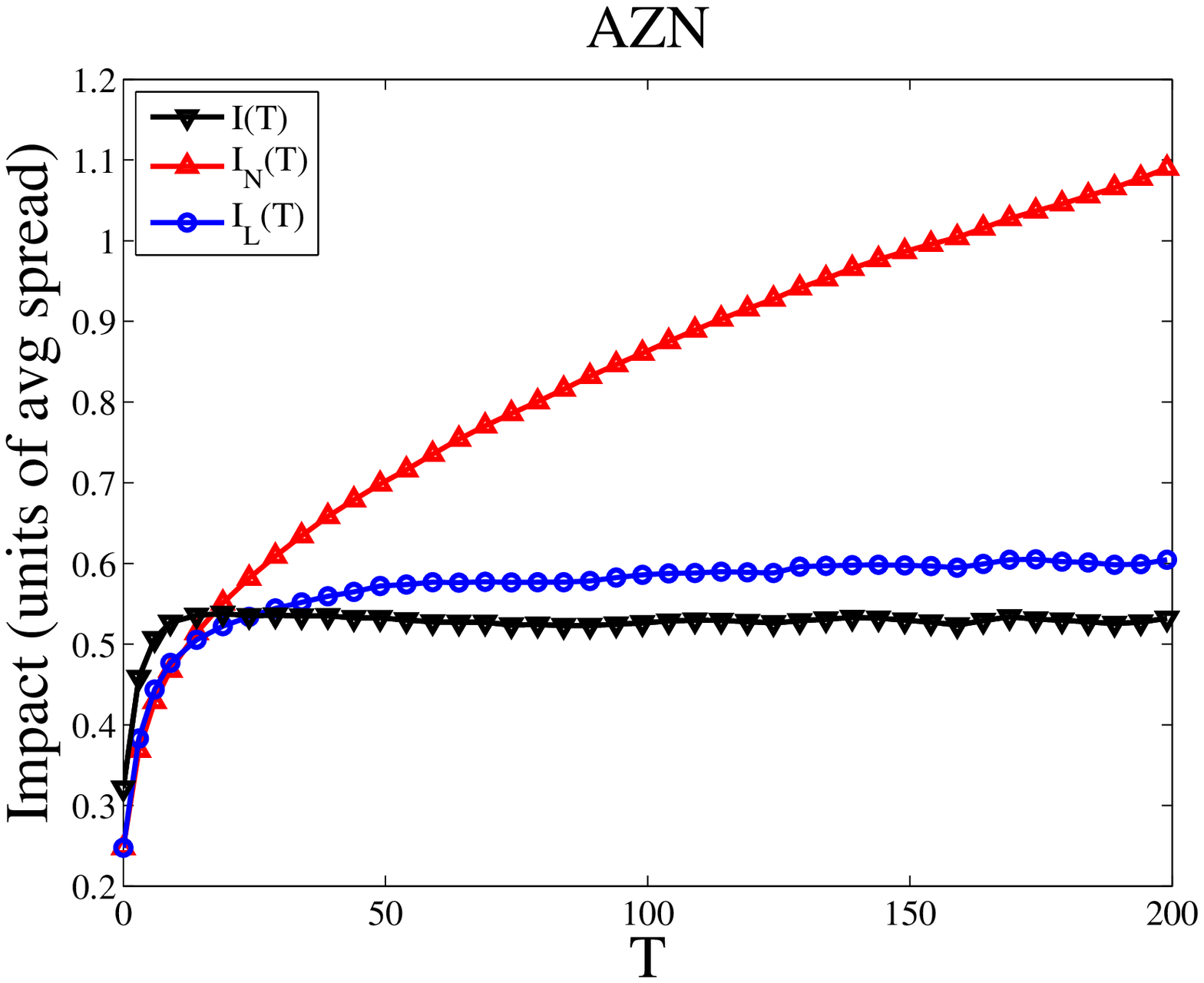}
\includegraphics[width=4.5in]{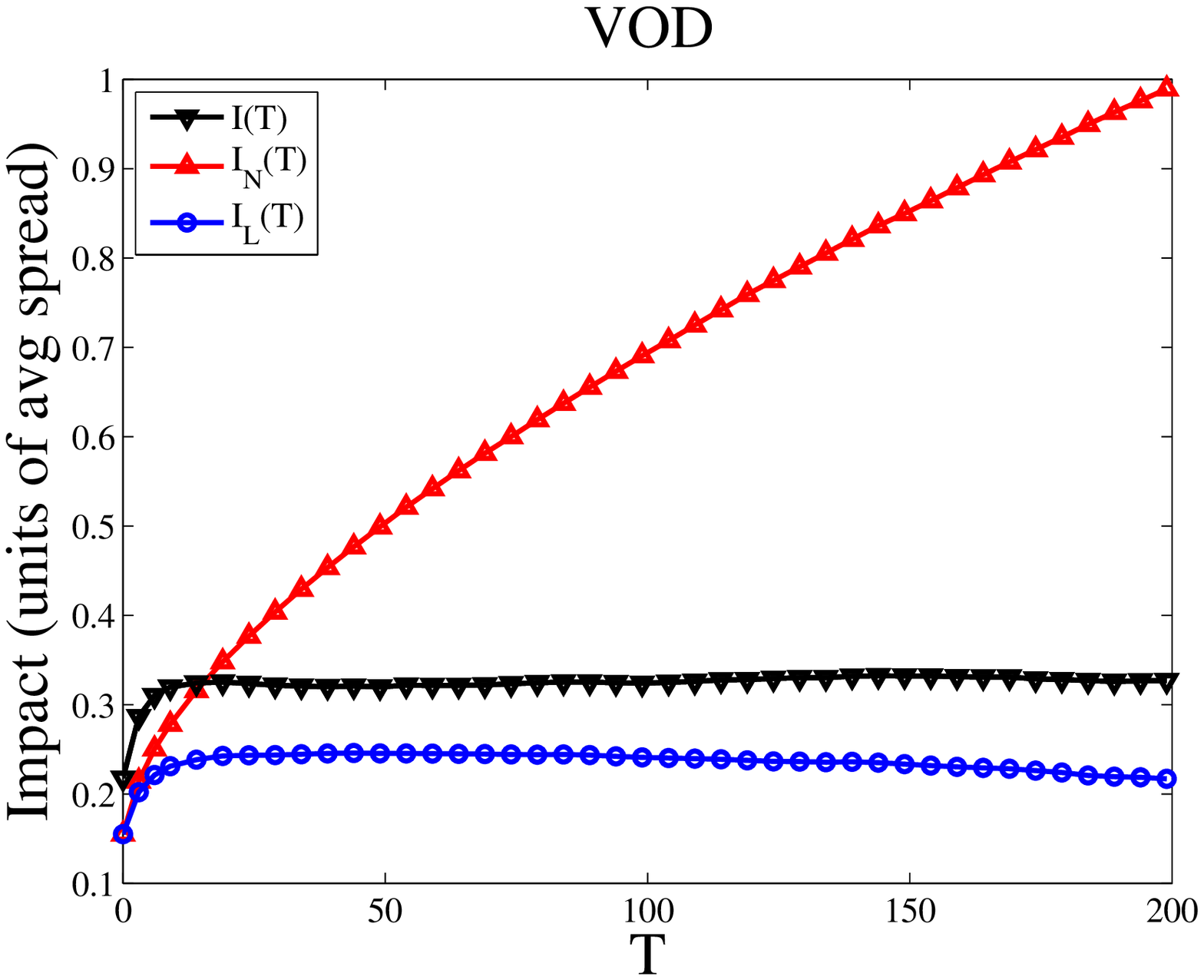}
\caption{The total price impact caused by a transaction at time $T=0$ measured as a function of time $T$ for the stocks AZN and VOD.  The $y$-axis is measured in units of the average spread.  $I(T)$ is the empirically measured impact.  $I_N(T)$ is the impact that would occur if the predictability of sign is not compensated for.  $I_L(T)$ is the impact that would occur if quote revisions are ignored and only fluctuating liquidity is included.  That $I_L(T)$ remains relatively flat suggests that quote revisions are unimportant for keeping returns efficient.}
\label{fig.BOTH_cumimpact}
\end{figure}
I plot in Fig.~\ref{fig.BOTH_cumimpact} the empirically measured values of $I(T)$, $I_N(T)$, and $I_L(T)$ for the stocks AZN and VOD as a function of time since the transaction.  The scale on the $y$-axis is set in units of the average spread, i.e., the average distance between the best bid price and best offer price in the order book.  As seen in the figure, the empirically measured impact increases until $T \approx 15$ for both stocks and thereafter remains constant.  $I_N(T)$ increases throughout and nicely exemplifies the efficiency puzzle - that returns would be predictable if not for compensation due to liquidity fluctuations and/or mean reverting quote revisions.  Notice that $I_L(T)$ closely matches $I$ and that it also remains relatively constant throughout - this again verifies that it is asymmetric liquidity alone that solves most, if not all, of the efficiency puzzle.

Notice also in Fig.~\ref{fig.BOTH_cumimpact}, that impacts do not have a transient component.  Even when decomposing the initial transaction return (i.e., $r_i$ at $T=0$) into the components $l_i$ and $q_i$, the initial impact $l_i$ is not reverted at all by $q_i$ - in fact the impact is enhanced such that $r_i$ is larger.  This is quite different than what is usually assumed in the microstructure literature.  It shows that on average, all transactions cause an initial price response $l_i$ that is enhanced by the future values, $q_i, q_{i+1}, \dots, q_{i+k}$ and $l_{i+1}, l_{i+2}, \dots, l_{i+k}$, until $k \approx 15$.  `The market' accepts this new price (which is actually slightly larger than the original impact) as the reference price for at least the next 200 transactions (about 2 hours for AZN).  It is possible that decay occurs at a much later time, but errors grow quickly\cite{Farmer06} and to discern this requires a much larger dataset.

\chapter{Theory of Asymmetric Liquidity\label{ch.asym_liq}}

\section{Introduction}
In Chapter~\ref{ch.intro}, I introduced the simple model for returns,
\begin{equation}
r_i = \epsilon_i f(v_i) + \eta_i.
\end{equation}
In Chapter~\ref{ch.long_mem}, we learned that this model is incorrect because of the strong autocorrelation of $\epsilon_i$, and I presented two modified versions of the model that were suggested by the first two papers on this subject,
\begin{eqnarray}
r_i & = & \frac{\epsilon_i f(v_i)}{\lambda_i}  + \eta_i,\label{eq.LF_form}\\
r_i & = & \epsilon_i f(v_i) - \tilde{\lambda}_i  + \eta_i. \label{eq.BGPW_form}
\end{eqnarray}
These models are equivalent with a simple change of variable.  It was shown that $\lambda_i$ and $\tilde{\lambda}_i$ are more appropriately considered liquidity terms rather than quote revision terms because their effects are felt at the instant of the transaction.

There are two unresolved issues:
\begin{itemize}
\item [(1)] Why does $\epsilon_i$ exhibit long memory, and who enforces that $r_i$ is uncorrelated, i.e., who is influencing $\lambda_i$ and $\tilde{\lambda}_i$?
\item [(2)] What is the correct relation between $\lambda_i$, $\tilde{\lambda}_i$ and the set of variables $\{\epsilon_i,f(v_i),\Omega\}$?
\end{itemize}
These questions are discussed in succession below.

\section{Theory for Autocorrelated Orderflow and Liquidity Response}
This section will discuss the first question from the Introduction of this chapter - Why does $\epsilon_i$ exhibit long memory, and who enforces that $r_i$ is uncorrelated?  Let us look at the influence of an individual transaction at time $i-k$ of sign $\epsilon_{i-k}$.  Given the model,
\begin{equation}
r_i = \epsilon_i f(v_i) - \tilde{\lambda}_i  + \eta_i,
\end{equation}
and using the weak form of the efficient market hypothesis (EMH), $E\left[r_{i}\middle|\epsilon_{i-k}\right] = 0$, we have that,
\begin{equation}
E\left[\tilde{\lambda}_{i} \middle|\epsilon_{i-k}\right] = E\left[\epsilon_{i} f(v_{i}) \middle|\epsilon_{i-k}\right].
\end{equation}
We know from empirical data that $E\left[\epsilon_{i} f(v_{i}) \middle|\epsilon_{i-k}\right]$ and therefore $E\left[\tilde{\lambda}_{i} \middle|\epsilon_{i-k}\right]$ are generally not equal to zero.  In fact, because $\epsilon_{i-k}$ and $\epsilon_i$ are positively correlated, we know that,
\begin{equation}
\epsilon_{i-k} E\left[\epsilon_{i} f(v_{i}) \middle|\epsilon_{i-k}\right] > 0,
\label{eq.pred_orderflow}
\end{equation}
and therefore,
\begin{equation}
\epsilon_{i-k} E\left[\tilde{\lambda}_{i} \middle|\epsilon_{i-k}\right] > 0.
\label{eq.pred_liquidity}
\end{equation}
I have so far offered no physical explanation for why these inequalities hold.  A theory for Eq.~\ref{eq.pred_orderflow} was developed by Lillo, Mike, and Farmer in 2005 (hereafter known as LMF)\cite{Lillo05b}.  It is still not understood how Eq.~\ref{eq.pred_liquidity} holds.

\subsection{Theory for Autocorrelated Orderflow\label{sec.LMF}}
\begin{figure}[htb]
\centering
\includegraphics[width=4in]{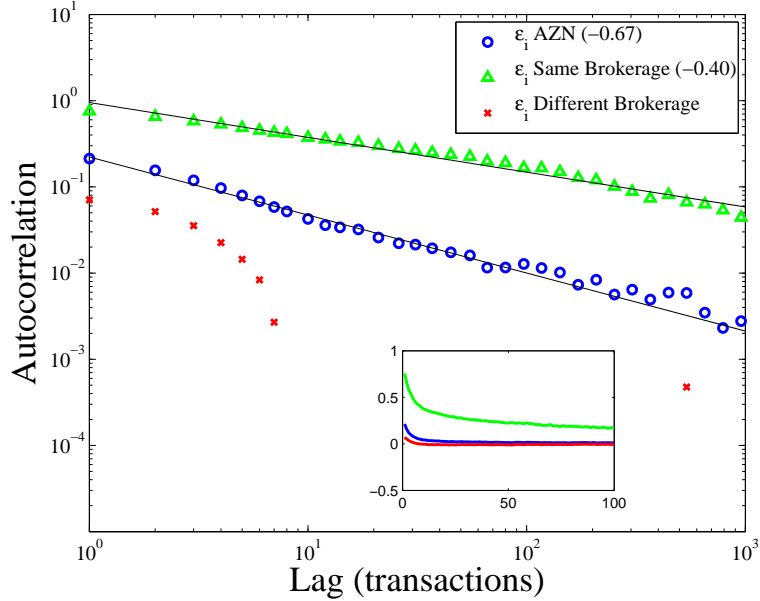}
\caption{The autocorrelation function of transaction sign $\epsilon_i$ for the stock AZN under various conditions.  The blue circles are the autocorrelation function when using the entire series, the green upward triangles are the autocorrelation function when using only transaction signs originating from the same brokerage, and the red x's are the autocorrelation function when using only transaction signs originating from different brokerages.  That order flow within a brokerage is highly autocorrelated (and order flow across brokerages is not) suggests that market participants produce autocorrelated order flow by chopping large hidden orders into smaller pieces.}
\label{fig.AZN_sign_broke_autocorr}
\end{figure}

LMF suggest that market participants have \emph{hidden orders} that are a true reflection of their intention to buy or sell a security, and these orders are chopped into equal sized pieces, called \emph{revealed orders}, that are then transacted.  LMF show that if the size of hidden orders are distributed as a power law with exponent $-(1+\alpha)$, then this produces an autocorrelation function for $\epsilon_i$ that decays as a power law with exponent $-\gamma$, where,
\begin{equation}
\gamma = \alpha - 1.
\end{equation}
The autocorrelation of $\epsilon_i$ is then due to the splitting up of much larger orders, and therefore we should expect Eq.~\ref{eq.pred_orderflow} to hold because there is a certain probability that $\epsilon_i$ and $\epsilon_{i-k}$ are parts of the same hidden order.  I show evidence for this interpretation in Fig.~\ref{fig.AZN_sign_broke_autocorr}.  This figure uses a unique feature of my dataset - that the brokerage codes are attached to all orders.  These codes are numbers that uniquely identify the member firms of the London Stock Exchange (in the dataset they are anonymous - so that I cannot identify the firm by name).  With these codes, I can discern if a group of transactions or other order flow is originating from the same brokerage or a different brokerage.  If from the same brokerage, this might be the action of two separate market participants trading through the same brokerage \dots or might be the action of one participant trading an order she has split into pieces.  I will talk more about these codes and how I use them to determine larger hidden orders in Chapter~\ref{ch.emp_results}. 
Notice in Fig.~\ref{fig.AZN_sign_broke_autocorr}, that the sign of transactions $\epsilon_i$ is most correlated within the same brokerage (power law exponent $-0.40$), is still heavily correlated but less so without considering the brokerage (power law exponent $-0.67$), and is not correlated at all across brokerages.  There are two critical results we can interpret from this, and these results will influence the rest of this thesis.
\begin{itemize}
\item[(R1)] Market participants exhibit autocorrelated orderflow (most likely because they have split up into pieces a much larger order).
\item[(R2)] Market participants do not correlate with each others orderflow.
\end{itemize}
Because I reference these extensively, I have also added them to the Appendix.

The concept of hidden orders will be used throughout the rest of this thesis, and I present here the notation I use for these orders.  All hidden orders are indexed by $j$.  The activity parameter, $A_{i,j}$, determines if a hidden order is currently active, i.e., if the hidden order has the potential to cause a transaction.  $A_{i,j}$ is equal to $1$ if hidden order $j$ is active at time $i$, and is equal to $0$ otherwise.  Hidden order $j$ has sign $\epsilon_j$ and size $V_j$ and begins at time $i=t_j$.  It is assumed this order is split into equal sized pieces of volume $v_j$ for a total of $N_j=V_j/v_j$ pieces.  These pieces, numbered as $n_j=1,2,\dots,N_j$ are transacted with probability $1/\theta_j$ at every timestep $i>t_j$ until the hidden order is finished.  This means that the order transacts on average every $\theta_j$ transactions.  The time at which piece $n_j$ transacts is $t(n_j)$.


\subsection{Theory for Liquidity Response \label{sec.theory_liq}}
We still have no physical explanation for why Eq.~\ref{eq.pred_liquidity} should hold.  There are two possibilities that can serve as starting points to a theory.  Suppose a market participant, identified as participant $p$ initiated the transaction at time $i-k$.
\begin{itemize}
\item[(P1)]  \textbf{There are a group of market participants that collectively condition liquidity on past orderflow such that Eq.~\ref{eq.pred_liquidity} holds}.  At time $i$, these participants force transactions with the same sign as $\epsilon_{i-k}$ to have lower impact and/or transactions with the opposite sign as $\epsilon_{i-k}$ to have larger impact.  They do so because there is a certain probability that participant $p$ will place another transaction at time $i$ with known sign $\epsilon_{i-k}$.  The participants are thought of as market makers who do not want the price to become superdiffusive.\footnote{See BGPW and the later paper by Bouchaud, Kockelkoren, and Potters \cite{Bouchaud04b} for a more detailed explanation.}
\item[(P2)]  \textbf{Market participant $p$ conditions her transactions on the liquidity parameter such that Eq.~\ref{eq.pred_liquidity} holds}.  If participant $p$ notices that her trades are causing large impacts, she may decide to postpone the trade to some later time or reduce the size of her transactions $v_i$ - in fact, she may not be willing to trade at all unless she does observe liquidity as dictated in Eq.~\ref{eq.pred_liquidity}.  Instead of market makers who work to keep the price diffusive, the participants that determine liquidity are patient position takers that are willing to wait for participant $p$ to cause transactions.\footnote{If this description is true, it would be quite interesting to study the dynamics of this order-matching \dots the long memory of $\epsilon_i$ would result from two players each showing and then matching a small portion of their true intentions at a time (they collectively decide which participant will be the initiator of the transactions).  They show only small portions at a time so not to scare the counterparty away if it happens that their supply/demand is much larger than the other participant.}

\end{itemize}
Because I reference these extensively, I have also added them to the Appendix.


\section{Determining $\lambda_i$ and $\tilde{\lambda}_i$\label{sec.E1E2}}
This section will discuss the second question from the Introduction of this chapter - What is the correct relation between $\lambda_i$, $\tilde{\lambda}_i$ and the set of variables $\{\epsilon_i,f(v_i),\Omega\}$?.  This can be decomposed into two parts.  First, what is the dependence of $\lambda_i$ and $\tilde{\lambda}_i$ on historical financial data and what historical information is included in determining these variables.  Second, how are $\lambda_i$ and $\tilde{\lambda}_i$ dependent on $\epsilon_i$ and $f(v_i)$? 
\subsection{Dependence of Liquidity on $\Omega$\label{sec.liq_omega}}
Given that the weak form of the EMH holds (A1), we have that,
\begin{eqnarray}
E\left[r_{i}\middle|\Omega\right] & = & E\left[\epsilon_{i} f(v_{i}) \middle|\Omega\right] - E\left[\tilde{\lambda}_{i} \middle|\Omega\right],\\
 & = & 0,
\end{eqnarray}
and therefore,
\begin{equation}
E\left[\tilde{\lambda}_{i} \middle|\Omega\right] = E\left[\epsilon_{i} f(v_{i}) \middle|\Omega\right].\label{eq.exp_lamtild}
\end{equation}
Using (A3), an autoregressive model for $\epsilon_{i} f(v_{i})$ determines $E\left[\epsilon_{i} f(v_{i}) \middle|\Omega\right]$ so that,
\begin{equation}
E\left[\tilde{\lambda}_{i} \middle|\Omega\right] = \sum_{k>0} a_k \epsilon_{i-k} f(v_{i-k})\label{eq.ar_model2}.
\end{equation}
There are two reasons why this equation, and therefore (A3), may be wrong.  First, taking results (R1) and (R2) and assuming that (P2) from above is correct, then $\tilde{\lambda}_{i}$ is not a response to historic values of $\epsilon_i$ and $f(v_i)$ as suggested by Eq.~\ref{eq.ar_model2} but, in fact, $\epsilon_{i} f(v_{i})$ is a response to $\tilde{\lambda}_{i}$.  The current and future values of $\tilde{\lambda}_{i}$ can therefore be completely decoupled from the history of $\epsilon_i$ and $f(v_i)$.  As a concrete example, suppose that participant $p$ experiences a favorable state of liquidity, $\tilde{\lambda}_i$ - favorable in the sense that she can place a series of transactions without causing a predictable return.  We can assume she will continue to place pieces of her larger hidden order until either her hidden order is completed, or until she finds that $\tilde{\lambda}_i$ is no longer favorable.  In either case, $\tilde{\lambda}_i$ is free to change at will and is not dependent on historical values of $\epsilon_i$ and $f(v_i)$ - it may correlate with these, but it is not dependent on them.  In this case, there is no enforcer of the EMH as is suggested by Eq.~\ref{eq.ar_model2}, but the EMH holds as a consequence of participant $p$'s reluctance to produce predictable returns.

There is a second reason why Eq.~\ref{eq.ar_model2} may be wrong.  Depending on the particulars of the $\epsilon_{i}$ and $f(v_{i})$ series, we might more accurately model $E\left[\epsilon_{i} f(v_{i}) \middle|\Omega\right]$ with something other than an autoregressive model.  Taking results (R1) and (R2) and assuming that (P1) from above is correct, market makers might somehow predict $E\left[\epsilon_{i} f(v_{i}) \middle|\Omega\right]$ more efficiently than with an autoregressive model.  They will then set liquidity accordingly.  Either of these reasons, or some combination of them both, mean that the validity of Eq.~\ref{eq.ar_model2} is up for debate.

Let me formalize two equations for $E\left[\tilde{\lambda}_{i} \middle|\Omega\right]$ motivated by the discussion so far.  These equations are an attempt to bound the dependence of $\tilde{\lambda}_{i}$ on $\Omega$ between two extremes.
\begin{itemize}
\item[(E1)]  Assume that (P1) is correct and that the best predictor of $\epsilon_i f(v_i)$ is an autoregressive model.
\begin{equation}
E\left[\tilde{\lambda}_{i} \middle|\Omega\right] = \sum_{k>0} a_k \epsilon_{i-k} f(v_{i-k}).
\end{equation}

\item[(E2)]  Take that (R1) and (R2) are correct.  Assume one of the following: (P2) is true, or alternatively, (P1) is true and market makers have information about who is initiating transactions and when they have finished placing a hidden order.
\begin{equation}
E\left[\tilde{\lambda}_{i}\middle|\Omega'\right] = \sum_{j} A_{i,j} \left(\frac{n_j}{n_j+1}\right)^\alpha \frac{\epsilon_j f(v_j)}{\theta_j},
\end{equation}
where $\Omega'$ is information about all of the active hidden orders (those with $A_{i,j}=1$) \dots this includes their sign $\epsilon_j$, their typical size $v_j$, the average number of timesteps between transactions for the order $\theta_j$, and how many pieces of the order have already been transacted $n_j$.  I use $\Omega'$ instead of $\Omega$ because it is not necessarily publicly discernable information.

\end{itemize}
Because I reference these extensively, I have also added them to the Appendix.

These equations bound $\tilde{\lambda}_i$ because they represent the extreme views of publicly discernable information.  On one extreme, assuming transaction order flow is a pure FARIMA process without any other structure - then the right hand side of the equation in (E1) is the best predictor of $\epsilon_i f(v_i)$ and liquidity should be set according to this equation.  On the other extreme, assuming there is complete structure to order flow such that individual hidden orders are discernable, then the right hand side of the equation in (E2) is the best predictor of $\epsilon_i f(v_i)$ and liquidity should be set according to this equation.

The equation in (E1) was derived in Chapter~\ref{ch.long_mem}, so I will not rederive it here.  The equation in (E2) results because of the following.  The expected value of the liquidity parameter at time $i$ is the sum of the liquidity required by participants currently placing orders, assuming (P2); or alternatively is the total liquidity necessary for market efficiency given information about who is initiating orders and when they have finished placing a hidden order, assuming (P1).  These are equivalent for determining $E\left[\tilde{\lambda}_{i} \middle|\Omega'\right]$

I will derive (E2) assuming (P2).  The main principle is that market participants will initiate transactions such that their impacts cannot be exploited by other participants.  Assume we have a background of noisy returns - completely uncorrelated.  The task for participant $p$ is to scatter the impacts of her hidden order amongst this series such that returns remain unpredictable.  Assume that participant $p$ is currently executing the hidden order $j$ with pieces of size $\left(v_j=v\right)$ each of sign $\left(\epsilon_j=\epsilon\right)$.  She initiates a transaction with probability $\left(1/\theta_j=1/\theta\right)$ at every timestep $i>t_j$ until all $\left(N_j=N\right)$ pieces have executed.  The impact of each piece $(n_j=n)$, measured from before the piece was executed until directly before the next piece executes, is labeled $r_n$ and its expected value is,
\begin{equation}
E\left[r_{n}\right] = \epsilon f(v) - \theta E\left[\tilde{\lambda}_i \middle| t(n)\le i < t(n+1)\right].
\end{equation}
Here, $E\left[\tilde{\lambda}_i \middle| t(n)\le i < t(n+1)\right]$ is the expected liquidity during the period $t(n)\le i < t(n+1)$.  It is multiplied by $\theta$, the expected number of timesteps until the next piece is executed.  If $\tilde{\lambda}_{i}=0$ throughout the hidden order, then a sophisticated trader would notice an autocorrelated structure of returns at and around lag $\theta$.  He could calculate the probability that this pattern will continue as follows:
\begin{eqnarray}
P\left(N \ge n+1 \middle| N \ge n \right) & = &  \frac{\sum_{x=n+1}^\infty Ax^{-(1+\alpha)}}{\sum_{x=n}^\infty Ax^{-(1+\alpha)}},\label{eq.prob_size}\\
 & \approx & \left(\frac{n}{n+1}\right)^\alpha, 
\end{eqnarray}
where we are assuming as in LMF, that hidden orders are distributed as a power law, $f(x)=Ax^{-(1+\alpha)}$.  Eq.~\ref{eq.prob_size} is motivated by the following: the sophisticated trader does not know $N$ and therefore must calculate the probability that it is at least $n+1$ given that he has observed it is at least $n$.  We can understand how participant $p$ is conditioning her orders by noting that the sophisticated trader would be unable to exploit her if his expected return is zero.
\begin{eqnarray}
E\left[r_{n+1}\right] & = & P\left(\epsilon f(v) - \theta E\left[\tilde{\lambda}_i \middle| t(n)\le i < t(n+1)\right]\right) - \nonumber \\
& & \qquad \left(1-P\right)\theta E\left[\tilde{\lambda}_i \middle| t(n)\le i < t(n+1)\right],\\
 & = & 0.
\end{eqnarray}
where $P$ is short for $P\left(N \ge n+1 \middle| N \ge n \right)$.  Solving for the expected value of $\tilde{\lambda}_i$,
\begin{equation}
E\left[\tilde{\lambda}_i \middle| t(n)\le i < t(n+1)\right] = \left(\frac{n}{n+1}\right)^\alpha \frac{\epsilon f(v)}{\theta},
\end{equation}
This is what is required for participant $p$ to continue initiating transactions.\footnote{Of course she would continue to initiate transactions if she found liquidity even more favorable than this - this equation represents the bound at which she changes her willingness to transact.}  Given that $E\left[\tilde{\lambda}_{i}\middle|\Omega'\right]$ is the sum of what is expected by the current active hidden orders, we have,
\begin{equation}
E\left[\tilde{\lambda}_{i}\middle|\Omega'\right] = \sum_{j} A_{i,j} \left(\frac{n_j}{n_j+1}\right)^\alpha \frac{\epsilon_j f(v_j)}{\theta_j},\label{eq.hid_efficiency}
\end{equation}
which is the result shown in (E2).  To quickly see that this is the same for (P1) under (E2), note that market makers are acting to ensure $E\left[r_i\middle|\Omega'\right]=0$.  This gives,
\begin{equation}
E\left[\tilde{\lambda}_{i} \middle|\Omega'\right] = E\left[\epsilon_{i} f(v_{i}) \middle|\Omega'\right].
\end{equation}
The right side of this equation can be calculated by summing the expected impacts of each hidden order \dots the expected impact of a hidden order is just the probability that a current hidden order continues multiplied times its expected impact at time $i$, $\left(\epsilon_j f(v_j)/\theta_j\right)$.  This reproduces Eq.~\ref{eq.hid_efficiency}.

As for the plausability of (E1) and (E2), there are reasons to believe (E1): Market participants may not be able or willing to condition their orderflow on $\tilde{\lambda}_i$, so that (P1) is correct.  Also, it seems unlikely that market makers have full information about who is initiating transactions.  There are also reasons to believe (E2): Participants may not initially condition orders on $\tilde{\lambda}_i$, but might realize they are being exploited and therefore incrementally adjust their trading algorithm so that it slowly converges to a full conditioning on $\tilde{\lambda}_i$.  Also, although it is unlikely that market makers know the full set of information about hidden orders, they can make educated guesses - especially if it turns out that only one dominant market participant is initiating transactions at any moment in time.

In the next chapter, I will derive the implications of (E1) and (E2) for hidden order market impact and show that one implies permanent impact and the other transient impact.  In Chapter~\ref{ch.emp_results} I will present several plots that suggest which is more accurate.


\subsection{Dependence of Liquidity on $\epsilon_i$ and $f(v_i)$\label{sec.liq_eps_fv}}
In this section, we will see that the dependence of $\lambda_i$, $\tilde{\lambda}_i$ on $\epsilon_i$ and $f(v_i)$ is not fully specified by the EMH (A1).  The relationship will need to be determined empirically.  
 
Starting with Eq.~\ref{eq.LF_form}
\begin{equation}
r_i = \frac{\epsilon_i f(v_i)}{\lambda_i}  + \eta_i.
\end{equation}
Given that $\Omega$ is a set of historical financial information, the weak form of the EMH states that $E[r_i|\Omega]=0$.  This gives,
\begin{equation}
E\left[\frac{\epsilon_i f(v_i)}{\lambda_i}\middle|\Omega\right] = 0.
\end{equation}
We can state,
\begin{equation}
E\left[\frac{\epsilon_i f(v_i)}{\lambda_i}\middle|\Omega\right] = \sum_{\epsilon_i} \epsilon_i E\left[\frac{f(v_i)}{\lambda_i}\middle|\epsilon_i,\Omega\right] P(\epsilon_i|\Omega),
\end{equation}
so that,
\begin{equation}
\sum_{\epsilon_i} \epsilon_i E\left[\frac{f(v_i)}{\lambda_i}\middle|\epsilon_i,\Omega\right] P(\epsilon_i|\Omega) = 0.
\end{equation}
Because $\epsilon_i$ is either $+1$ or $-1$, this sum is easily computed.  Using that,
\begin{equation}
P(\epsilon_i|\Omega) = \frac{1 + \epsilon_i \hat{\epsilon}_i}{2}, \label{eq.prob_eps}
\end{equation}
where $\hat{\epsilon}_i \equiv E\left[\epsilon_i\middle|\Omega\right]$, produces the following,
\begin{equation}
\frac{E\left[\frac{f(v_i)}{\lambda_i}\middle|\epsilon_i=+1,\Omega\right]}{E\left[\frac{f(v_i)}{\lambda_i}\middle|\epsilon_i=-1,\Omega\right]} = \frac{1-\hat{\epsilon}_i}{1+\hat{\epsilon}_i}.
\label{eq.gen_result}
\end{equation}
I define the following terms,
\begin{eqnarray}
r_i^+ & \equiv & E\left[\frac{f(v_i)}{\lambda_i}\middle|\epsilon_i=+1,\Omega\right],\label{eq.r_plus}\\
r_i^- & \equiv & E\left[\frac{f(v_i)}{\lambda_i}\middle|\epsilon_i=-1,\Omega\right],\label{eq.r_minus}
\end{eqnarray}
so that,
\begin{equation}
\frac{r_i^+}{r_i^-} = \frac{1-\hat{\epsilon}_i}{1+\hat{\epsilon}_i}. \label{eq.ratio}
\end{equation}
Therefore, the ratio of the expected return of buyer vs seller initiated transactions must satisfy the specific ratio in Eq.~\ref{eq.ratio} for the EMH to hold.  There are a multitude of ways this can happen, for example:
\begin{equation}
\lambda_i = 1 + \epsilon_i \hat{\epsilon}_i,
\end{equation}
or in terms of $\tilde{\lambda}_i$,
\begin{equation}
\tilde{\lambda}_i = \frac{\hat{\epsilon}_i f(v_i)}{1+\epsilon_i \hat{\epsilon}_i}.
\end{equation}
This hypothesized form for $\lambda_i$ (and therefore $\tilde{\lambda}_i$) satisfies the ratio.  We can look back at the form postulated by BGPW and verify that this also satisfies the ratio (this form results from (A2), i.e., that $\tilde{\lambda}_i$ is independent of $\epsilon_i$ and $f(v_i)$)   Using Eq.~\ref{eq.bouch_autoregress},
\begin{equation}
r_i = \epsilon_i f(v_i) - E\left[\epsilon_i f(v_i)\middle|\Omega\right] + \eta_i,
\end{equation}
which incorporates (A1) and (A2) together, we see that,
\begin{eqnarray}
r_i^+ & = & E\left[f(v_i)\middle|\epsilon_i=+1,\Omega\right] - E\left[\epsilon_i f(v_i)\middle| \Omega\right],\\
r_i^- & = & E\left[f(v_i)\middle|\epsilon_i=-1,\Omega\right] + E\left[\epsilon_i f(v_i)\middle| \Omega\right],
\end{eqnarray}
Given that,
\begin{equation}
E\left[\epsilon_i f(v_i)\middle| \Omega\right] = \sum_{\epsilon_i} \epsilon_i E\left[f(v_i)\middle|\epsilon_i,\Omega\right] P(\epsilon_i|\Omega),
\end{equation}
and with a little bit of algebra, we have,
\begin{eqnarray}
r_i^+ & = & \frac{1-\hat{\epsilon}_i}{2}\left(E\left[f(v_i)\middle|\epsilon_i=+1,\Omega\right] + E\left[f(v_i)\middle|\epsilon_i=-1,\Omega\right]\right),\\
r_i^- & = & \frac{1+\hat{\epsilon}_i}{2}\left(E\left[f(v_i)\middle|\epsilon_i=+1,\Omega\right] + E\left[f(v_i)\middle|\epsilon_i=-1,\Omega\right]\right).
\end{eqnarray}
Notice that Eq.~\ref{eq.ratio} holds, as it should.

The point is that Eq.~\ref{eq.ratio} underspecifies a relationship between $\lambda_i$, $\tilde{\lambda}_i$ and $\{\epsilon_i, f(v_i)\}$.  We must look to empirical data to tell us about this relationship - this is done in Chapter~\ref{ch.emp_results}.  Whatever this relationship, it must satisfy (A1) and therefore Eq.~\ref{eq.ratio}.

\chapter{Theory of Market Impact\label{ch.price_impact}}

\section{Introduction}
In Chapter~\ref{ch.long_mem} I presented the two modified return models suggested by LF and BGPW,
\begin{eqnarray}
r_i & = & \frac{\epsilon_i f(v_i)}{\lambda_i}  + \eta_i,\\
r_i & = & \epsilon_i f(v_i) - \tilde{\lambda}_i  + \eta_i.\label{eq.bouch_model_pi}
\end{eqnarray}
These models are equivalent, with the liquidity parameters $\lambda_i$ and $\tilde{\lambda}_i$ related as follows,
\begin{equation}
\frac{1}{\lambda_i} \equiv 1 - \frac{\tilde{\lambda}_i}{\epsilon_i f(v_i)}.
\end{equation}
In Section~\ref{sec.liq_omega} I calculated the dependence of liquidity on $\Omega$, assuming the EMH holds, and found that under two competing sets of assumptions, (E1) and (E2), the respective results were,
\begin{eqnarray}
E\left[\tilde{\lambda}_{i} \middle|\Omega\right] & = & \sum_{k>0} a_k \epsilon_{i-k} f(v_{i-k}),\label{eq.E1_autoreg}\\
E\left[\tilde{\lambda}_{i}\middle|\Omega'\right] & = & \sum_{j} A_{i,j} \left(\frac{n_j}{n_j+1}\right)^\alpha \frac{\epsilon_j f(v_j)}{\theta_j}.\label{eq.E2_hidord}
\end{eqnarray}
The first equation suggests liquidity is set by market makers using an autoregressive model for $\epsilon_i f(v_i)$.  The second equation suggests that liquidity depends on the current state of all active hidden orders - this is either set by market makers that know this information or exists because it is required for participants to trade the active hidden orders observed.  Because the current state of active hidden orders may or may not be discernable with public data, the second equation uses $\Omega'$ instead of $\Omega$.


In this chapter, I develop a quantitative theory for the market impact of individual participants' order flow under the assumptions of (E1) and (E2).  This is formulated as a price impact function of hidden orders.  (E1) produces a power law price impact function that can be arbitrarily scaled to zero by slowing the speed of trading.  (E2) produces a logarithmic price impact function that is independent of the speed of trading.  At the end of this chapter, I determine that under (E1) market impact decays to zero and under (E2) it is permanent. 


\section{Determination of Price Impact Function}
The price impact function, $f(\cdot)$, is defined as the average price response due to a transaction as a function of the transaction's volume,
\begin{equation}
f(v_i) \equiv E\left[\epsilon_i r_i\middle|v_i\right].
\end{equation}
We can also define a price impact function for hidden orders, $F(\cdot)$,
\begin{equation}
F(V) \equiv E\left[\epsilon \sum_{i=t}^{t(N)} r_{i} \middle| V \right], 
\end{equation}
where the index $j$ has been removed.  There have been many studies that measure the revealed and hidden order price impact functions for stocks\cite{Lillo03d,Potters03,Kempf99,Chen02,Yuen07,Almgren05} and some papers that suggest price impact should theoretically scale as a square root of the volume\cite{Torre97,Gabaix03}.  The empirical papers are in agreement that the price impact is a monotonically increasing and concave function of volume, but they disagree on the functional form (some posit a power law, others posit a logarithm).

\subsection{Revealed Order Price Impact Function\label{sec.f_fit}}

\begin{figure}[htb]
\centering
\includegraphics[width=4in]{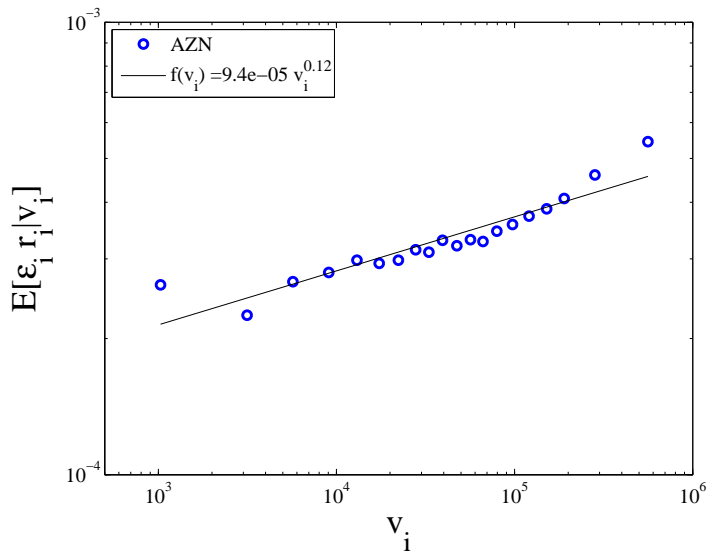}
\caption{One-transaction price impact function, $f(\cdot)$, for the stock AZN.  Empirical data is binned by volume and the expected return (measured in the direction of the transaction) is plotted on the $y$-axis.  The curve is fit by a power law with the estimated parameters given in the legend.}
\label{fig.AZN_f_fit}
\end{figure}

\begin{figure}[htb]
\centering
\includegraphics[width=4in]{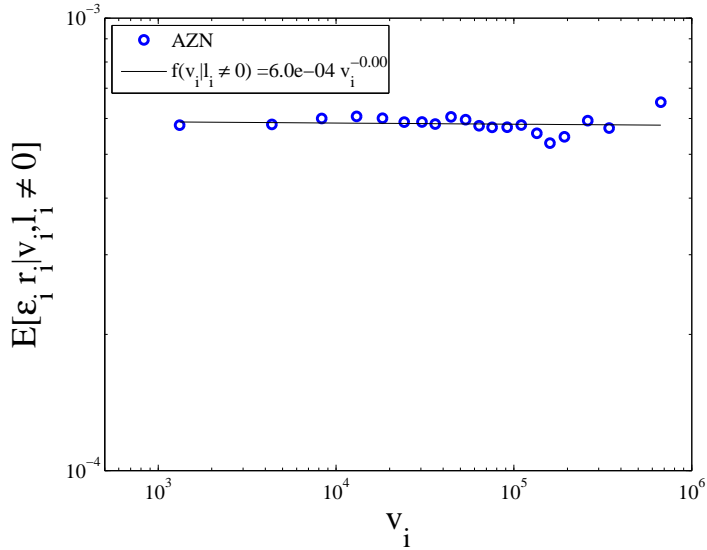}
\caption{One-transaction price impact function for the stock AZN when using only transactions that cause nonzero initial impact.  This function is completely flat, suggesting that the curvature of $f(\cdot)$ is entirely due to the probability that a transaction of size $v_i$ causes an initial impact.}
\label{fig.AZN_fnz_fit}
\end{figure}

In Fig.~\ref{fig.AZN_f_fit} I plot the empirical one transaction price impact function, $f(v_i)$, for AZN.
This is fit with a power law and the resulting exponent is $.12$, which is typical for the stocks that I study.  This suggests that larger sized transactions have a larger absolute impact than smaller sized transactions but a much smaller relative impact (the function is highly concave).  In Fig.~\ref{fig.AZN_fnz_fit} I plot the estimate of $f(v_i)$ using only transactions that cause nonzero impacts, i.e., $l_i\neq0$.
This curve is a constant.  It suggests that the size of a transaction does not determine the magnitude of the impact per se, but only determines whether an impact occurs at all.  The curvature of $f(v_i)$ in Fig.~\ref{fig.AZN_f_fit} is then entirely due to the probability that a transaction causes a nonzero impact - the larger the size of the transaction, the larger this probability.  This result was anticipated in the paper by Farmer et al.\cite{Farmer04b} where it was shown that participants condition the size of their transactions to the size that is currently offered at the best price in the orderbook, and rarely (about $15\%$ of the time) submit transactions with a larger size.


\subsection{Hidden Order Price Impact Function}

So far, I have assumed that each active hidden order $j$ is split into smaller pieces of equal size, and that the size of the pieces, $v_j$, is independent of the size of the hidden order, $V_j$.  This assumption is motivated by the results in the previous section - that participants initiating transactions do not determine the size of their orders, but take what liquidity providers offer at the current best price.\footnote{Sometimes they take less, as seen with the large proportion of non-penetrating orders.  I am mainly interested in hidden orders of more than one or two transactions, and there is no incentive for the revealed parts of these orders to be less than what is currently offered at the best price - so I assume their size should equal what is offered by liquidity providers at the current best price.}  I also have assumed that the pieces of each hidden order are executed, on average, every $\theta_j$ transactions.\footnote{This is similar to a VWAP trade (Volume Weighted Average Price).  This is a standard execution strategy and requires participation in a percentage of all transactions.}


If we take the assumptions above as true, then we can calculate the expected price response due to a hidden order.  Disregarding the index $j$, the expected return of a hidden order with parameters $\Psi = \{\epsilon, v, \theta, N\}$ is,
\begin{eqnarray}
E_\Psi\left[R\right] & = & E_\Psi\left[\sum_{i=t}^{t(N)} r_{i} \right],\\
 & = & E_\Psi\left[ \sum_{i=t}^{t(N)} \epsilon_{i} f(v_{i}) - \tilde{\lambda}_{i}\right],\\
 & = & N\epsilon f(v) - E_\Psi\left[\sum_{i=t}^{t(N)} \tilde{\lambda}_{i}\right],\label{eq.hid_impact}
\end{eqnarray}
Here, I'm using the shorthand, $E_\Psi\left[R\right]\equiv E\left[R \middle| \Psi \right]$.  The impact $R$ is defined as the full return caused by the hidden order measured from directly before it was started to immediately after it finishes.  Eq.~\ref{eq.hid_impact} results from the following: the expected sign, $E_\Psi\left[\epsilon_i\right]$, is zero for all transactions other than those that are part of the larger hidden order - this assumes (R2).  We can now use Eq.~\ref{eq.hid_impact} to derive the expected price response due to hidden orders under the assumptions of (E1) and (E2).

\subsubsection{Impact Under (E1)}
Recall the equation for $E_\Psi\left[\tilde{\lambda}_i\middle|\Omega\right]$ in (E1),
\begin{equation}
E_\Psi\left[\tilde{\lambda}_{i} \middle|\Omega\right] = \sum_{k>0} a_k \epsilon_{i-k} f(v_{i-k}).
\end{equation}
Because Eq.~\ref{eq.hid_impact} calculates the \emph{unconditional} expected return of a hidden order, the full information set $\Omega$ is not included \dots only information about the parts of that hidden order are included.  The expected value of $\tilde{\lambda}_{i}$ during the hidden order is then,
\begin{eqnarray}
E_\Psi\left[\tilde{\lambda}_{i} \middle| t \le i < t(N)\right] & = & E_\Psi\left[\sum_{k>0} a_k \epsilon_{i-k} f(v_{i-k})\middle| t \le i < t(N)\right],\\
& = & E_\Psi\left[\sum_{k=1}^{i-t} a_k \frac{\epsilon f(v)}{\theta}\middle| t \le i < t(N)\right].\label{eq.hid_autoreg}
\end{eqnarray}
Eq.~\ref{eq.hid_autoreg} results from the following: the expected sign, $E_\Psi\left[\epsilon_{i-k}\right]$, is zero for all transactions other than those that are part of the larger hidden order - again, this assumes (R2).  $\left(a_k \epsilon f(v)/\theta\right)$ is the expected contribution of the hidden order to the autoregressive model at each lag, $k$, while the order is active.  Putting Eq.~\ref{eq.hid_autoreg} into Eq.~\ref{eq.hid_impact} gives,
\begin{eqnarray}
E_\Psi\left[R\right] & = & N\epsilon f(v) - E_\Psi\left[\sum_{i=t}^{t(N)} \sum_{k=1}^{i-t} a_k \frac{\epsilon f(v)}{\theta}\right],\\
 & \approx & N\epsilon f(v) -  \frac{\epsilon f(v)}{\theta} E_\Psi\left[ (t(N)-t) - \frac{(t(N)-t)^{1-\phi}}{1-\phi}  \right],\label{eq.R_tmp_E1}\\
 & \approx  & \frac{\epsilon f(v) \theta^{-\phi} N^{1-\phi}}{1-\phi}.\label{eq.R_E1}
\end{eqnarray}
Eq.~\ref{eq.R_tmp_E1} is a result of approximating the sums with integrals and by approximating $a_k$,
\begin{equation}
a_k \approx \phi k^{-1-\phi},
\end{equation}
which uses Eqs.~\ref{eq.ar_scaling} and \ref{eq.bb_gam} and that $\int_1^\infty a_k = 1$ for a FARIMA process.  Eq.~\ref{eq.R_E1} uses that $\theta N = E[(t(N)-t)]$.

The final result is:
\begin{eqnarray}
E\left[R \middle| \epsilon, v, \theta, N \right] & = & \frac{\epsilon f(v)}{1-\phi} \  \theta^{-\phi} N^{1-\phi},\label{eq.R_E1_final}\\
 & = & \frac{\epsilon f(v)}{v^{1-\phi}(1-\phi)} \  \theta^{-\phi} V^{1-\phi}.\label{eq.R_E1_final_V}
\end{eqnarray}
We know from Chapter~\ref{ch.long_mem} that $\phi = H-1/2$, where $H$ is the Hurst exponent of the $\epsilon_i$ series.  For stocks, $1/2 < H < 1$, so that $ 0 < \phi < 1/2$.  Eq.~\ref{eq.R_E1_final_V} then has very specific implications for trading.  It suggests that the impact of a hidden order is a concave function of its total volume (a power law with exponent greater than $1/2$ but less than $1$) and that the expected total impact can arbitrarily be scaled by changing the speed of trading (changing $\theta$).  In the limit that $\theta$ is infinitely large, the expected total impact is zero.

We can now determine $F(\cdot)$ under (E1):
\begin{eqnarray}
F(V) & = & \epsilon E\left[R\middle|V\right],\\
 & = & \frac{E\left[f(v) v^{\phi-1} \theta^{-\phi} \middle| V \right]}{1-\phi} V^{1-\phi}.\label{eq.R_scaling_E1}
\end{eqnarray}
In general, $v$ and $\theta$ may depend on $V$, so they stay within the expectation.

\subsubsection{Impact Under (E2)}
Recall the equation for $E_\Psi\left[\tilde{\lambda}_i\middle|\Omega'\right]$ in (E2),
\begin{equation}
E_\Psi\left[\tilde{\lambda}_{i}\middle|\Omega'\right] = \sum_{j} A_{i,j} \left(\frac{n_j}{n_j+1}\right)^\alpha \frac{\epsilon_j f(v_j)}{\theta_j}.
\end{equation}
Because Eq.~\ref{eq.hid_impact} calculates the \emph{unconditional} expected return of a hidden order, the full information set $\Omega'$ is not included \dots only information about the parts of that hidden order are included.  The expected value of $\tilde{\lambda}_{i}$ during the hidden order is then,
\begin{eqnarray}
E_\Psi\left[\tilde{\lambda}_{i} \middle| t \le i < t(N)\right] & = & E_\Psi\left[\sum_{j} A_{i,j} \left(\frac{n_j}{n_j+1}\right)^\alpha \frac{\epsilon_j f(v_j)}{\theta_j}\middle| t \le i < t(N)\right],\\
& = & E_\Psi\left[\left(\frac{n}{n+1}\right)^\alpha \frac{\epsilon f(v)}{\theta}\middle| t \le i < t(N)\right],\label{eq.hid_tmp_nimpact}\\
& = & \left(\frac{(i-t)/\theta}{(i-t)/\theta+1}\right)^\alpha \frac{\epsilon f(v)}{\theta}.\label{eq.hid_nimpact}
\end{eqnarray}
Eq.~\ref{eq.hid_tmp_nimpact} results from the following: the expected sign, $E_\Psi\left[\epsilon_{j}\right]$, is zero for all transactions other than those that are part of the larger hidden order - again, this assumes (R2).  Eq.~\ref{eq.hid_nimpact} uses that $E_\Psi\left[n\middle|t \le i < t(N)\right] = (i-t)/\theta$.  Putting Eq.~\ref{eq.hid_nimpact} into Eq.~\ref{eq.hid_impact} gives,
\begin{eqnarray}
E_\Psi\left[R\right] & = & N\epsilon f(v) - E_\Psi\left[\sum_{i=t}^{t(N)} \left(\frac{(i-t)/\theta}{(i-t)/\theta+1}\right)^\alpha \frac{\epsilon f(v)}{\theta} \right],\\
 & = & N\epsilon f(v) - \epsilon f(v) \sum_{x=0}^{N} \left(\frac{x}{x+1}\right)^\alpha,\label{eq.x_change}\\
 & = & \epsilon f(v) \sum_{x=0}^{N} \left(1+\left(\frac{x}{x+1}\right)^\alpha \right)
\end{eqnarray}
Eq.~\ref{eq.x_change} uses a change of variable $x\equiv (i-t)/\theta$.  In the limit $N>>1$,
\begin{eqnarray}
E_\Psi\left[R\right] & \approx & \epsilon f(v) \sum_{x=0}^{N} \frac{\alpha}{1+x},\\
 & \approx & \alpha \epsilon f(v) \log{\left(1+N\right)}. 
\end{eqnarray}

The final result is:
\begin{eqnarray}
E\left[R \middle| \epsilon, v, \theta, N \right] & = & \alpha \epsilon f(v) \log{\left(1+N\right)},\label{eq.R_E2_final}\\
& = & \alpha \epsilon f(v) \log{\left(1+\frac{V}{v}\right)}.\label{eq.R_E2_final_V}
\end{eqnarray}
Eq.~\ref{eq.R_E2_final_V} suggests that the impact of a hidden order is a concave function of its total volume, specifically a logarithm.  Notice that the impact function does not depend on $\theta$, so that the speed of trading does not influence the total impact of the transaction.

We can now determine $F(\cdot)$ under (E2):
\begin{eqnarray}
F(V) & = & \epsilon E\left[R\middle|V\right],\\
 & = & \alpha E\left[f(v) \log{\left(1+\frac{V}{v}\right)} \middle| V \right].\label{eq.R_scaling_E2}
\end{eqnarray}
Again, $v$ may depend on $V$, so the expectation remains.

\section{Decaying or Permanent Impact}
In Chapter~\ref{ch.long_mem} we were left with two interpretations, (I1) and (I2), for the model,
\begin{equation}
r_i = \epsilon_i f(v_i) - \tilde{\lambda}_i + \eta_i.
\end{equation}
(I1) states that transaction impacts are $\left(\epsilon_i f(v_i) - \tilde{\lambda}_i\right)$ and are permanent.  (I2) states that transaction impacts are $\epsilon_i f(v_i)$ and are transient.  I argue that the most natural interpretation should correspond with what market participants observe for their own initiated transactions - others might disagree, but I believe this is how most people use the terminology.  As shown at the end of Chapter~\ref{ch.long_mem}, at the finest level of measurement, a transaction causes an impact approximately equal to $\left(\epsilon_i f(v_i) - \tilde{\lambda}_i\right)$ - thus a market participant observes this impact and (I1) is correct for the form of the initial impact.  In this section, I will study the second part of these interpretations \dots whether a market participant observes her impact as permanent or transient.  We will see that (I1) is correct if (E2) is correct and that (I2) is correct if (E1) is correct. 

I introduce here the term, $\Phi_k$, an additional information term such that $\Phi_k=1$ when the larger hidden order that was partially transacted at time $i-k$ has finished and $\Phi_k=0$ otherwise.  Starting with Eq.~\ref{eq.bouch_model_pi},
\begin{equation}
r_i = \epsilon_i f(v_i) - \tilde{\lambda}_i  + \eta_i,
\end{equation}
we have,
\begin{eqnarray}
E\left[r_{i}\middle|\epsilon_{i-k},\Phi_k\right] & = & E\left[\epsilon_{i} f(v_{i}) \middle|\epsilon_{i-k},\Phi_k\right] - E\left[\tilde{\lambda}_{i} \middle|\epsilon_{i-k},\Phi_k\right],\\
 & \ne & 0. \label{eq.ne_zero}
\end{eqnarray}
Eq.~\ref{eq.ne_zero} holds in general, such that $E\left[r_{i}\middle|\epsilon_{i-k},\Phi_k\right]$ may or may not be equal to zero.  We can argue that, 
\begin{equation}
E\left[\epsilon_{i} f(v_{i}) \middle|\epsilon_{i-k},\Phi_k=1\right]  =  0.
\end{equation}
We have evidence that transactions are correlated only because of the splitting of orders (R1,R2).  Because the order that produced $\epsilon_{i-k}$ has ended, i.e. $\Phi_k=1$, then we expect it to have no predictive power for future order signs.  This gives,
\begin{equation}
E\left[r_{i}\middle|\epsilon_{i-k},\Phi_k=1\right] = - E\left[\tilde{\lambda}_{i} \middle|\epsilon_{i-k},\Phi_k=1\right].
\end{equation}
This is an important equation, because it measures the expected return given that the larger order partially transacted at time $i-k$ has ended \dots if,
\begin{equation}
\epsilon_{i-k} E\left[r_{i}\middle|\epsilon_{i-k},\Phi_k=1\right] < 0,
\end{equation}
the impact of the transaction at time $i-k$ decays, and if,
\begin{equation}
\epsilon_{i-k} E\left[r_{i}\middle|\epsilon_{i-k},\Phi_k=1\right] = 0,
\end{equation}
the impact of the transaction at time $i-k$ is permanent.  This is entirely determined by the term,
\begin{equation}
-\epsilon_{i-k} E\left[\tilde{\lambda}_{i} \middle|\epsilon_{i-k},\Phi_k=1\right].
\end{equation}

\subsection{Decay or Permanence under (E1)}
Under the assumptions of (E1), 
\begin{eqnarray}
-\epsilon_{i-k} E\left[\tilde{\lambda}_{i} \middle|\epsilon_{i-k},\Phi_k=1\right] & = & -\epsilon_{i-k} E\left[\tilde{\lambda}_{i} \middle|\epsilon_{i-k}\right],\\
 & < & 0,\label{eq.E1_decay}
\end{eqnarray}
and the impact of the transaction decays when the larger hidden order completes.  The first equation holds because the autoregressive model does not distinguish between transactions that are part of orders that have completed and transactions that are part of orders that have not completed.  Eq.~\ref{eq.E1_decay} comes from Eq.~\ref{eq.pred_liquidity}, and is justified because under (E1), Eq.~\ref{eq.E1_autoreg}, the order at time $i-k$ always contributes to liquidity in the $\epsilon_{i-k}$ direction.

Interpreting this as in BGPW (I2), the original impact of the transaction is $\epsilon_{i-k} f(v_{i-k})$ and decays by a fraction, $a_k$, at each timestep into the infinite future (this is its contribution to the autoregressive model at each timestep).  In the infinite future, the impact approaches zero - this is because the sum of the autoregressive coefficients $a_k$ is $1$.

We now have an explanation for the qualitative results of Eq.~\ref{eq.R_E1_final_V}.  Hidden order impact is a concave function of $N$ and $V$ because the impacts caused by the first pieces of the order have partially decayed by the end of the hidden order - keeping $\theta$ constant, the larger the $N$ the more the decay.  Also, hidden order impacts can arbitrarily be scaled by changing the speed of trading $\theta$ because the longer a hidden order takes to trade, the longer the first impacts of the order have had to decay. 

\subsection{Decay or Permanence under (E2)}
Under the assumptions of (E2), 
\begin{equation}
-\epsilon_{i-k} E\left[\tilde{\lambda}_{i} \middle|\epsilon_{i-k},\Phi_k=1\right] = 0,\label{eq.E2_perm}\\
\end{equation}
and the impact of the transaction is permanent when the larger hidden order completes.  Eq.~\ref{eq.E2_perm} holds because under (E2), Eq.~\ref{eq.E2_hidord}, an individual transaction no longer contributes to liquidity, $\tilde{\lambda}_i$, when the larger hidden order has completed.

Interpreting this as in LF (I1), the impact of the transaction at time $i-k$ is $\left(\epsilon_{i-k} f(v_{i-k}) - \tilde{\lambda}_{i-k}\right)$ and is permanent.  The transaction influences future impacts by contributing to $\tilde{\lambda}_{i}$ until the larger hidden order it was part of has completed.  Once the hidden order completes, the revealed orders no longer contribute to $\tilde{\lambda}_{i}$ and their impacts remain constant, i.e., are permanent.

There is a slight complication to this.  It is possible that under (E2) a hidden order continues to influence liquidity after it has completed, and therefore causes some amount of impact decay.  If market makers are determining hidden orders, then it can be argued that the hidden order should continue to influence liquidity, even after it completes, up until the market makers have reasonably determined that the order is no longer transacting (the time this takes might be some low multiple of $\theta$).  We will discuss this again when looking at empirical data in the next chapter. 

\chapter{Empirical Results \label{ch.emp_results}}

\section{Introduction}
In Chapters 3 and 4 I assumed that the weak form of the efficient market hypothesis (EMH) holds, 
\begin{equation}
E\left[r_i\middle|\Omega\right]=0,
\end{equation}
where $\Omega$ is any set of publicly available historical financial data.  From this assumption, I derived several results for the following equivalent return models:
\begin{eqnarray}
r_i & = & \frac{\epsilon_i f(v_i)}{\lambda_i}  + \eta_i,\\
r_i & = & \epsilon_i f(v_i) - \tilde{\lambda}_i  + \eta_i.\label{eq.bouch_ch5}
\end{eqnarray}
Specifically, I was interested in deriving the dependence of the liquidity parameters $\lambda_i$ and $\tilde{\lambda}_i$ on the order flow variables $\Omega$, $\epsilon_i$, and $v_i$.  In this chapter I use empirical data to qualitatively test the developed theory.

\begin{figure}[t]
\centering
\includegraphics[width=4in]{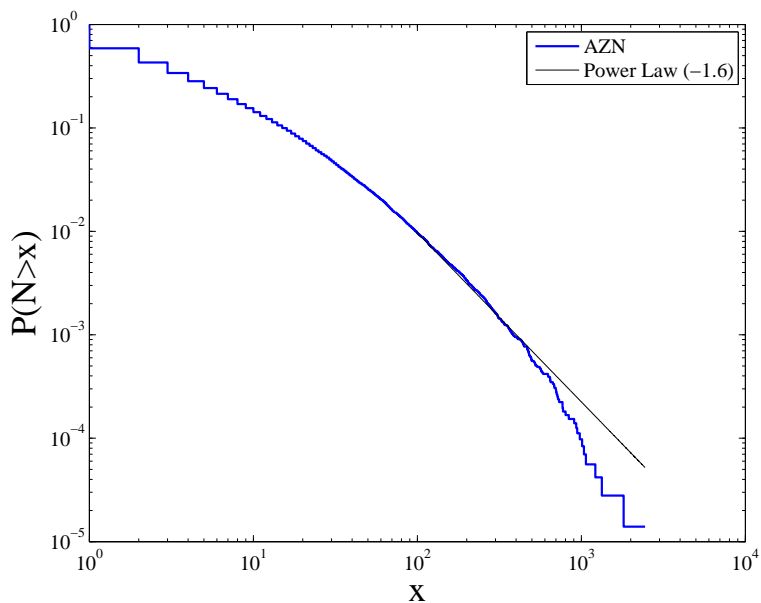}
\caption{The cumulative distribution function of hidden order size $N$ for the stock AZN.  The tail of this distribution decays as a power law with exponent $-1.6$ (although it appears to decrease faster than a power law for $N>50$).  The measured tail exponent of $1.6$ is evidence the LMF theory of hidden orders is correct.}
\label{fig.AZN_N_cdf}
\end{figure}

\section{Hidden Order Determination}
\begin{figure}[t]
\centering
\includegraphics[width=4in]{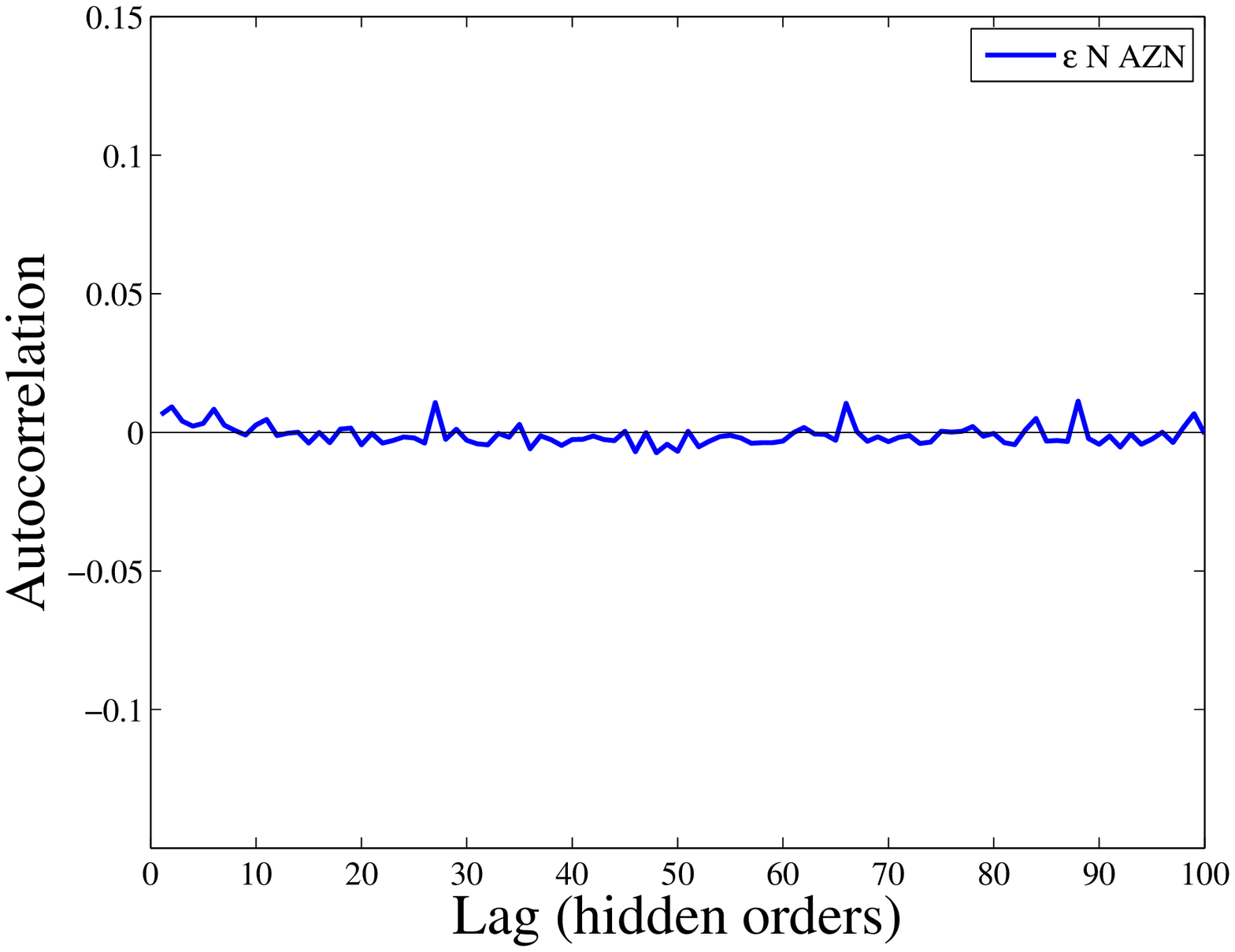}
\caption{Autocorrelation function of the signed hidden order size series $\epsilon N$ for the stock AZN.  Signed hidden order size is not autocorrelated - this suggests the algorithm I use to determine hidden orders is working.}
\label{fig.AZN_eN_autocorr}
\end{figure}
As mentioned earlier, the dataset I use has a unique feature - brokerage codes are attached to all orders.  These codes are numbers that uniquely identify the member firms of the London Stock Exchange (in the dataset they are anonymous - so that I cannot identify the firm by name).  Although I cannot determine if two orders originating from the same brokerage belong to the same participant or belong to two different participants using the same brokerage, it might still be possible to determine individual participant order flow.  As a crude and first attempt to determine the order flow of participants\footnote{I am currently participating in a study that uses more sophisticated methods to determine participant order flow for LSE stocks.}, I do the following: if two transactions are initiated by the same brokerage, are within 100 transactions of each other, and have the same sign $\epsilon_i$ \dots I consider these two transactions to originate from the same participant (and to be part of the same larger hidden order).  Using this ruleset, I group all transactions into hidden orders.  Some hidden orders will be of size 1 because either there were no transactions initiated by the brokerage for 100 transactions before and after the original transaction, or the activity has the opposite sign (in fact, most hidden orders are of size 1 - which is expected if hidden order size is distributed as a pure power law).  In Fig.~\ref{fig.AZN_N_cdf} I plot the cumulative distribution of hidden order sizes for the stock AZN using the method just described.  The tail exponent is estimated as $1.6$.  Looking back at Section~\ref{sec.LMF}, LMF suggest that,
\begin{equation}
\gamma = \alpha - 1,
\end{equation}
where $\gamma$ is the decay exponent of the autocorrelation function for $\epsilon_i$ and $\alpha$ is the tail exponent of the hidden order size cumulative distribution.  From Fig.~\ref{fig.AZN_N_cdf}, $\alpha=1.6$ and from Fig.~\ref{fig.AZN_sign_autocorr}, $\gamma=.67$.  This is an accord with the LMF theory and suggests my algorithm for determining hidden orders is working.

For further evidence that the algorithm correctly determines hidden orders, I plot the autocorrelation function of $\epsilon N$ in Fig.~\ref{fig.AZN_eN_autocorr}.  By aggregating transactions of the same sign initiated under the same brokerage code, the autocorrelated structure of transactions as seen in Fig.~\ref{fig.AZN_sign_autocorr} is suppressed in Fig.~\ref{fig.AZN_eN_autocorr}.

\section{Dependence of Liquidity on $\Omega$ or $\Omega'$ \label{sec.liq_om}}
When determining the dependence of liquidity on $\Omega$, I stated two extreme sets of assumptions about market participants that bound the solution between two equations.
\begin{eqnarray}
E\left[\tilde{\lambda}_{i} \middle|\Omega\right] & = & \sum_{k>0} a_k \epsilon_{i-k} f(v_{i-k}),\label{eq.E1_autoreg_ch5}\\
E\left[\tilde{\lambda}_{i}\middle|\Omega'\right] & = & \sum_{j} A_{i,j} \left(\frac{n_j}{n_j+1}\right)^\alpha \frac{\epsilon_j f(v_j)}{\theta_j}.\label{eq.E2_hidord_ch5}
\end{eqnarray} 
The first equation holds under the set of assumptions (E1): that market makers condition liquidity on the predictability of transaction order flow, and that the best predictor they can use is an autoregressive model for $\epsilon_i$ and $v_i$.  The second equation holds under the set of assumptions (E2): that either market participants condition their order flow so as not to produce predictable returns, or alternatively, that market makers can determine who is initiating transactions and when they have finished placing a hidden order.  In this section, I will try to determine which of the two equations, Eq.~\ref{eq.E1_autoreg_ch5} or Eq.~\ref{eq.E2_hidord_ch5}, is better supported by empirical data.  I first look at the expected return of a hidden order as a function of its size.  (E1) and (E2) offer two competing and testable equations for this relationship.  We will see that (E2) fits the data much better than (E1).  I then look at the \emph{transaction imbalance} and \emph{return imbalance} under (E1) and (E2), and show that after a certain delay the return imbalance matches the transaction imbalance for both (E1) and (E2) so that returns remain unpredictable with a certain lag.  This means that the market is approximately efficient under both (E1) and (E2) - because (E2) includes a much stronger information set, the market appears to be responding as suggested by (E2) with efficiency under (E1) as only a consequence of this.

\subsection{Expected Return}
In Chapter~\ref{ch.price_impact}, I derived two competing equations for the expected return of a hidden order.  Under (E1), this was:
\begin{equation}
E\left[R \middle| \epsilon, v, \theta, N \right] = \frac{\epsilon f(v)}{1-\phi} \  \theta^{-\phi} N^{1-\phi},
\end{equation}
and under (E2), this was:
\begin{equation}
E\left[R \middle| \epsilon, v, \theta, N \right] = \alpha \epsilon f(v) \log{\left(1+N\right)}.
\end{equation}

\begin{figure}[!ht]
\centering
\includegraphics[width=4in]{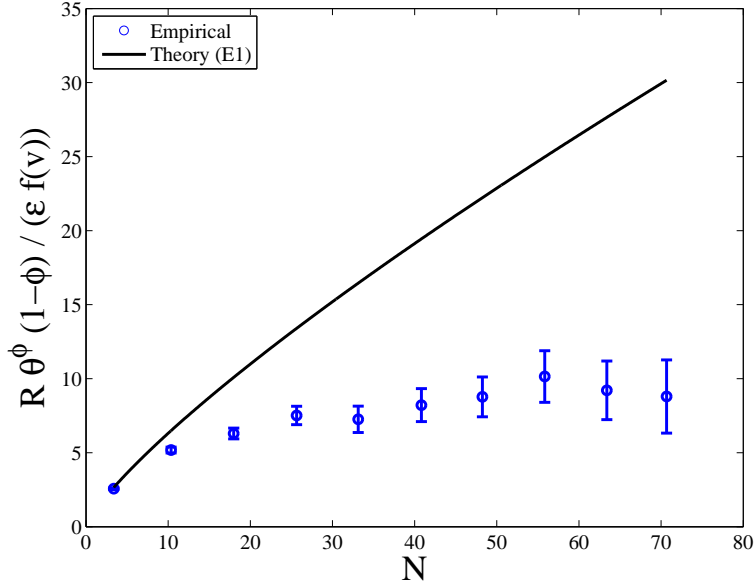}
\caption{Scaled hidden order return as a function of hidden order size $N$.  The empirical data includes all hidden orders for all 6 stocks.  This data is binned by $N$ and the mean value of $R\theta^\phi(1-\phi)/(\epsilon f(v))$ is plotted for each bin with the error bars showing the standard error of this average.  The theory curve is what is predicted under the assumptions of (E1), i.e., $N^{1-\phi}$.  I use $\phi=.2$ to generate this curve, which is the typical value of $\phi$ for the 6 stocks (see the table in the Appendix).}
\label{fig.TOT_retvsN_E1_bin}
\end{figure}
\begin{figure}[!ht]
\centering
\includegraphics[width=4in]{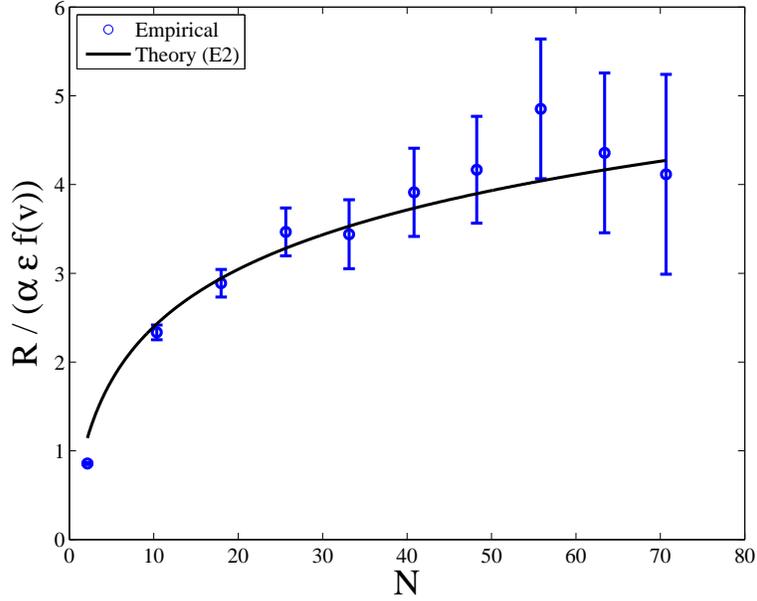}
\caption{Scaled hidden order return as a function of hidden order size $N$.  The empirical data includes all hidden orders for all 6 stocks.  This data is binned by $N$ and the mean value of $R/(\alpha \epsilon f(v))$ is plotted for each bin with the error bars showing the standard error of this average.  The theory curve is what is predicted under the assumptions of (E2), i.e., $\log(1+N)$.}
\label{fig.TOT_retvsN_E2_bin}
\end{figure}
\begin{figure}[!ht]
\centering
\includegraphics[width=4in]{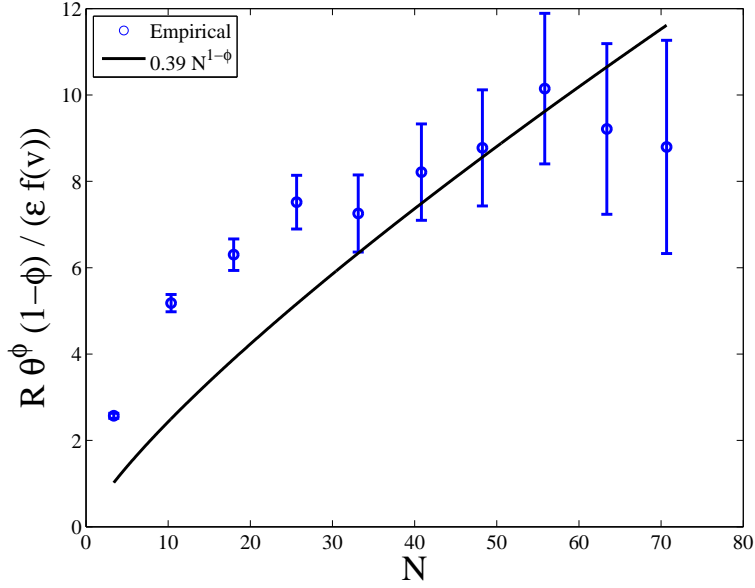}
\caption{Scaled hidden order return  as a function of hidden order size $N$.  The empirical data includes all hidden orders for all 6 stocks.  This data is binned by $N$ and the mean value of $R\theta^\phi(1-\phi)/(\epsilon f(v))$ is plotted for each bin with the error bars showing the standard error of this average.  The scale parameter for the theory curve is fit using least squares and the result is $0.39 N^{1-\phi}$, where $\phi=.2$, the typical value for these stocks.  Even with this free parameter, the theoretic curve (derived under the assumptions of (E1)) poorly fits the data.}
\label{fig.TOT_retvsN_E1_fit_bin}
\end{figure}

In Fig.~\ref{fig.TOT_retvsN_E1_bin} and Fig.~\ref{fig.TOT_retvsN_E2_bin}, I test the dependence on $N$ by dividing the measured $R$ by all other variables found on the right hand side of the above equations (also empirically measured - see the Appendix for details).  In the case of (E1), this should scale with $N^{1-\phi}$, and in the case of (E2), this should scale with $\log{\left(1+N\right)}$.  Both figures include data from all hidden orders for all 6 stocks.  The data is binned by $N$ and the average taken for each bin with the error bars showing the standard error of this average.  As seen in the figures, (E2) fits the data much better than (E1).  To show that this is not a problem with the scale parameter for (E1), I allow the scale to be a free parameter and I fit it using least squares in Fig.~\ref{fig.TOT_retvsN_E1_fit_bin}.  Even with this free parameter, (E1) still fails to produce a good fit to the data.

\subsection{Transaction and Return Imbalance}
\begin{figure}[!ht]
\centering
\includegraphics[width=4.5in]{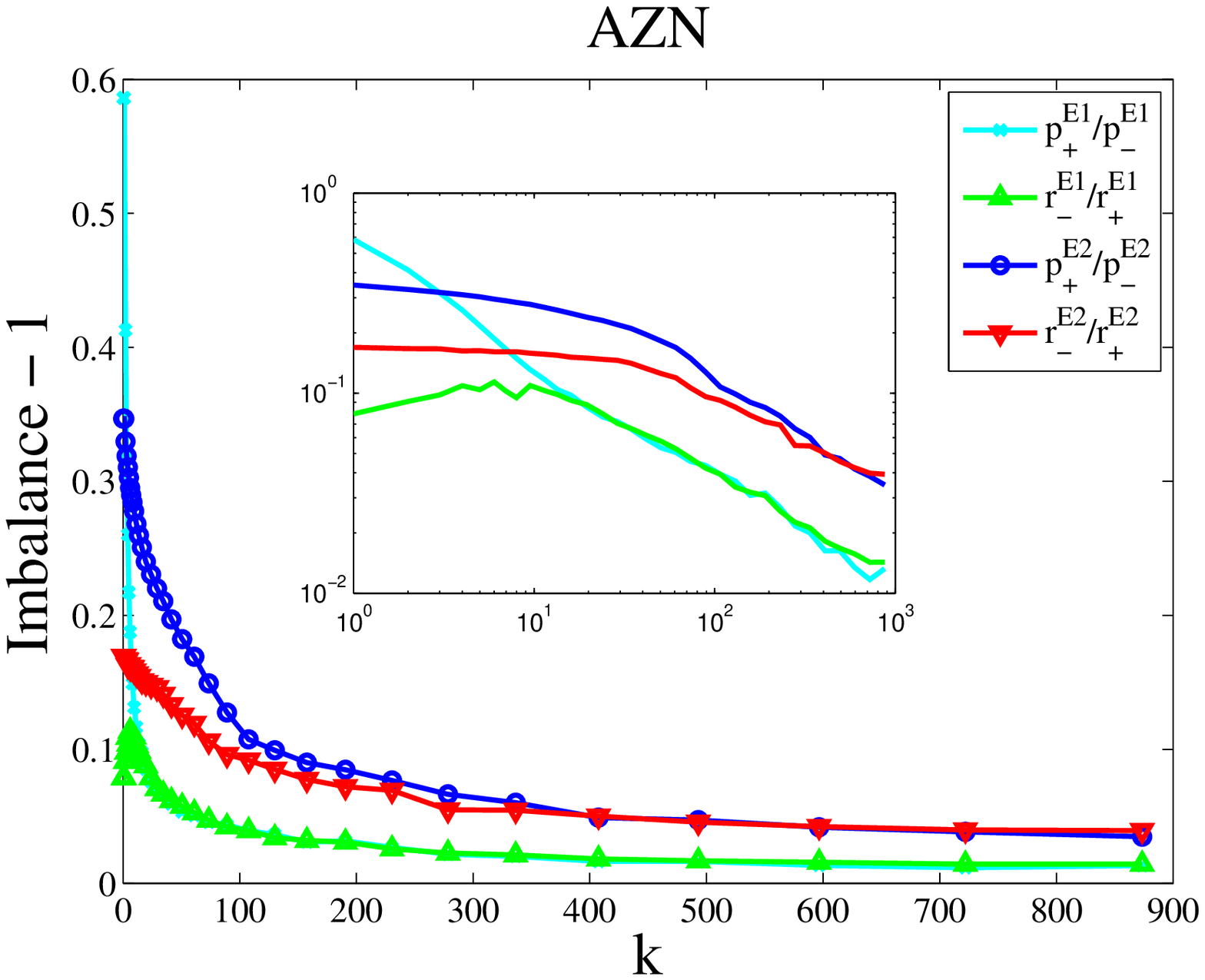}
\includegraphics[width=4.5in]{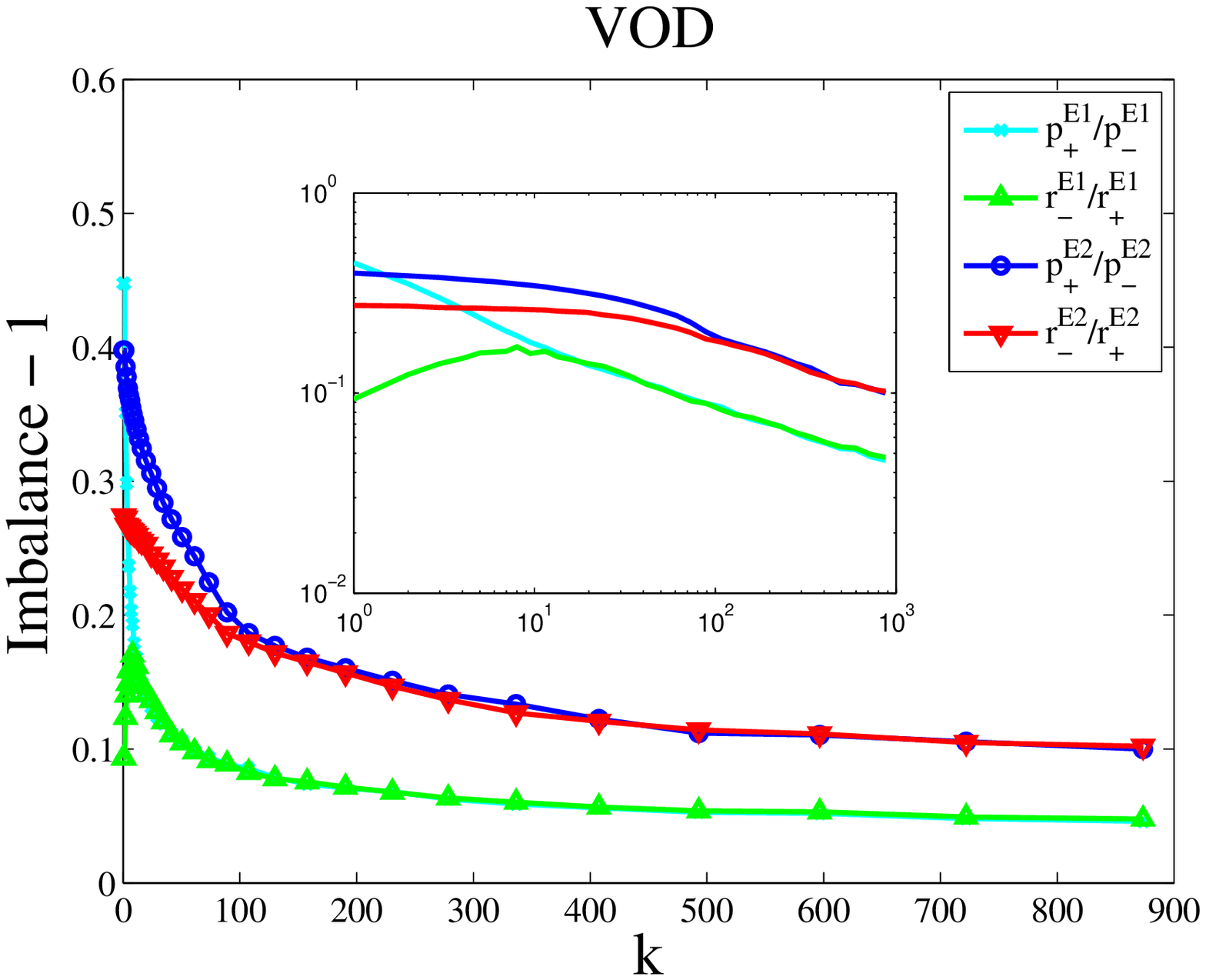}
\caption{The transaction imbalance ($p^{E1}_+/p^{E1}_-$ and $p^{E1}_+/p^{E1}_-$) and return imbalance ($r^{E1}_-/r^{E1}_+$ and $r^{E2}_-/r^{E2}_+$) under the assumptions of (E1) and (E2) respectively, as a function of lag $k$ for the stocks AZN and VOD.  The $y$-axis is measured in units of the average spread.  The transaction imbalance measures the predictability of transaction sign - to the extent that the return imbalance matches the transaction imbalance, this cannot be used to predict returns.  In both cases, the return imbalance matches the transaction imbalance with a lag, although the lag is larger for (E2) than for (E1).}
\label{fig.BOTH_sE_imb_avg}
\end{figure}

In Section~\ref{sec.imbalance} I looked at the \emph{transaction imbalance} and the \emph{return imbalance} caused by the existence of a transaction at time $i$ of known sign, $\epsilon_i$.  Analogous to that section, I look here at the transaction and return imbalance caused by a \emph{predicted} sign at time $i$.  The predicted sign under (E1) and (E2) is the following,
\begin{eqnarray}
\hat{\epsilon}_{i}^{E1} & = & \sum_{k>0} a_k \epsilon_{i-k},\\
\hat{\epsilon}_{i}^{E2} & = & \sum_{j} A_{i,j} \left(\frac{n_j}{n_j+1}\right)^\alpha \frac{\epsilon_j}{\theta_j}.\label{eq.E2_signp}
\end{eqnarray}
The probability that future transactions have this same sign or have the opposite sign is defined:
\begin{eqnarray}
p_+^{E1}(k) & = & P\left(\epsilon_{i+k} = \hat{\epsilon}_{i}^{E1} \middle|\hat{\epsilon}_{i}^{E1}\right),\\
p_-^{E1}(k) & = & P\left(\epsilon_{i+k} \neq \hat{\epsilon}_{i}^{E1} \middle|\hat{\epsilon}_{i}^{E1}\right),\\
p_+^{E2}(k) & = & P\left(\epsilon_{i+k} = \hat{\epsilon}_{i}^{E2} \middle|\hat{\epsilon}_{i}^{E2}\right),\\
p_-^{E2}(k) & = & P\left(\epsilon_{i+k} \neq \hat{\epsilon}_{i}^{E2} \middle|\hat{\epsilon}_{i}^{E2}\right).
\end{eqnarray}
The expected return of future transactions (measured in the direction of the transaction) with the same and opposite sign are defined analogously:
\begin{eqnarray}
r_+^{E1}(k) & = & E\left[\epsilon_{i+k} r_{i+k} \middle| \epsilon_{i+k} = \hat{\epsilon}_{i}^{E1} \right],\\
r_-^{E1}(k) & = & E\left[\epsilon_{i+k} r_{i+k} \middle| \epsilon_{i+k} \neq \hat{\epsilon}_{i}^{E1} \right],\\
r_+^{E2}(k) & = & E\left[\epsilon_{i+k} r_{i+k} \middle| \epsilon_{i+k} = \hat{\epsilon}_{i}^{E2} \right],\\
r_-^{E2}(k) & = & E\left[\epsilon_{i+k} r_{i+k} \middle| \epsilon_{i+k} \neq \hat{\epsilon}_{i}^{E2} \right].
\end{eqnarray}
Just as in Section~\ref{sec.imbalance}, the following ratios determine if the market is efficient under (E1) and (E2),
\begin{eqnarray}
\frac{r_-^{E1}(k)}{r_+^{E1}(k)} & = & \frac{p_+^{E1}(k)}{p_-^{E1}(k)},\\
\frac{r_-^{E2}(k)}{r_+^{E2}(k)} & = & \frac{p_+^{E2}(k)}{p_-^{E2}(k)}.\label{eq.E2_ratio}
\end{eqnarray}
As a reminder, the term on the left hand side is called the \emph{return imbalance} and the term on the right hand side is called the \emph{transaction imbalance}.  In Fig.~\ref{fig.BOTH_sE_imb_avg}, I plot all these imbalances for AZN and VOD.
Notice that the transaction imbalance is much stronger under (E2) than under (E1) (and both are stronger than the simple use of $\epsilon_i$ in Fig.~\ref{fig.BOTH_s_imb_avg} from Section~\ref{sec.imbalance}).  This confirms that information about who is initiating transactions and when they have finished increases the predictability of order flow  - a nontrivial result, but expected under (R1) and (R2).  Much more surprising is that the liquidity response (return imbalance) is much higher under (E2) than (E1).  This confirms that the information available in (E2) is a better predictor of liquidity, and that the market is responding to this information.

As before, there is a delay before the return imbalance meets the transaction imbalance.  The delay is longer for (E2) than (E1), suggesting the market does not respond as quickly to the information assumed available in (E2).  The market does, however, eventually respond to this information (typically at $k\approx100$) - a quite amazing result given how much detailed information is used in (E2).  This result offers more support that liquidity responds to hidden order information (E2), and not just an autoregressive model for order flow (E1).

\subsection{Conclusion}
Given the results of this section, (E2) gives a more accurate description of the market than (E1).  This means that either market makers can discern hidden order flow, or that market participants condition their orders on available liquidity.  I will leave the determination of this to future work.  In the next section, I will assume that (E2) is valid when determining the dependence of liquidity on $\epsilon_i$ and $v_i$ - this means that all predictors will be based on the assumptions of (E2).

\section{Dependence of Liquidity on $\epsilon_i$ and $v_i$\label{sec.liq_eps}}

When determining the dependence of liquidity on $\epsilon_i$ and $f(v_i)$, I found that the relationship is underspecified but the following ratio must be satisfied,
\begin{equation}
\frac{r_i^+}{r_i^-} = \frac{1-\hat{\epsilon}_i}{1+\hat{\epsilon}_i},\label{eq.ratio_ch5}
\end{equation}
where,
\begin{eqnarray}
r_i^+ & \equiv & E\left[\frac{f(v_i)}{\lambda_i}\middle|\epsilon_i=+1,\Omega\right],\\
r_i^- & \equiv & E\left[\frac{f(v_i)}{\lambda_i}\middle|\epsilon_i=-1,\Omega\right],\\
\hat{\epsilon}_i & \equiv & E\left[\epsilon_i\middle|\Omega\right].
\end{eqnarray}
Given that empirical data suggests the validity of (E2), I will assume that the ratio in Eq.~\ref{eq.ratio_ch5} holds when $\Omega=\Omega'$.  Rewriting the equations above with this assumption and explicitly adding dependence on the lag $k$, produces the following, 
\begin{eqnarray}
r_i^+(k) & = & E\left[\frac{f(v_{i+k})}{\lambda_{i+k}}\middle|\epsilon_i=+1,\Omega'\right],\\
 & = & E\left[r_{i+k}\middle|\epsilon_i=+1,\Omega'\right],\\
r_i^-(k) & = & E\left[\frac{f(v_{i+k})}{\lambda_{i+k}}\middle|\epsilon_i=-1,\Omega'\right],\\
 & = & E\left[r_{i+k}\middle|\epsilon_i=-1,\Omega'\right],\\
\hat{\epsilon}_i(k) & = & E\left[\epsilon_{i+k}\middle|\Omega'\right].
\end{eqnarray}
Eq.~\ref{eq.ratio_ch5} is then,
\begin{equation}
\frac{r_i^+(k)}{r_i^-(k)} = \frac{1-\hat{\epsilon}_i(k)}{1+\hat{\epsilon}_i(k)},\label{eq.ratiok_ch5}
\end{equation}

\begin{figure}[!ht]
\centering
\includegraphics[width=4.7in]{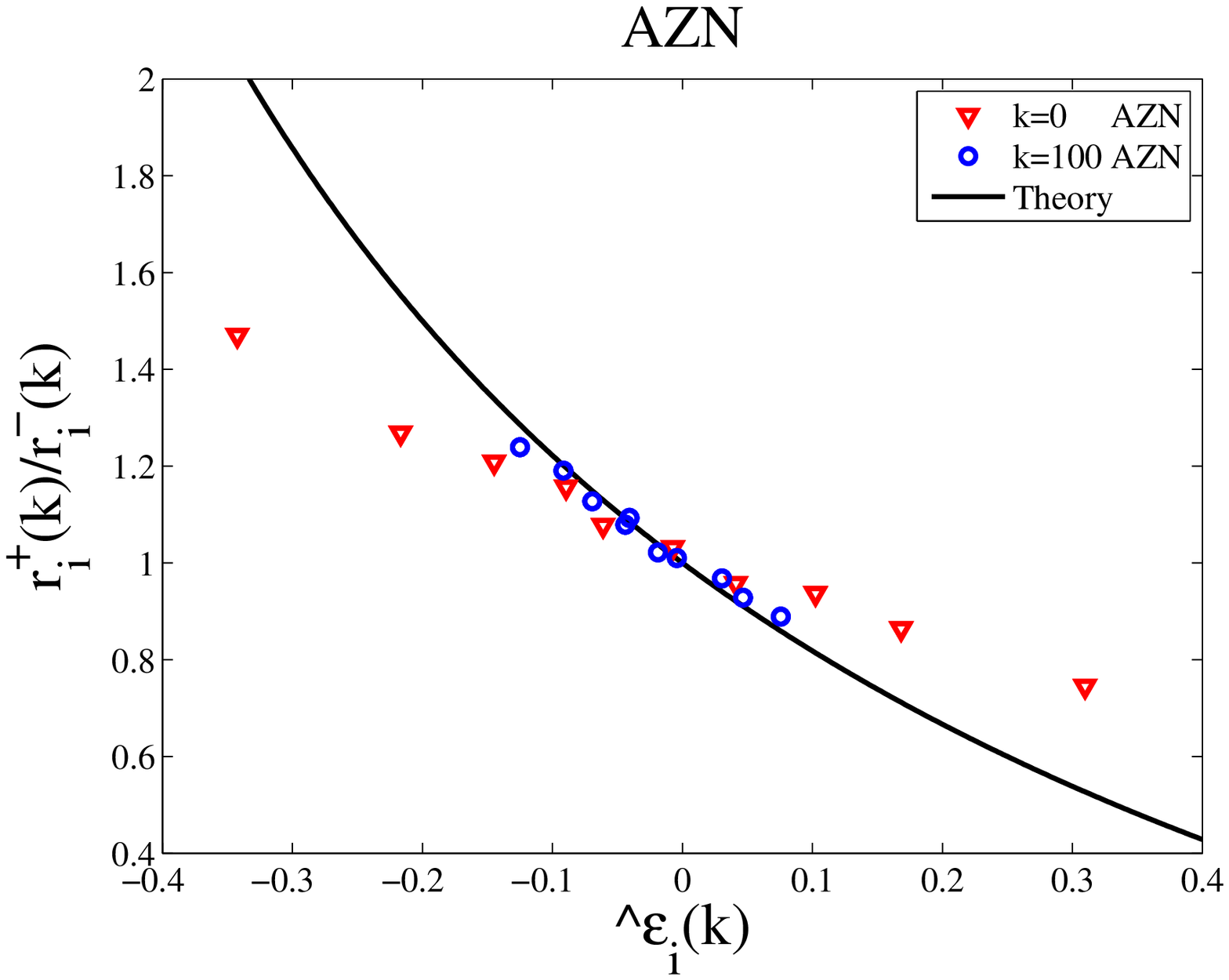}
\includegraphics[width=4.7in]{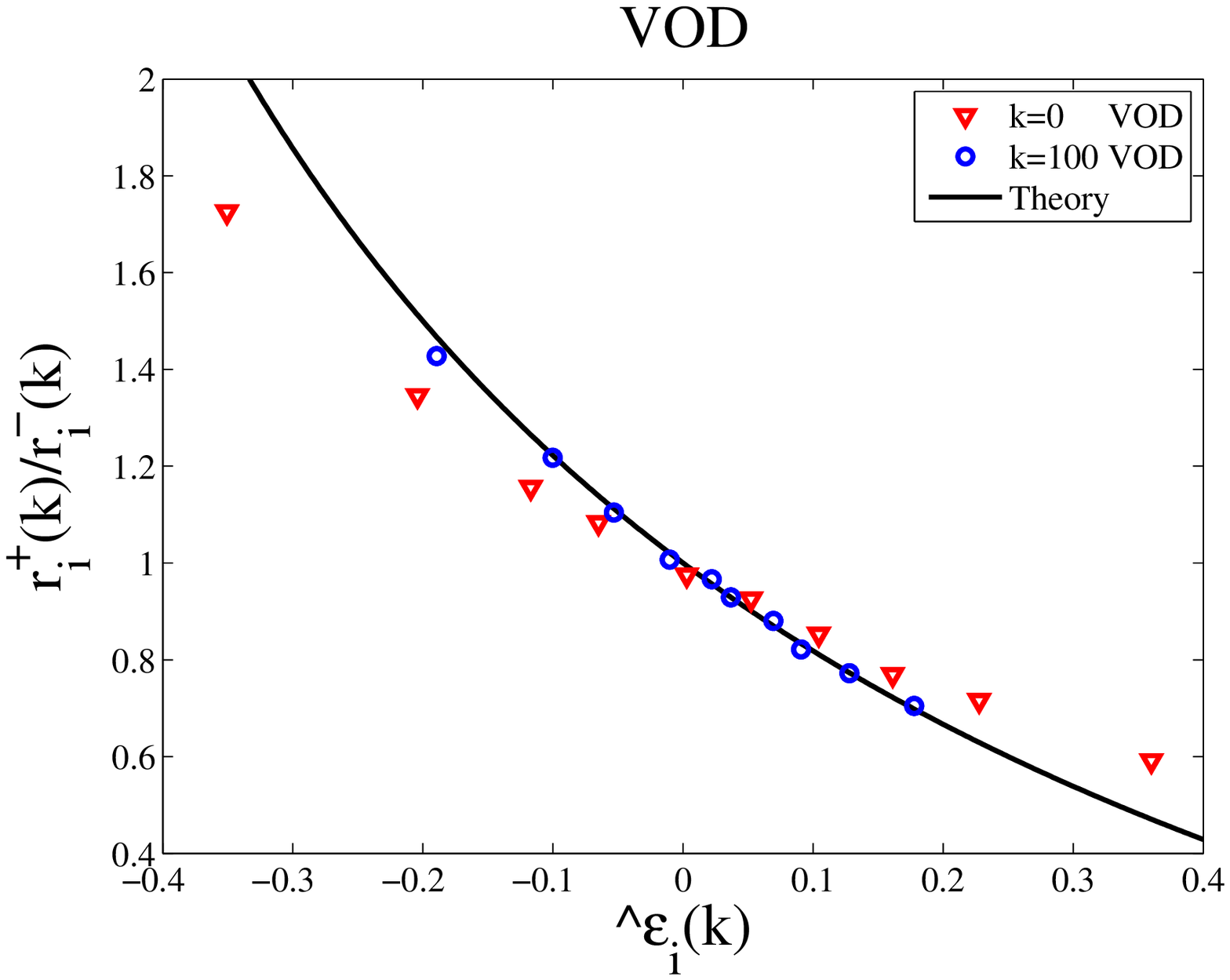}
\caption{The ratio $\left(r_i^+(k)/r_i^-(k)\right)$ as a function of the sign predictor $\hat{\epsilon}_i(k)$ for $k=0$ and $k=100$ for the stocks AZN and VOD.  The data is binned by $\hat{\epsilon}_i(k)$ such that each point contains an equal number of observations, and the mean value of $\left(r_i^+(k)/r_i^-(k)\right)$ is plotted for each bin. For returns to be unpredictable, these points must lie along the curve $\left(1-\hat{\epsilon}_i(k)\right)/\left(1+\hat{\epsilon}_i(k)\right)$ (shown in black). }
\label{fig.BOTH_ratio_signp}
\end{figure}
To show how well Eq.~\ref{eq.ratiok_ch5} holds for empirical data, I plot in Fig.~\ref{fig.BOTH_ratio_signp} the left hand side of the equation as a function of $\hat{\epsilon}_i(k)$ for AZN and VOD.  I show two values of $k$, $k=0$ and $k=100$, and all of the data is included for both.  $\epsilon_i$ can be predicted much better for $k=0$ than for $k=100$, this is why there is more spread on the $x$-axis for the data when $k=0$ than when $k=100$.  As seen in the figure, the theoretic result does not hold for $k=0$, but holds (at least approximately) for $k=100$.  This is the same result shown in the previous section, that under (E2) the return imbalance approaches the transaction imbalance with a lag of $k\approx100$.

Fig.~\ref{fig.BOTH_ratio_signp} shows that it will be quite difficult to measure the exact dependence of liquidity on $\epsilon_i$.  Most of the data is centered  on the $x$-axis around $0$ where it is difficult to distinguish linear from nonlinear behavior.  This means that when looking at the separate dependence of $r_i^+(k)$ and $r_i^-(k)$ on $\hat{\epsilon}_i(k)$, it will be difficult to distinguish between the following possibilities:
\begin{eqnarray}
r_i^+(k) & = & 1 - \hat{\epsilon}_i(k),\\
r_i^-(k) & = & 1 + \hat{\epsilon}_i(k).
\end{eqnarray}
or,
\begin{eqnarray}
r_i^+(k) & = & \frac{1} {1 + \hat{\epsilon}_i(k)},\\
r_i^-(k) & = & \frac{1} {1 - \hat{\epsilon}_i(k)}.
\end{eqnarray}
This difficulty is compounded by several other issues.  First, $r_i^+(k)$ and $r_i^-(k)$ depend on $k$ and therefore must be specified for all $k$.  Second, for large values of $k$, where (E2) is most valid, the results are noisier and even more clustered around $0$.  Third, this relationship is stock dependent.  Fourth and finally, the relationship is not always symmetric for $r_i^+(k)$ and $r_i^-(k)$ (in the above equations, the relationships are symmetric).  Because of these problems, I will be satisfied here with a qualitative understanding of the dependence and will leave as future work the exact specification.
\begin{figure}[!ht]
\centering
\includegraphics[width=4.7in]{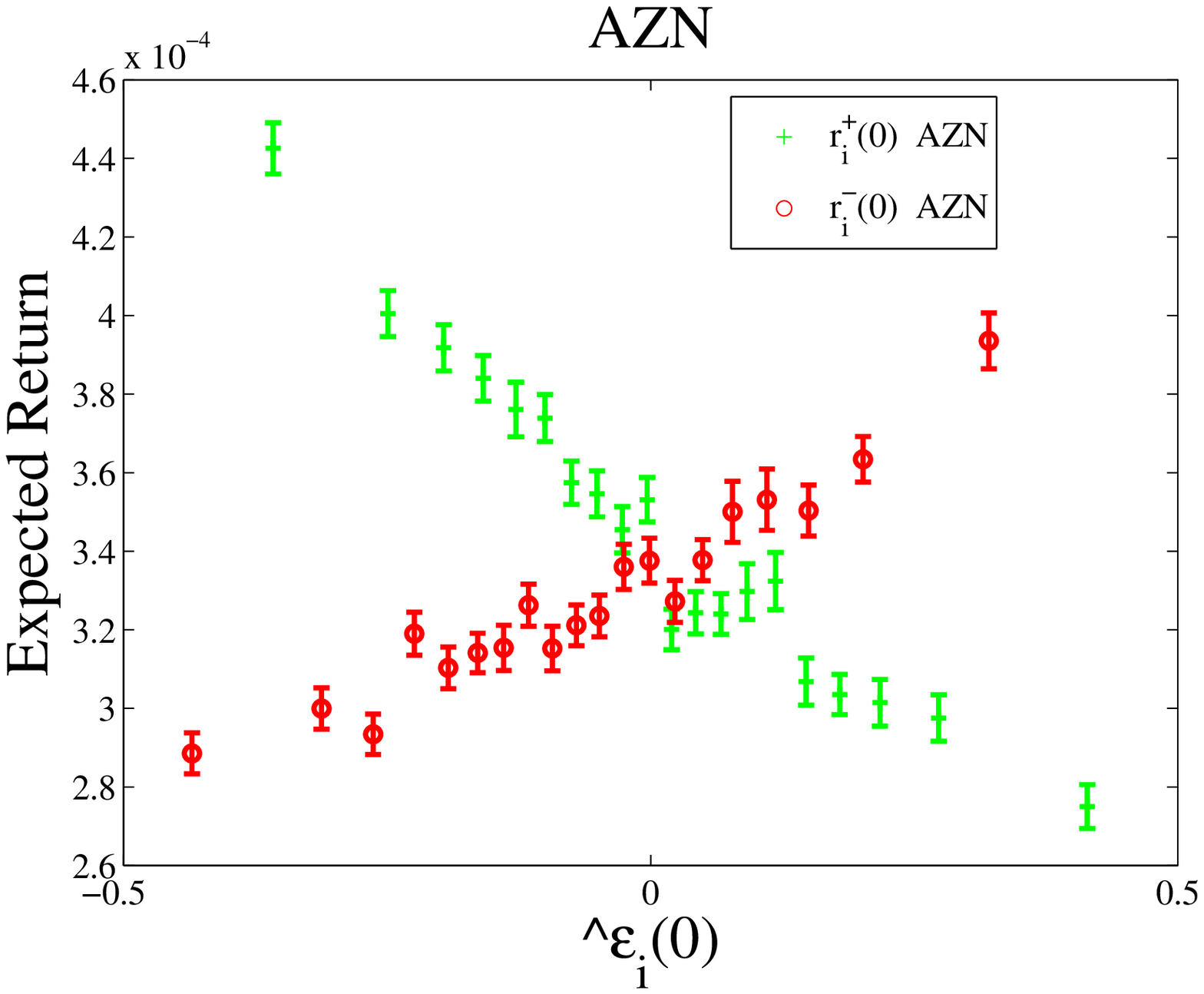}
\includegraphics[width=4.7in]{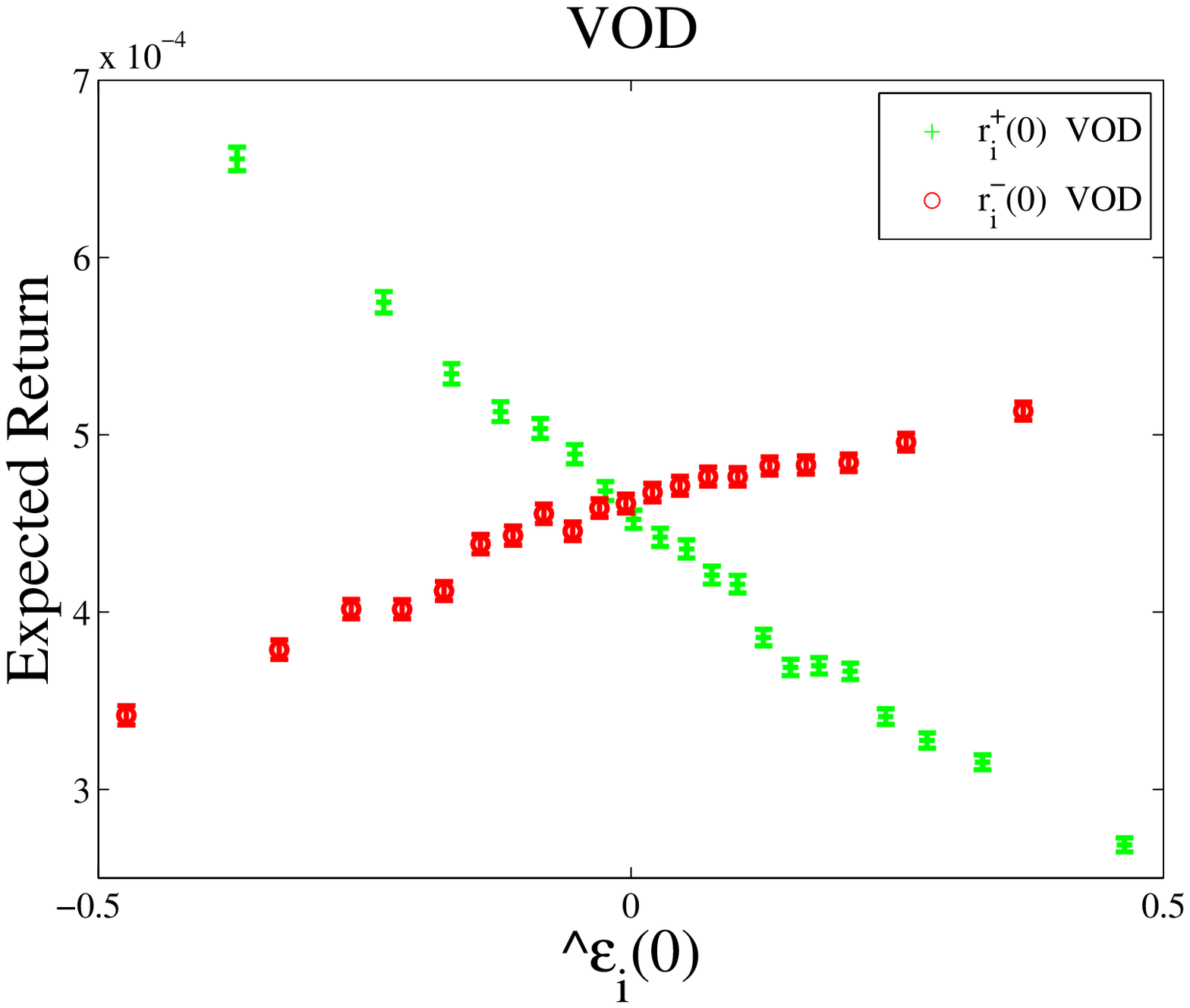}
\caption{The expected return of buyer $\left(r_i^+(0)\right)$ and seller $\left(r_i^-(0)\right)$ initiated transactions as a function of the sign predictor $\hat{\epsilon}_i(0)$.  $k=0$ in these plots.  The data is binned by $\hat{\epsilon}_i(0)$ such that each point contains an equal number of observations, and the mean values of $r_i^+(0)$ and $r_i^-(0)$ are plotted for each bin with the error bars showing the standard error of this average.}
\label{fig.BOTH_r_signp}
\end{figure}
In Fig.~\ref{fig.BOTH_r_signp}, I plot $r_i^+(0)$ and $r_i^-(0)$ vs. $\hat{\epsilon}_i(0)$ for the stocks AZN and VOD.  The dependence is relatively linear for both $r_i^+(0)$ and $r_i^-(0)$ and appears symmetric for $r_i^+(0)$ and $r_i^-(0)$ (although for VOD, buyer initiated transactions tend to respond to predictability more than seller initiated transactions).  This suggests that the dependence is close to,
\begin{eqnarray}
r_i^+(0) & \sim & 1 - \hat{\epsilon}_{i}(0),\label{eq.rpl}\\
r_i^-(0) & \sim & 1 + \hat{\epsilon}_{i}(0).\label{eq.rmi}
\end{eqnarray}
As shown in Section~\ref{sec.liq_eps_fv}, these equations result from one of the assumptions made by BGPW, specifically (A2) - that $\tilde{\lambda}_i$ is independent of $\epsilon_i$ and $f(v_i)$.  This suggests the following,
\begin{equation}
\tilde{\lambda}_i(\Omega') = E\left[\tilde{\lambda}_i \middle| \Omega'\right].\label{eq.lambda_ind}
\end{equation}
Another possibility that also reproduces Eqs.~\ref{eq.rpl} and \ref{eq.rmi} is that,
\begin{equation}
\frac{1}{\lambda_i} = 1-\epsilon_i\hat{\epsilon}_{i}(0),\label{eq.1overlambda_dep}
\end{equation}
which means,
\begin{equation}
\tilde{\lambda}_i = \hat{\epsilon}_{i}(0) f(v_i).\label{eq.lambda_dep}
\end{equation}
Although I do not show it here, the results of the model are nearly identical when using Eq.~\ref{eq.lambda_ind} or Eq.~\ref{eq.lambda_dep}.


\subsection{Conclusion}
Qualitatively, $\tilde{\lambda}_i$ seems independent of $\epsilon_i$ and $v_i$.  This suggests that the final form of the model is just,
\begin{equation}
r_i = \epsilon_i f(v_i) - \tilde{\lambda}_i(\Omega')  + \eta_i,
\end{equation}
where $\tilde{\lambda}_i$ is determined by the set of information $\Omega'$ as follows,
\begin{equation}
\tilde{\lambda}_i = \sum_{j} A_{i,j} \left(\frac{n_j}{n_j+1}\right)^\alpha \frac{\epsilon_j f(v_j)}{\theta_j}.
\end{equation}
I will look at the results of this model in the next section.

\section{Replicating the Properties of Returns}
In this section, I encorporate the results above into the BGPW form of the modified return model,
\begin{equation}
r_i = \epsilon_i f(v_i) - \tilde{\lambda}_i + \eta_i.
\end{equation}
The results of Section~\ref{sec.liq_om} suggest that (E2) is a more accurate description of the stock market than (E1).  The expected value of liquidity, therefore, can be approximated by the following,
\begin{equation}
E\left[\tilde{\lambda}_{i}\middle|\Omega'\right] = \sum_{j} A_{i,j} \left(\frac{n_j}{n_j+1}\right)^\alpha \frac{\epsilon_j f(v_j)}{\theta_j}.
\end{equation}
The results of Section~\ref{sec.liq_eps}, although qualititative and not decisive, suggest that the liquidity term $\lambda_i$ is independent of $\epsilon_i$ and $f(v_i)$,
\begin{equation}
\tilde{\lambda}_i(\Omega') = E\left[\tilde{\lambda}_{i}\middle|\Omega'\right].
\end{equation}
The final result for the model is then,
\begin{equation}
r_i = \epsilon_i f(v_i) - \sum_{j} A_{i,j} \left(\frac{n_j}{n_j+1}\right)^\alpha \frac{\epsilon_j f(v_j)}{\theta_j} + \eta_i.\label{eq.final_model}
\end{equation}
Below, I use the right hand side of Eq.~\ref{eq.final_model} to generate a return series for the stocks AZN and VOD.  All hidden order information is determined using the algorithm discussed at the beginning of this chapter.  The noise term $\eta_i$ is modeled as Gaussian noise with the variance scaled such that the autocorrelation of absolute returns is of the same magnitude in the model as in the empirical data.  For information about fitting the other parameters, see the Appendix.

\begin{figure}[!ht]
\centering
\includegraphics[width=4.7in]{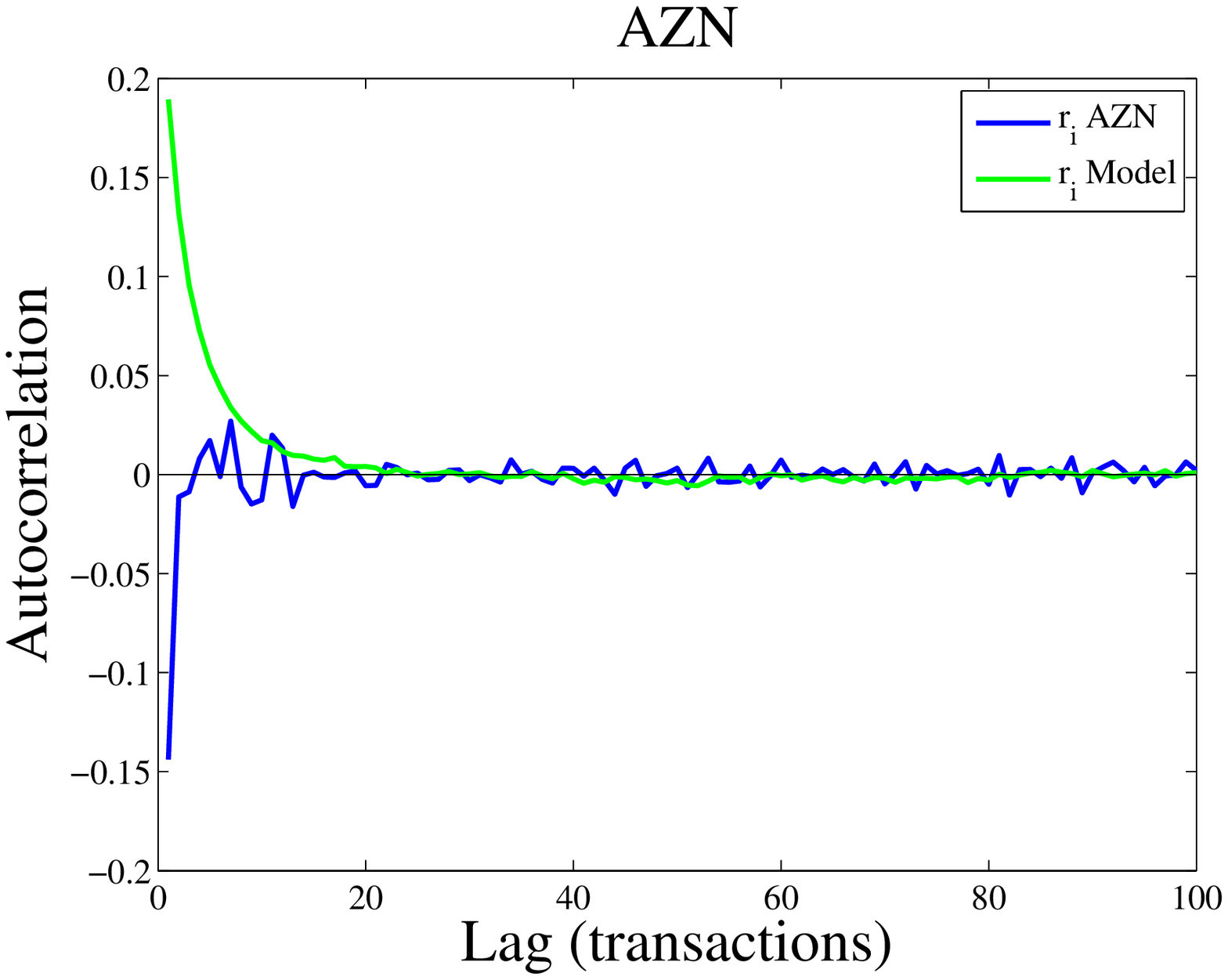}
\includegraphics[width=4.7in]{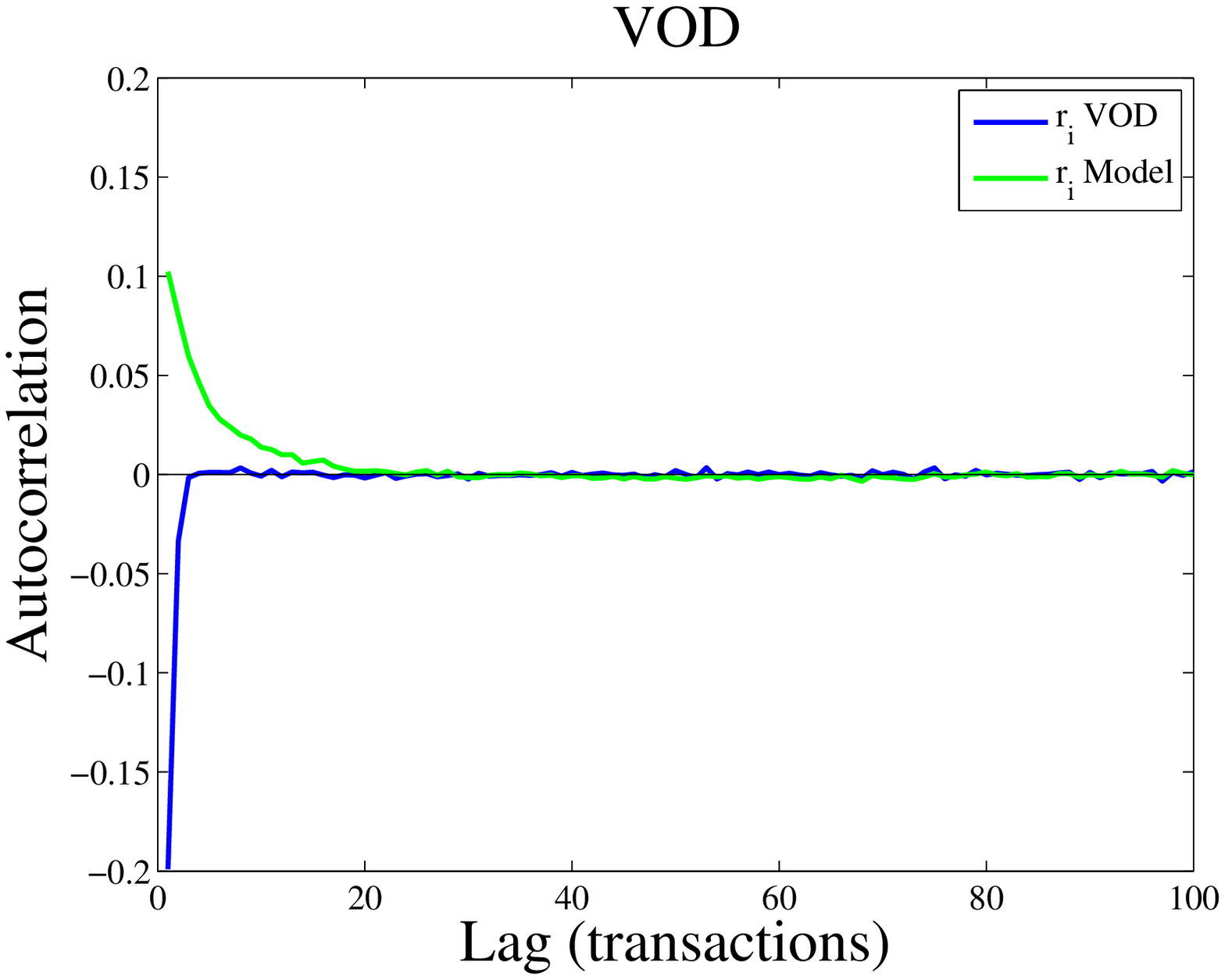}
\caption{The autocorrelation function of one-transaction returns generated by the model of Eq.~\ref{eq.final_model} for the stocks AZN and VOD.  The result is compared to the autocorrelation function of empirical returns.  The model produces returns that are correlated at low lags, but this quickly dies away so that the result matches the empirical result after lag $\approx20$ for both stocks.}
\label{fig.BOTH_model_ret_autocorr}
\end{figure}
\subsection{Not Autocorrelated / Efficient}
In Section~\ref{sec.efficient} I showed that returns are uncorrelated for the stock AZN.  In Fig.~\ref{fig.BOTH_model_ret_autocorr}, I compare the autocorrelation function of one-transaction returns generated by the model of Eq.~\ref{eq.final_model} to the autocorrelation function of empirical returns.  The figure contains results for the stocks AZN and VOD.  Returns generated by the model are autocorrelated at low lags - this is a discrepancy with the empirical returns.  However, this quickly dies away so that after a lag of $\approx20$ transactions, the autocorrelation function of the generated returns matches the empirical result.

\begin{figure}[!ht]
\centering
\includegraphics[width=4.7in]{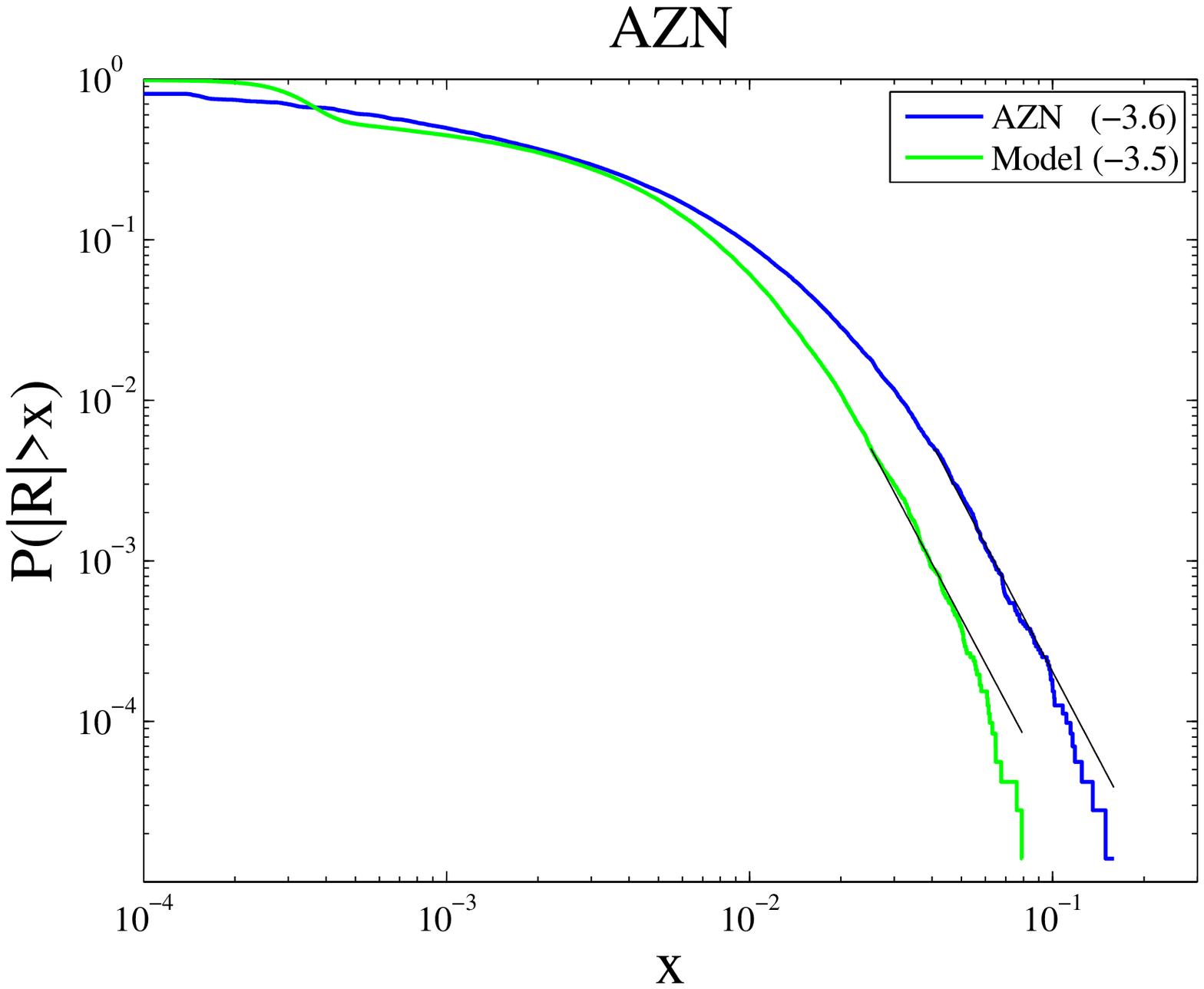}
\includegraphics[width=4.7in]{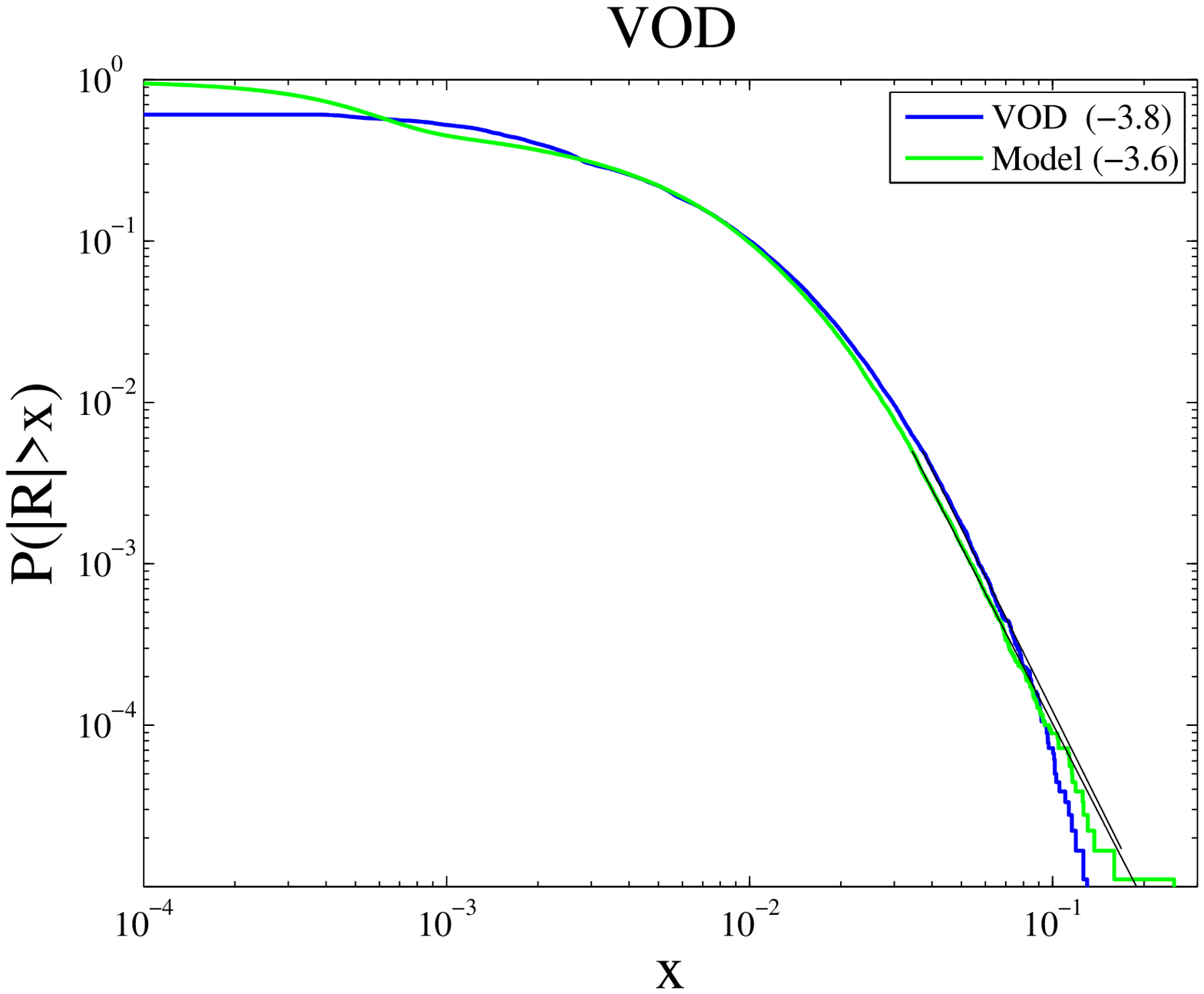}
\caption{The cumulative distribution function of hidden order returns produced by the model of Eq.~\ref{eq.final_model} for the stocks AZN and VOD.  The result is compared to the empirically determined CDF of hidden order returns.  The tail exponents of the compared distributions match well (3.6 vs. 3.5 for AZN, and 3.8 vs. 3.6 for VOD).}
\label{fig.BOTH_model_rhid_ecdf}
\end{figure}

\subsection{Distributed with Power Law Tails}
In Section~\ref{sec.power_tails} I showed that the cumulative distribution of one-transaction returns exhibits a power law tail.  Without showing the results, the model fails to reproduce this property of one-transaction returns.  This is not altogether surprising.  It is known that liquidity fluctuations dominate over volume fluctuations at short timescales and that these fluctuations are distributed with a power law tail\cite{Farmer04b,Weber04}  The model in Eq.~\ref{eq.final_model} does not allow large fluctuations in liquidity at timescales of one-transaction because the liquidity term is a sum over active hidden orders.  This sum slowly varies through time.  To correctly model one-transaction liquidity fluctuations, it is necessary to use a noise term distributed with a power law tail.

To show that the model does reproduce the distribution of returns when aggregated, I plot the cumulative probability distribution of $R$ (hidden order returns) determined empirically and generated by the model for AZN and VOD.  These plots are shown in Fig.~\ref{fig.BOTH_model_rhid_ecdf}.  The tail exponent of all distributions are determined and printed in the figure legend.

\subsection{Clustered Volatility}
\begin{figure}[!ht]
\centering
\includegraphics[width=4.6in]{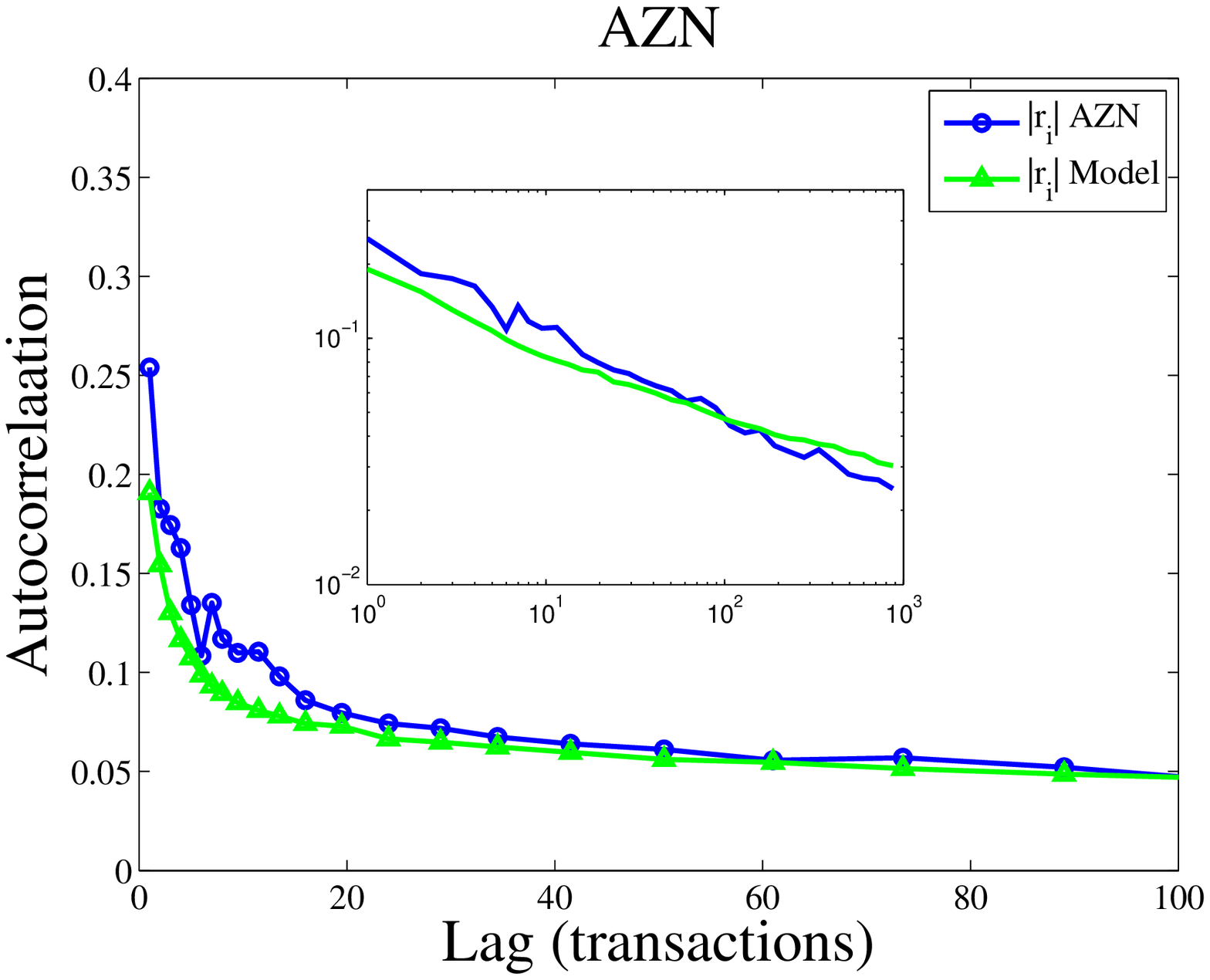}
\includegraphics[width=4.6in]{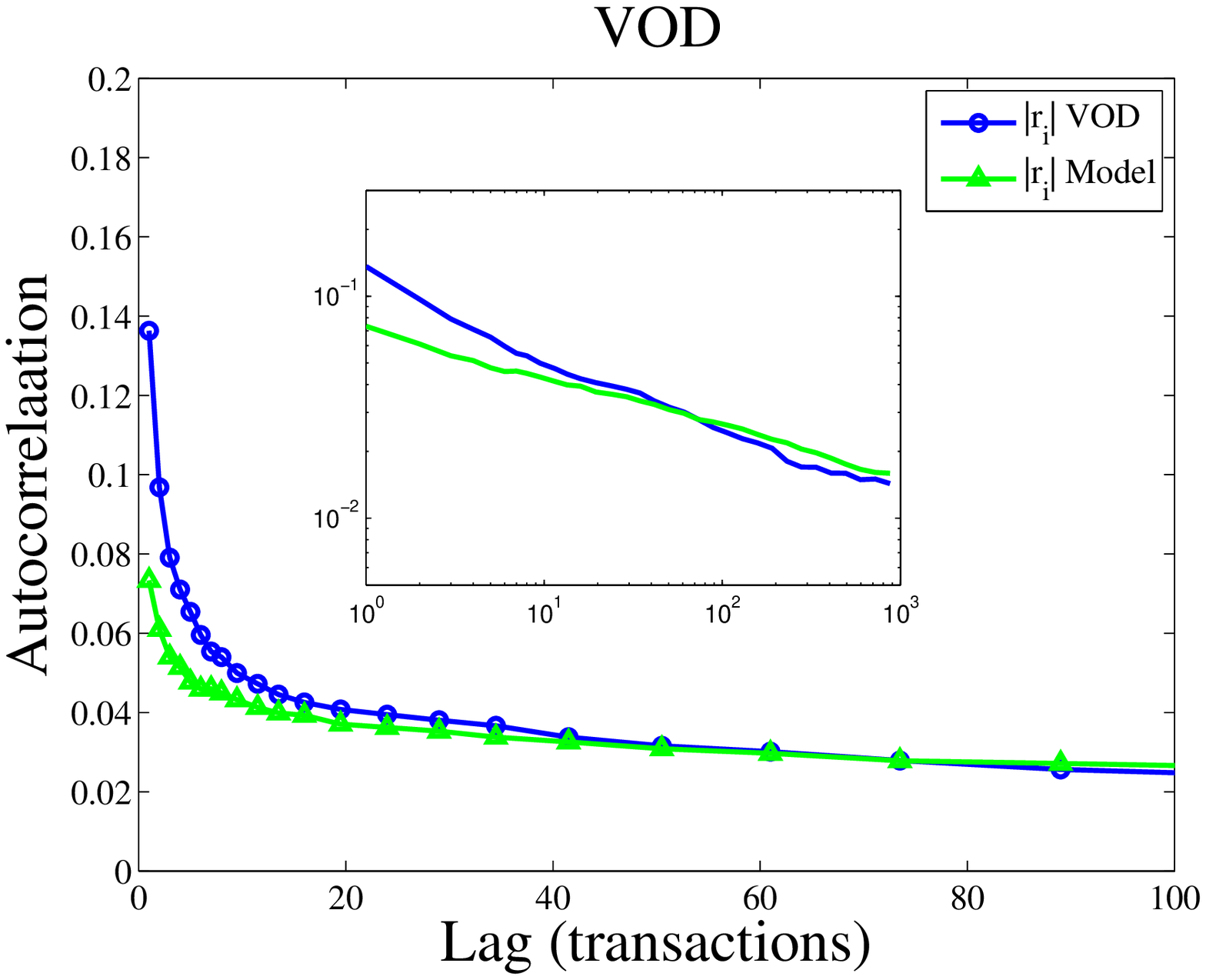}
\caption{Autocorrelation function of the magnitude of one-transaction returns produced by the model of Eq.~\ref{eq.final_model} for the stocks AZN and VOD.  The result is compared to the autocorrelation function using empirical returns.  The plots show that the model is producing autocorrelated return magnitudes and that the autocorrelation function scales similarly to the empirical result - however, the match is not exact.}
\label{fig.BOTH_model_absret_autocorr}
\end{figure}

In Section~\ref{sec.clust_vol} I showed that the absolute value of one-transaction returns is highly autocorrelated.  This property has been called \emph{clustered volatility}, because volatility, or the magnitude of returns, tends to cluster in time.  In Fig.~\ref{fig.BOTH_model_absret_autocorr}, I compare the autocorrelation function of absolute returns determined using the empirical return series to that generated by the model for the stocks AZN and VOD.
The model produces returns with autocorrelated magnitude and the autocorrelation function scales similarly for generated returns and empirical returns, although in both cases the slope for generated returns is smaller in magnitude than for empirical returns.  It will take more extensive study to fully determine if return magnitudes (and their autocorrelated structure) are correctly generated by the model.

\section{Decay or Permanence of Hidden Order Impact}
\begin{figure}[!ht]
\centering
\includegraphics[width=4.4in]{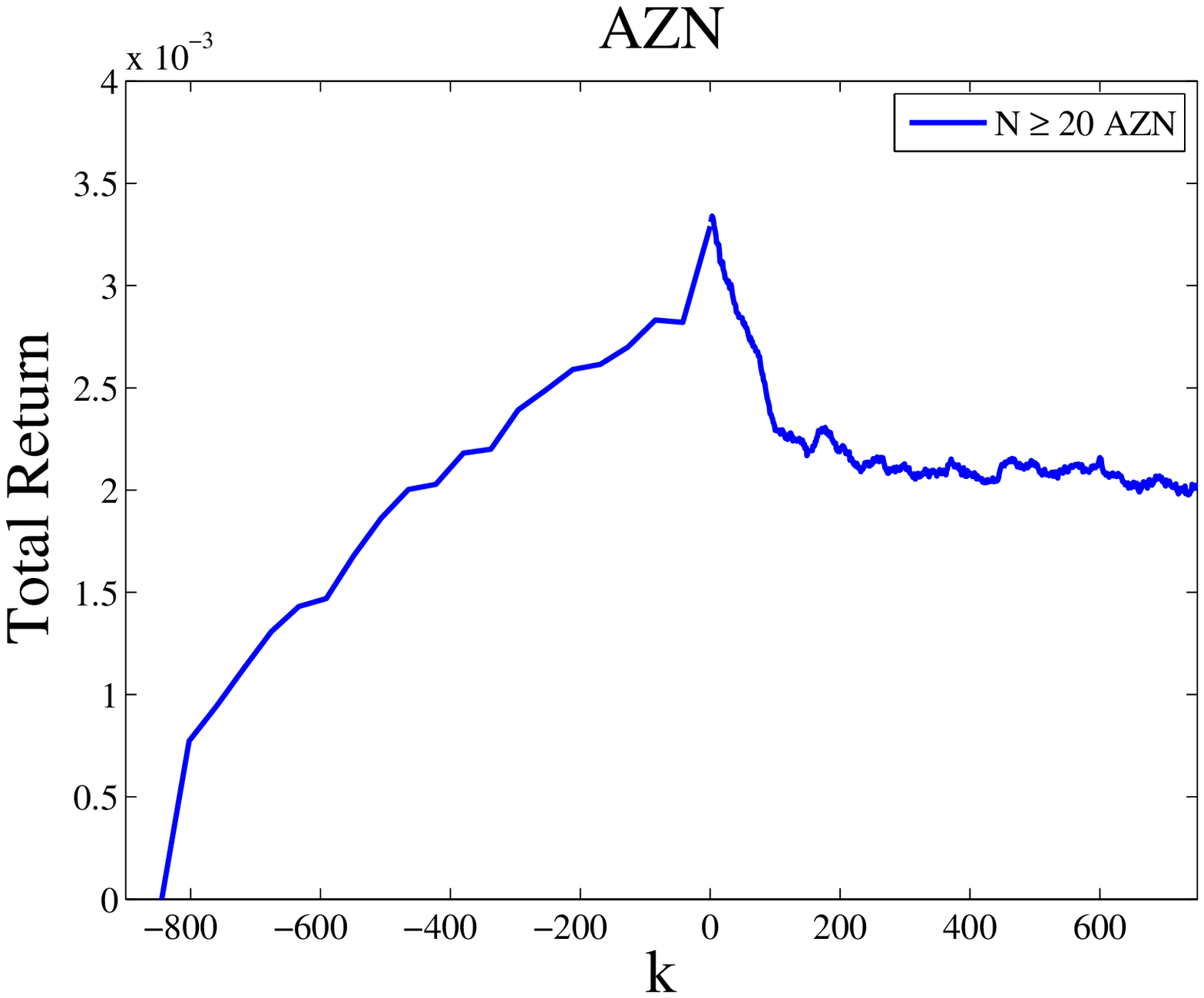}
\includegraphics[width=4.4in]{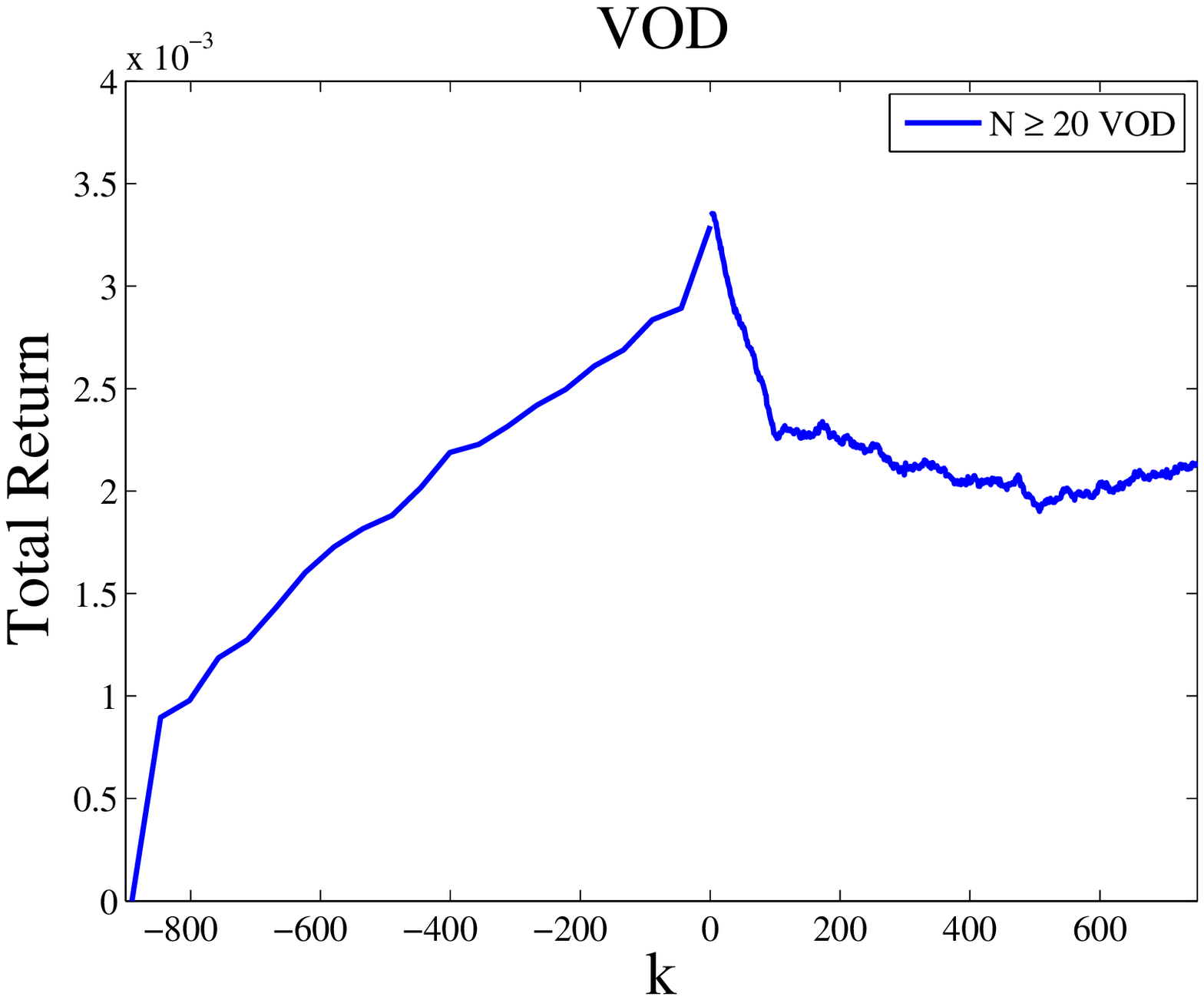}
\caption{The average cumulative impact of a hidden order with size $N\geq20$ as a function of lag $k$ for the stocks AZN and VOD.  $k=0$ corresponds to the completion time of the hidden order and $k\approx-875$ corresponds to the start time of the hidden order.  To make the plot, the execution time for all hidden orders are rescaled such that they start and stop at the times just specified.  For lags $k\leq0$, the average impact along this rescaled time is computed.  For lags $k>0$, the the average impact is computed in normal transaction time such that $k$ is the number of transactions since the hidden order has completed.  Notice that for both AZN and VOD, there exists a permanent and transient component to impact, and the permanent component is approximately $2/3$ of the total.}
\label{fig.BOTH_decay}
\end{figure}

At the end of Chapter~\ref{ch.price_impact}, I showed that transaction impacts should be completely transient under the assumptions of (E1) and they should be completely permanent under the assumptions of (E2).  In this section, I test this result.  In Fig.~\ref{fig.BOTH_decay}, I plot the average cumulative impact of a hidden order with size $N\geq20$ as a function of lag $k$ from its completion time.
To make these plots, I treated $k>0$ differently than $k\leq0$.  For $k>0$, I plotted on the $y$-axis the average of the price response at the specific lag $k$.  For $k\leq0$, I rescaled the total execution time for each hidden order such that all hidden orders began at the same lag (the start lag corresponds to the average total time that hidden orders of size $N\geq20$ take to complete).  I then sampled the impact at 20 equidistant intervals for each hidden order and averaged this impact over all hidden orders.  The result is an increasing impact until $k=0$ that rapidly decays until $k=100$ and thereafter remains relatively constant.  This suggests that (E1) and (E2) are both partially correct.  As suggested at the end of Chapter~\ref{ch.price_impact}, it is possible that under (E2) a hidden order continues to influence liquidity after it has completed, and therefore causes some amount of impact decay.  This might explain both the abrupt decay seen in the figures and also the permanency afterwards.

We can now look back at the two interpretations (I1) and (I2) of the modified return model,
\begin{equation}
r_i = \epsilon_i f(v_i) - \tilde{\lambda}_i + \eta_i.
\end{equation}
(I1) states that transaction impacts are $\left(\epsilon_i f(v_i) - \tilde{\lambda}_i\right)$ and are permanent.  (I2) states that transaction impacts are $\epsilon_i f(v_i)$ and are transient.  Because I have argued that the most natural interpretation should correspond with what market participants observe for their own initiated transactions, it appears neither interpretation is fully correct.  Fig.~\ref{fig.BOTH_decay} suggests that market participants observe both a transient and permanent component to their impact - about $2/3$ is permanent and $1/3$ is transient.  A complete study of this phenomenon is beyond the scope of this thesis. 
\chapter{Summary}

In Chapter~\ref{ch.intro}, I looked at a very simple model for stock returns,
\begin{equation}
r_i = \epsilon_i f(v_i) + \eta_i,\label{eq.simple_conc}
\end{equation}
where $r_i$ is the one-transaction return, $\epsilon_i$ is the sign of the transaction at time $i$ (whether it is buyer or seller initiated), $v_i$ is the volume of the transaction, $f(\cdot)$ is the price impact function - an empirically measured function, and $\eta_i$ is a noise term.  This equation was motivated by a simple premise: that the stock market can be modeled in a mechanical way - as a deterministic translating device that transforms order flow into a price stream.  Eq.~\ref{eq.simple_conc} fails because the transaction sign series, $\epsilon_i$, is highly autocorrelated and exhibits long memory.

In Chapter~\ref{ch.long_mem}, I studied two modified versions of the simple model in Eq.~\ref{eq.simple_conc},
\begin{eqnarray}
r_i & = & \frac{\epsilon_i f(v_i)}{\lambda_i}  + \eta_i,\label{eq.LF_form_conc}\\
r_i & = & \epsilon_i f(v_i) - \tilde{\lambda}_i  + \eta_i. \label{eq.BGPW_form_conc}
\end{eqnarray}
These models were suggested by Lillo and Farmer\cite{Lillo03c} (LF) and Bouchaud, Gefen, Potters, and Wyart\cite{Bouchaud04} (BGPW).  The $\lambda_i$ and $\tilde{\lambda}_i$ terms exist in the models to compensate for the autocorrelated structure of $\epsilon_i$.  In this general form, the two models are equivalent with a simple change of variable,
\begin{equation}
\frac{1}{\lambda_i} \equiv 1 - \frac{\tilde{\lambda}_i}{\epsilon_i f(v_i)}.
\end{equation}
This means that all differences between the papers by LF and BGPW are due to the assumed structure of $\lambda_i$ and $\tilde{\lambda}_i$.  LF posit that these terms model asymmetric fluctuating liquidity, and BGPW posit that these terms model mean reverting quote revisions.  By decomposing returns into two components, one component including only liquidity effects and another component including only quote revision effects, I used empirical data to show that returns are efficient when disregarding quote revisions.  This suggests that $\lambda_i$ and $\tilde{\lambda}_i$ model asymmetric liquidity rather than mean reverting quote revisions.

In Chapter~\ref{ch.asym_liq}, I assumed that the efficient market hypothesis (EMH) is valid,
\begin{equation}
E[r_i|\Omega] = E[r_i] \approx 0,
\end{equation}
where $\Omega$ is any set of publicly available historical financial data.  From this, I derived the dependence of the liquidity parameters $\lambda_i$ and $\tilde{\lambda}_i$ on the order flow variables $\Omega$, $\epsilon_i$, and $v_i$.  Under two extreme views of publicly discernable information ((E1) and (E2)), I bound the dependence of $\lambda_i$ and $\tilde{\lambda}_i$ on $\Omega$ between two equations,
\begin{eqnarray}
E\left[\tilde{\lambda}_{i} \middle|\Omega\right] & = & \sum_{k>0} a_k \epsilon_{i-k} f(v_{i-k}),\label{eq.E1_autoreg_conc}\\
E\left[\tilde{\lambda}_{i}\middle|\Omega'\right] & = & \sum_{j} A_{i,j} \left(\frac{n_j}{n_j+1}\right)^\alpha \frac{\epsilon_j f(v_j)}{\theta_j}.\label{eq.E2_hidord_conc}
\end{eqnarray}
Eq.~\ref{eq.E1_autoreg_conc} is an autoregressive model for $\epsilon_i f(v_i)$ and Eq.~\ref{eq.E2_hidord_conc} uses hidden order information to predict $\epsilon_i$ and $f(v_i)$ (hidden orders are large orders that have been split into pieces and then transacted on the market).  Eq.~\ref{eq.E1_autoreg_conc} holds under the set of assumptions (E1): that market makers condition liquidity on the predictability of transaction order flow, and that the best predictor they can use is an autoregressive model for $\epsilon_i$ and $v_i$.  Eq.~\ref{eq.E2_hidord_conc} holds under the set of assumptions (E2): that either market participants condition their order flow so as not to produce predictable returns, or alternatively, that market makers can determine who is initiating transactions and when they have finished placing a hidden order.

When determining the dependence of liquidity on $\epsilon_i$ and $f(v_i)$, I found that the relationship is underspecified but the following ratio must be satisfied,
\begin{equation}
\frac{r_i^+}{r_i^-} = \frac{1-\hat{\epsilon}_i}{1+\hat{\epsilon}_i},\label{eq.ratio_conc}
\end{equation}
where the left hand side is the ratio of expected returns for buyer $r_i^+$ and seller $r_i^-$ initiated transactions, and $\hat{\epsilon}_i$ is the predicted value of $\epsilon_i$.

In Chapter~\ref{ch.price_impact}, I derived two competing equations for the expected return of a hidden order.  Under (E1), this was:
\begin{equation}
E\left[R \middle| \epsilon, v, \theta, N \right] = \frac{\epsilon f(v)}{1-\phi} \  \theta^{-\phi} N^{1-\phi}\label{eq.e1_impact_conc},
\end{equation}
and under (E2), this was:
\begin{equation}
E\left[R \middle| \epsilon, v, \theta, N \right] = \alpha \epsilon f(v) \log{\left(1+N\right)}\label{eq.e2_impact_conc}.
\end{equation}
Eq.~\ref{eq.e1_impact_conc} suggests that the impact of a hidden order is a concave function of its total volume (a power law with exponent greater than $1/2$ but less than $1$) and that the expected total impact can arbitrarily be scaled by changing the speed of trading (changing $\theta$).  In the limit that $\theta$ is infinitely large, i.e., the hidden order is traded infinitely slow, the expected total impact is zero.  Eq.~\ref{eq.e2_impact_conc} suggests that the impact of a hidden order is a concave function of its total volume, specifically a logarithm, and that the impact is independent of the speed of trading, $1/\theta$.

At the end of Chapter~\ref{ch.price_impact}, I showed that transaction impacts should be completely transient under the assumptions of (E1) and they should be completely permanent under the assumptions of (E2).  The impact is transient under (E1) because it decays by a fraction, $a_k$, at each timestep into the infinite future and the sum of these fractions approaches $1$ at infinity.  The impact is permanent under (E2) because liquidity providers are assumed to know when a hidden order completes - therefore, the hidden order no longer influences liquidity when finished and its impact remains constant.

In Chapter~\ref{ch.emp_results}, I used data to test the theory developed in the previous chapters.  I found that (E2) is supported by empirical data and that (E1) is not.  This result is surprising considering that information about hidden orders is not explicitly available to the market participants setting liquidity.  It suggests that either market makers can discern this information or that market participants condition their order flow on available liquidity such that they do not produce predictable returns.  When studying the response of buyer and seller initiated transactions to their predictability, I qualitatively determined that $\tilde{\lambda}_i$ is independent of $\Omega$.  Putting these results together, I formulated a final version of the return (or price impact) model,
\begin{equation}
r_i = \epsilon_i f(v_i) - \sum_{j} A_{i,j} \left(\frac{n_j}{n_j+1}\right)^\alpha \frac{\epsilon_j f(v_j)}{\theta_j} + \eta_i.\label{eq.final_model_conc}
\end{equation}
Using empirical data, I generated returns using the model and found that (1) the returns were uncorrelated at lags larger than 20, (2) hidden order returns were distributed with a power law tail that matched the empirical data, and (3) the magnitude of the returns were autocorrelated in a similar way to the empirical data.

Finally, using data from hidden orders of size $N\geq20$, I showed that the total impact of these orders builds continuously until they complete, decays abruptly to $2/3$ of its maximum value over the next $100$ transactions, and thereafter remains relatively constant.

\appendix
\chapter{}

\section{Stock Market Preliminaries}
Modern electronic markets follow the continuous double auction trading mechanism.  This is structured such that buyers and sellers anonymously submit orders into an \emph{order book}.  Once submitted, the order is either immediately matched with another order (if both parties find the price agreeable), or is stored in the order book until either cancelled or matched with an incoming order.  There are two basic order types that are placed into the order book.  The first is a \emph{limit order}.  Limit orders are requests to buy or sell a certain volume (quantity) of stock at a specified limit price - the worst price at which the participant is willing to transact.  Limit orders to buy are often called \emph{bids} and limit orders to sell are often called \emph{offers}; and the intention to buy or sell is often called the \emph{sign} of the order.  If a limit order is placed and there exists no order of the opposite sign with a price agreeable to the limit price, then the limit order is stored into the order book until either cancelled or transacted with an incoming order.  The other basic type of order is a \emph{market order}.  These orders are requests to buy or sell a certain volume (quantity) of stock with no limit price, meaning the order will accept the best available price currently showing in the order book (called the \emph{best price}).  Market orders are immediately matched (transacted) with the order of opposite sign located at the best price.  If two orders specify the best price, then the market order transacts with the order that has been stored in the order book longer.  It is possible for an order to transact with more than one other order.  For example, if a market order requests 100 shares and the order at the best price is only for 50 shares, then the market order will transact 50 shares with the first order and the remaining with each order sequentially showing the best price until all 100 shares have transacted.

\begin{figure}[htb]
\centering
\includegraphics[width=4in]{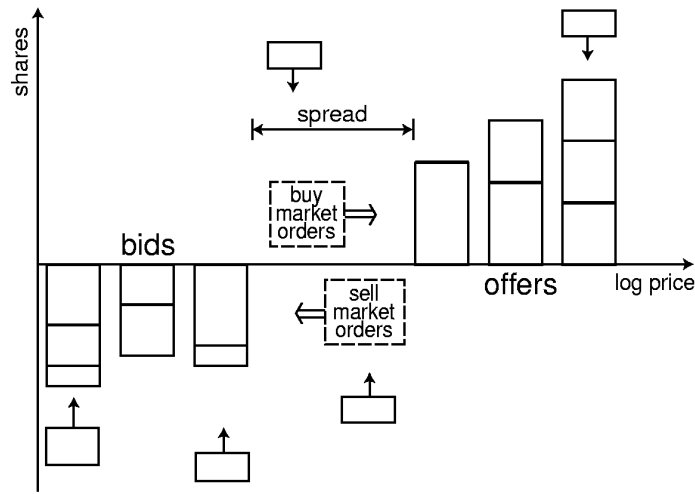}
\caption{A schematic of the order book used in modern electronic markets.  This figure is taken from the paper by Smith et al.\cite{Smith03} }
\label{fig.double_auction}
\end{figure}

In Fig.~\ref{fig.double_auction} I show a schematic of the orderbook.
Notice that there always exists two best prices.  The best price willing to buy is called the \emph{best bid} price.  The best price willing to sell is called the \emph{best offer} price (sometimes called the best ask price).  The midpoint between these two prices is called the \emph{midpoint price} and is a standard reference for the current price of the stock.  The difference between the best bid and best offer price is called the \emph{spread}.  Limit orders (and the prices they specify) are often referred to as \emph{quotes}, especially if the limit order exists at the best price.  Prices for electronic markets are discrete, meaning that limit orders must specify prices in increments, the minimum sized increment of price for a stock is called the \emph{tick size}.

\section{Description of Data}
Unless otherwise specified, all results presented in this thesis are from 6 stock traded on the London Stock Exchange over the period May 20, 2000 to Dec 31, 2002.  These stocks are Astrazeneca (AZN), British Sky Broadcasting Group (BSY), Lloyds Tsb Group (LLOY), Prudential (PRU), Rentokil Initial (RTO), and Vodafone Group (VOD).  The London Stock Exchange contains two markets, an upstairs market and an electronic orderbook exchange called SETS.  The dataset I use contains all order flow information for the electronic exchange, but does not include any upstairs information.  The electronic exchange contains roughly $60\%$ of all traded volume during the period I study, and is universally used to determine the current price of a stock.

Because I am only interested in transactions in this thesis, I do not consider any of the various order types that exist in the dataset.  I am only concerned with whether an order causes a transaction or not.  If it does, I use only information about the volume (measured in British Pounds $\pounds$) and the sign (whether initiated by a buyer or seller) of the transaction.  Prices are determined using a reconstruction of the orderbook and referencing the midpoint between the current best bid and best ask price.  Returns are then measured as the difference in the logarithm of this midpoint price measured from directly before a transaction occurs to directly before the next transaction occurs.  The initial impact of a transaction, when used in this thesis, is measured as the difference in the logarithm of the midpoint price from directly before to directly after the transaction.  

\subsection{Properties of the Individual Stocks}
The following table contains a list of measured parameters that are used in this thesis.
\begin{table}[htb]
\centering
\begin{tabular}{ccccccc}
Security	&Transactions		&$H$		&$f_1$		&$f_2$  &$\alpha$ &$\phi$ \\
\hline
AZN		&569321		&0.68		&$9.4\times10^{-5}$		&.12		&1.64		&.18\\
BSY		&359479		&0.68   &$1.9\times10^{-4}$		&.12		&1.64		&.18\\
LLOY	&599739		&0.69   &$9.6\times10^{-5}$		&.14		&1.62		&.19\\
PRU		&392020		&0.70   &$1.9\times10^{-4}$		&.11		&1.60		&.20\\
RTO		&213474		&0.73   &$1.6\times10^{-4}$		&.16		&1.54		&.23\\
VOD		&1047833	&0.67   &$2.7\times10^{-5}$		&.25		&1.66		&.17\\
\hline
\end{tabular}
\caption{Table of parameters for the six stocks studied in this thesis.}
\end{table}
The Hurst exponent of the transaction sign series, $\epsilon_i$, is measured using the periodogram method.  The parameters of the one transaction price impact function, $f(v_i) = f_1 v_i^{f_2}$, are measured by first binning the data by volumes and measuring the expected return for each bin; then the log of volumes and log of expected returns is fit by a line using least squares.  The parameters of this fit determine $f_1$ and $f_2$.  The parameters $\alpha$ and $\phi$ are both determined using the Hurst exponent, $\alpha=3-2H$ and $\phi=H-1/2$.

\section{Referenced Assumptions, Interpretations, etc.}

\noindent From Section~\ref{sec.bouch_assumptions},
\begin{itemize}
\item[(A1)] The weak form of the efficient market hypothesis (EMH) holds.  That is,
\begin{equation}
E[r_i|\Omega] = 0.
\end{equation}
\item[(A2)] $\tilde{\lambda}_i$ is independent of $\epsilon_i$ and $f(v_i)$.
\item[(A3)] $E\left[\epsilon_i f(v_i)\middle|\Omega\right]$ can be approximated by an infinite autoregressive model for $\epsilon_i$ and $f(v_i)$ with coefficients determined by treating $\epsilon_i$ as a FARIMA process.
\end{itemize}

\noindent From Section~\ref{sec.I1I2},
\begin{itemize}
\item[(I1)] The impact of the transaction at time $i$ is $\left(\epsilon_i f(v_i) - \tilde{\lambda}_i\right)$ with noise and is permanent.
\item[(I2)] The impact of the transaction at time $i$ is $\epsilon_i f(v_i)$ with noise and is transient.  $\tilde{\lambda}_i$ is the instantaneous decay at time $i$ of past transactions.
\end{itemize}

\noindent From Section~\ref{sec.LMF},
\begin{itemize}
\item[(R1)] Market participants exhibit autocorrelated orderflow (most likely because they have split up into pieces a much larger order).
\item[(R2)] Market participants do not correlate with each others orderflow.
\end{itemize}

\noindent From Section~\ref{sec.theory_liq},
\begin{itemize}
\item[(P1)]  \textbf{There are a group of market participants that collectively condition liquidity on past orderflow such that Eq.~\ref{eq.pred_liquidity} holds}.  At time $i$, these participants force transactions with the same sign as $\epsilon_{i-k}$ to have lower impact and/or transactions with the opposite sign as $\epsilon_{i-k}$ to have larger impact.  They do so because there is a certain probability that participant $p$ will place another transaction at time $i$ with known sign $\epsilon_{i-k}$.  The participants are thought of as market makers who do not want the price to become superdiffusive.
\item[(P2)]  \textbf{Market participant $p$ conditions her transactions on the liquidity parameter such that Eq.~\ref{eq.pred_liquidity} holds}.  If participant $p$ notices that her trades are causing large impacts, she may decide to postpone the trade to some later time or reduce the size of her transactions $v_i$ - in fact, she may not be willing to trade at all unless she does observe liquidity as dictated in Eq.~\ref{eq.pred_liquidity}.  Instead of market makers who work to keep the price diffusive, the participants that determine liquidity are patient position takers that are willing to wait for participant $p$ to cause transactions.

\end{itemize}


\noindent From Section~\ref{sec.E1E2},
\begin{itemize}
\item[(E1)]  Assume that (P1) is correct and that the best predictor of $\epsilon_i f(v_i)$ is an autoregressive model.
\begin{equation}
E\left[\tilde{\lambda}_{i} \middle|\Omega\right] = \sum_{k>0} a_k \epsilon_{i-k} f(v_{i-k}).
\end{equation}

\item[(E2)]  Take that (R1) and (R2) are correct.  Assume one of the following: (P2) is true, or alternatively, (P1) is true and market makers have information about who is initiating transactions and when they have finished placing a hidden order.
\begin{equation}
E\left[\tilde{\lambda}_{i}\middle|\Omega'\right] = \sum_{j} A_{i,j} \left(\frac{n_j}{n_j+1}\right)^\alpha \frac{\epsilon_j f(v_j)}{\theta_j},
\end{equation}
where $\Omega'$ is information about all of the active hidden orders (those with $A_{i,j}=1$) \dots this includes their sign $\epsilon_j$, their typical size $v_j$, the average number of timesteps between transactions for the order $\theta_j$, and how many pieces of the order have already been transacted $n_j$.  I use $\Omega'$ instead of $\Omega$ because it is not necessarily publicly discernable information.

\end{itemize}
\chapter{Low-Dimensional Open Chaotic Systems}

In addition to my econophysics research, I have also researched low-dimensional open chaotic systems.  In the following, I present my work in this area.

\section{Introduction}

Flows are a frequent topic of research among physicists - fluid flow \cite{Sreenivasan99}, traffic flow \cite{Helbing01}, crowd movement \cite{Helbing00}, and granular flow \cite{Jaeger96} are just a few examples. They are of particular interest because the particles that constitute the flow can exhibit complex behavior at certain flow parameters.  There is a large body of work focused on understanding the cause of this motion and predicting the patterns and structures that these flows produce \cite{Sreenivasan99, Helbing01, Helbing00, Jaeger96, Gollub99}.

Most flows consist of many mutually interacting degrees of freedom and a complete description of the dynamics is often impossible.  It is not surprising then, that theories have historically focused on a statistical description of complex flow \cite{Aref96}. Several studies, however, suggest that we can understand flows at a more fundamental level.  Cellular automata fluid models are an example - for certain types of flows, the scale and particulars of collisions seem unimportant to the overall structure of the flow\cite{Wolfram86,Frisch86}.  The study of bifurcations in Taylor-Couette flow \cite{Brandstater83} suggest that the complicated motion in large scale flows can result from the interplay of a few chaotic degrees of freedom.  Several authors have successfully extended these ideas to the general study of large coherent structures in fluid flow \cite{Holmes96}.  Finally, it has recently been shown that a one-dimensional series of nonlinear oscillators, when driven, can behave quite similar to larger scale turbulent flow \cite{Peyrard02}

We seek to study complex open-boundary flow using a bottom up approach - by studying the simplest possible flows that exhibit complex motion.  The low-dimensional model we present here is unique because particles interact with linear forces and the system has open boundaries.  There are many studies of low-dimensional, \emph{closed}-boundary dissipative systems in the literature.  Examples include driven, damped pendulums, the Lorenz equations, and driven Frenkel-Kontorova models \cite{Sagdeev88, Braun98}.  Like these driven, dissipative systems, the system we present continually gains and depletes energy, but it also allows particles to pass across boundaries into and out of the flow.  Research of complex, yet very low-dimensional, open-boundary systems is scarce in the literature.

We present below a simple one-dimensional flow with open boundaries that exhibits chaotic dynamics.  The system consists of a line of point particles interacting with nearest neighbors according to a linear force law (Hooke's law). These particles travel towards an inlet at constant velocity, pass into a region where they are free to move according to their nearest neighbor interactions, and then pass an outlet where they are driven so that they have a sinusoidally varying velocity.  This outlet driving force continually supplies energy to the system and energy is dissipated when particles exit at any point away from their equilibrium position.  As the amplitude of the outlet oscillations is increased, we find that the resident time of particles between the inlet and outlet follows a bifurcating route to chaos.  We discuss the resulting dynamics of this system and suggest possible implications for larger dimensional flows.

\section{A Simple One-Dimensional Discrete Flow}
\begin{figure}[htb]
\begin{center}
\includegraphics[width=4in]{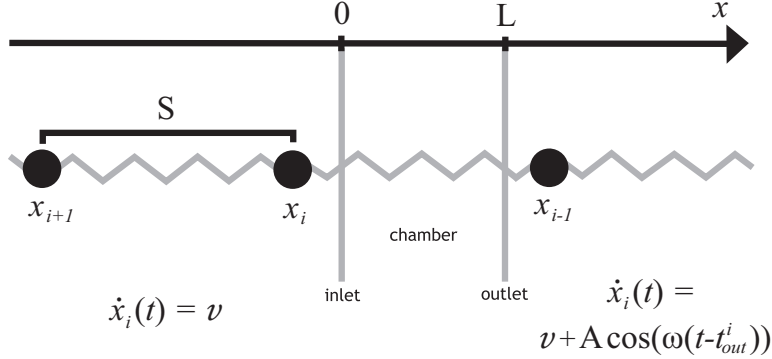}
\end{center}
\caption{\label{fig.system} Diagram of the system.  A series of point particles are connected by ideal springs and initially spaced $S$ apart.  Before reaching $x=0$ (the inlet) and after passing $x=L$ (the outlet), each particle is constrained to the velocities shown.  A particle moves according to its nearest neighbor interactions when between these points.}
\end{figure}
The system consists of a one-dimensional chain of identical point particles spaced a distance $S$ apart and connected by ideal springs.  The particles travel in the $\hat{x}$-direction and their positions are labeled $x_{i}$, where $i$ is the index $i=1,2,\dots,N$.  The number of particles $N$ is chosen large enough so that the flow is sustained throughout our simulations. Particles undergo different dynamics as they pass certain points along the flow. Fig.~\ref{fig.system} is a schematic of the system.

The inlet is located at $x=0$, the outlet at $x=L$, and the region between is referred to as the chamber.  The initial spacing between particles, $S$, is chosen large enough so that only one particle is located in the chamber at any moment in time.  The time when particle $i$ reaches $x=0$ is labeled $t^{i}_{in}$ and is calculated $t^{i}_{in}=i S/v$.  The time when particle $i$ reaches $x=L$ is labeled $t^{i}_{out}$ and is determined implicitly from the equation $x_{i}(t^{i}_{out})=L$.  The velocity of a particle before and after these times is constrained as given in Eqs.~(\ref{eq.vin},\ref{eq.vout}). Particle velocities are not constrained between $t^{i}_{in}$ and $t^{i}_{out}$.
\begin{subequations}
\begin{eqnarray}
\label{eq.vin}
\dot{x}_{i}=v &  & t < t^{i}_{in},\\
\label{eq.vout}
\dot{x}_{i}=v+A \cos\left[\omega (t-t^i_{out})\right] & & t\geq t^{i}_{out}.
\end{eqnarray}
\end{subequations}
Here $A$ and $\omega$ are the amplitude and frequency of velocity oscillations after reaching the outlet.  

If we consider nearest neighbor interactions for all particles, then the forcing functions required for these constraints are nontrivial.  This is simplified greatly by assuming that particles before the inlet and after the outlet are not influenced by their neighbors; the result is that there is no forcing for particles before the inlet, and an oscillatory forcing for particles, $F_i=-A\omega\sin\left[\omega(t-t^i_{out})\right]$, after the outlet.  This asymmetry in interaction might exist in real flows in two ways.  First, if the flow consists of agents and not particles, then the chamber simply represents a region where agents (who normally follow simple velocity patterns) suddenly become concerned about nearest neighbors.  A physical example of this might be the boundary between two regions of traffic flow, where drivers who are normally unconcerned about the cars around them change behavior to coincide with neighboring vehicles at this boundary.  Second, for continuous flow, the regions of flow directly before the inlet and directly after the outlet might contain much more mass than the flow within the chamber.  This means that flow within the chamber would react to the motion of neighboring regions, but once outside, would move with little regard for what is occurring in the chamber.

The system as a whole could be replicated experimentally by placing the string of particles on two separate conveyors that are separated by a distance $L$ and that lock the particles in place - the first conveyor operates at constant velocity as in Eq.~\ref{eq.vin} and the second oscillates such that particles are driven according to Eq.~\ref{eq.vout}.

The following are the complete equations of motion for the particles.
\begin{subequations}
\begin{eqnarray}
x_{i}=v t-i S & & t < t^{i}_{in},\\
m \ddot{x}_{i}+2k x_{i}=k(x_{i-1}+x_{i+1}) & & t^{i}_{in}\leq t < t^{i}_{out},\\
\dot{x}_{i}=v+A \cos\left[\omega (t-t^i_{out})\right] & & t\geq t^{i}_{out}.
\end{eqnarray}
\end{subequations}

All particles have the same mass, $m$, and the linear restoring force, $k$, is the same for all springs. The solution for the position of the $i^{th}$ particle for times $t^{i}_{in}\leq t \leq t^{i}_{out}$ is given in Eq.~(\ref{eq.xsol}). It is used to solve $x_{i}(t^{i}_{out})=L$ implicitly for $t^{i}_{out}$ using Newton's method.
\begin{eqnarray}
\label{eq.xsol}
x_{i}(t) & = & B_1 \cos(\sqrt{2/\alpha}t)+B_2 \sin(\sqrt{2/\alpha}t) + \\
 & & \frac{A}{\omega(2-\alpha\omega^2)}\sin\left[\omega(t-t^{i-1}_{out})\right]+vt+\frac{L-vt^{i-1}_{out}-S}{2},
\end{eqnarray}
where,
\begin{subequations}
\begin{eqnarray}
\alpha \; & = & m/k \\
B_1 & = & -\frac{L-vt^{i-1}_{out}-S}{2}+\frac{A}{\omega(2-\alpha\omega^2)}\sin(\omega t^{i-1}_{out}) \\
B_2 & = & -\frac{A}{\sqrt{2/\alpha}(2-\alpha\omega^2)}\cos(\omega t^{i-1}_{out}).
\end{eqnarray}
\end{subequations}

\section{Chaotic Dynamics of the System}
In all of the following results, we use the following parameters: $S=100$, $L=2$, $v=10$, $\alpha=.06$, $\omega=13.2$, $A=0$ to $40$.  These parameter values ensure that, at most, only one particle is located in the chamber at any moment in time.

\begin{figure}[htb]
\begin{center}
\includegraphics[width=4in]{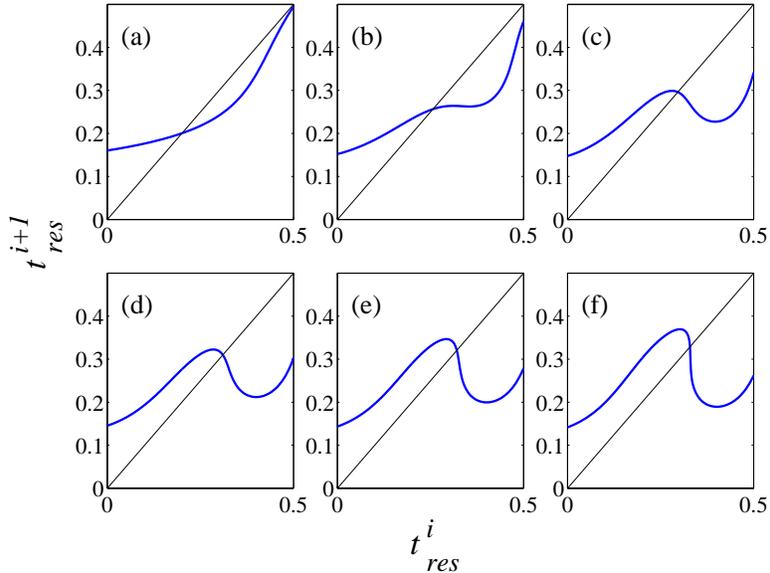}
\end{center}
\caption{\label{fig.returnmap} Return map for $t^{i}_{res}$ for several values of the amplitude, (a) $A=0$, (b) $A=15$, (c) $A=25$, (d) $A=30$, (e) $A=35$, (f) $A=40$.  Parameters for this simulation were: $S=100$, $L=2$, $v=10$, $\alpha=.06$, $\omega=13.2$.}
\end{figure}

For each particle we calculate the resident time within the chamber. This quantity is defined,
\begin{equation}
t^{i}_{res} = t^{i}_{out}-t^{i}_{in}.
\end{equation}
$t^i_{in}$ is determined from Eq.~\ref{eq.vin} and $t^{i}_{out}$ is determined by numerically solving Eq.~\ref{eq.xsol} for $x_{i}(t^{i}_{out})=L$.
Fig.~\ref{fig.returnmap} shows several return maps for the resident time (these map $t^{i}_{res}$ to $t^{i+1}_{res}$).  In general $t^{i+1}_{res}=f(t^{i}_{res},t^{i-1}_{res},...)$, but for this system, 
\begin{equation}
t^{i+1}_{res}=f(t^{i}_{res}), 
\end{equation}
i.e., the return map is one-dimensional.  This results from constraining particle velocities before $x=0$ and after $x=L$ - reducing the system to only one degree of freedom.

\begin{figure}[htb]
\begin{center}
\includegraphics[width=5in]{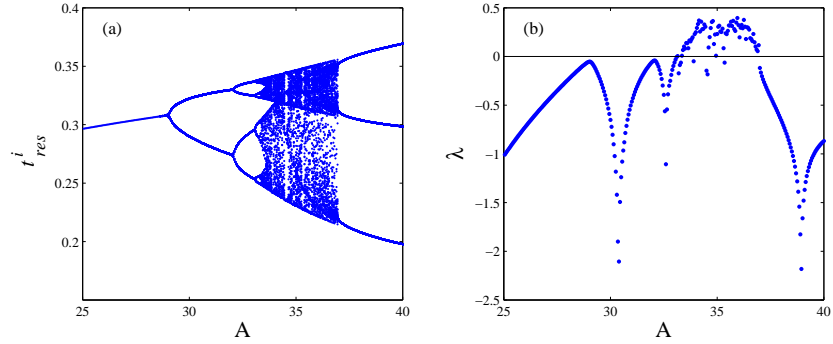}
\end{center}
\caption{\label{fig.bifurcation_lyapunov} (a) Bifurcation diagram of $t^{i}_{res}$ plotted for values of $A$ from 25 to 40. (b) The Lyapunov exponent, $\lambda$, as a function of the amplitude of outlet oscillations, $A$, also from 25 to 40.  Parameters for this simulation were: $S=100$, $L=2$, $v=10$, $\alpha=.06$, $\omega=13.2$.  The first three bifurcation points are located at $A=29.0$, $A=32.1$, and $A=33.1$; and $\lambda>0$ first at $A=33.4$}
\end{figure}

As $A$ is increased, a peak develops in the return map and the fixed point eventually becomes unstable.  The one-hump map that develops is likely to exhibit the features of many other unimodal return maps - specifically a bifurcation route to chaos \cite{Feigenbaum78}. As Fig.~\ref{fig.bifurcation_lyapunov}a shows, this is indeed the case. The system is iterated onto the attractor and the next 100 values for $t^{i}_{res}$ are plotted for values of the amplitude, $A$, from 25 to 40. The resident time bifurcates several times and eventually becomes chaotic before settling back to a period three dynamics.  The first three bifurcation points are located at $A=29.0$, $A=32.1$, and $A=33.1$.

In Fig.~\ref{fig.bifurcation_lyapunov}b we plot the Lyapunov exponent $\lambda$ for the resident time as a function of the amplitude $A$.  The Lyapunov exponent is a measure of the separation of infinitesimally close trajectories and in this case is calculated numerically from the following equation,
\begin{equation}
\lambda=\lim_{n \rightarrow \infty}\frac{1}{n}
\sum_{i=0}^{n-1}{\log\left|f'(t^{i}_{res})\right|}.
\end{equation}
When $\lambda>0$ trajectories exponentially diverge, which produces chaos when the trajectories remain bounded.  The system becomes chaotic at $A=33.4$ where $\lambda$ first turns positive.

\section{Conclusions}
The system we present above is quite different than other one-dimensional particle models in the literature \cite{Peyrard02, Braun98}.  Instead of using nonlinear interactions between particles, the particles in our system interact with linear forces and constraints are applied abruptly at the boundaries.  This shows that complex motion can arise in a flow at the boundary between simple constrained motions without the need for nonlinear interactions between particles.  Many large scale flows contain regions where the dynamics are tightly constrained to regular motion, with complex motion occuring at the boundaries between these regions.  Simple models such as the one we have presented can provide insight into how this behavior develops.

Summarizing, we have presented a fully describable one-dimensional flow of point particles connected by ideal springs.  Particle motion is constrained before reaching an inlet and after passing an outlet, and the system is shown to exhibit chaotic dynamics when particles are driven sinusoidally after crossing the outlet.  The outlet driving force continually adds energy to the system.  No drag force is present, but energy is dissipated when particles exit at any point away from their equilibrium positions.  The model can be reduced to a one-dimensional map that produces chaotic dynamics, showing that chaos can occur in flows at the boundary between simple constrained motion, even when particles in the flow interact with linear forces.

\backmatter

\chapter{Author's Biography}

Austin Nathaniel Gerig was born on March 16, 1978 in Fort Wayne, Indiana and lived in the surrounding area until 1996 when he enrolled at the University of Notre Dame.  He completed a bachelors degree in physics at Notre Dame in the Spring of 2000 and entered the physics graduate program at the University of Illinois at Urbana-Champaign the following year.  In 2002, he married his college sweetheart, Molly Kathryn Syron, and in 2004 they had their first child, a son, Elliot Lloyd Gerig.  While attending UIUC, Austin completed a masters in physics in 2002 and a masters in finance in 2006.  He started working with Alfred H\"{u}bler in 2002 researching simple models of mechanical systems.  In 2005 he traveled with Alfred to the Santa Fe Institute to assist with a summer school course.  While at the Santa Fe Institute, he collaborated with J. Doyne Farmer on several econophysics projects.  He was awarded a Graduate Fellowship at SFI for 2006-2007 and continued working with Doyne, finishing his thesis in the area of econophysics.

\end{document}